\documentclass{jfm}

\usepackage{graphicx}
\usepackage{newtxtext}
\usepackage{newtxmath}

\usepackage[caption=false,listofformat=parens,subrefformat=subsimple,labelformat=parens]{subfig}
\usepackage{floatrow} 

\usepackage{multirow} 

\usepackage{natbib}
\usepackage{hyperref}
\usepackage[noabbrev]{cleveref}
\hypersetup{
    colorlinks = true,
    urlcolor   = blue,
    citecolor  = black,
}

\newcommand{\RomanNumeralCaps}[1]
\linenumbers

\newcommand{\pcref}[1]{(\cref{#1})}
\newcommand{\pcreftwo}[2]{(figures \ref{#1},\subref*{#2})}

\newcommand{\lapp}{\nabla^2_\mathbf{p}}

\newcommand{\pavg}{\langle \mathbf{p} \rangle}

\newcommand{\psth}[1]{\langle p_{#1} \rangle}

\newcommand{\delp}{\bnabla_\mathbf{p}}
\newcommand{\delx}{\bnabla_\mathbf{x}}

\newcommand\Pe{\mbox{\textit{Pe}}}  

\title{A local approximation model for macroscale transport of biased active Brownian particles in a flowing suspension}

\author{Lloyd Fung\aff{1}
  \corresp{\email{lloyd.fung12@imperial.ac.uk}},
  Rachel N. Bearon\aff{2}
 \and Yongyun Hwang\aff{1}}

\affiliation{\aff{1}Department of Aeronautics, Imperial College London, London SW7 2AZ, UK
\aff{2}Department of Mathematical Sciences, University of Liverpool, Liverpool L69 7ZL, UK}

\begin{document}
\maketitle

\begin{abstract}
	A dilute suspension of motile microorganisms subjected to a strong ambient flow, such as algae in the ocean, can be modelled as a population of non-interacting, orientable active Brownian particles (ABPs).
	Using the Smoluchowski equation (i.e. Fokker-Planck equation in space and orientation), one can describe the non-trivial transport phenomena of ABPs such as taxis and shear-induced migration.
	This work transforms the Smoluchowski equation into a transport equation, in which the drifts and dispersions can be further approximated as a function of the local flow field. The new model can be applied to any global flow field due to its local nature, unlike previous methods such as those utilising the generalised Taylor dispersion theory.
	The transformation shows that the overall drift includes both the biased motility of individual particles in the presence of taxis and the shear-induced migration in the absence of taxis. In addition, it uncovers other new drifts and dispersions caused by the interactions between the orientational dynamics and the passive advection-diffusion of ABPs.
	Finally, the performance of this model is assessed using examples of gyrotactic suspensions, where the proposed model is demonstrated to be most accurate when the biased motility of particles (i.e. taxis) is weak.
\end{abstract}



\section{Introduction}
\label{sec:intro}

\noindent The transport of orientable microorganisms or particles in a suspension is of fundamental importance for many ecological, medical and engineering applications.
For example, the non-trivial macroscopic transport of motile species of phytoplankton is responsible for the formation of an ecological hotspot \citep{Durham2009,Durham2013}. The shape-dependent sedimentation of non-motile species in turbulent water \citep{Voth2017} may also be a significant factor affecting the carbon sequestration process in the ocean, a crucial process in the Earth's carbon cycle. On the medical front, modelling the transport of bacteria helps us to understand how they spread or propagate collectively on surfaces through swarming \citep[see review by][]{Koch2011}. In engineering, controlling the transport of bottom-heavy algal species in bioreactors may improve biofuel harvesting \citep{Croze2013}.

The macroscopic transport of particles is a key component in modelling the complex dynamics and rich collective behaviour of a suspension. While individual particles can be modelled with the Stokes equations given their small size (1-10 $\mu$m for bacteria and 10-100 $\mu$m for algae), the emergent behaviour and the flow environment are often at a larger length scale than the individuals \citep[see review by][]{Koch2011,Elgeti2015,Clement2016,Bees2020}. The difference in length scale poses a significant challenge to the modelling of their collective behaviour.

Some of these emergent behaviours are the result of many-body interactions between particles. For example, bacterial turbulence and spontaneous self-organisation of bacterial suspension in confinement \citep{Dombrowski2004,Wioland2013} are usually found in dense suspension where near-field hydrodynamics and nematic alignment are the driving mechanisms \citep{Costanzo2012,Wioland2013,Lushi2014}. Phenomenological models \citep[e.g.][]{Wensink2012,Dunkel2013,Slomka2017} are often used to describe the collective phenomenon in these dense suspensions when it is challenging to derive a model from the microscopic dynamics via a `bottom-up' approach. 
On the other hand, in dilute or semi-dilute suspensions where interactions between particles are predominantly hydrodynamic, Stokesian Dynamics \citep[e.g.][]{Brady1988,Sierou2001} or other Stokesian-based simulations \citep[e.g.][]{Ishikawa2008,Delmotte2015,Schoeller2018} are often used to model the collective dynamics of self-propelling particles.
However, these numerical tools do not apply to large-scale fluid systems where the inertial force becomes important in the dynamics (e.g. bioconvection and turbulence).

This work focuses on the collective phenomena in the dilute limit, i.e.~when the length scale of a particle is much smaller than the distance between particles such that the volume fraction is vanishing. Emergent behaviours in dilute suspensions are usually the result of the non-trivial transport caused by the biased motility (e.g. `taxis' for microorganisms) or sedimentation of orientable particles stemming from the flow-field-dependent orientation of individual particles. For example, bioconvection \citep[see review by][]{Bees2020}, unmixing \citep{Durham2013} and gyrotactic shear trapping \citep{Durham2009} are the results of the balancing influence of external fields (e.g. gravity, light or chemical gradient) and the flow field on the particles' orientation. Other phenomena such as shear trapping \citep{Ezhilan2015,Bearon2015,Vennamneni2020} and enhanced sedimentation \citep{Clifton2018} are the results of the alignment of elongated particles with the flow field. These macroscopic phenomena in the dilute limit can be derived from the microscopic trajectories of particles. Here, a `bottom-up' derivation is preferred because it offers an explicit link between the micro- and macroscale phenomena, helping us to better understand the physics behind the phenomena.

Microscopically, a Langevin equation is often used to describe a particle's positional and rotational dynamics in the presence of thermal and/or biochemical noise. The typical approach to model the rotational dynamics of particles is to combine Jeffery's Orbit \citep{Jeffery1922,Bretherton1962,Hinch1972a,Hinch1972}, which governs how the local vorticity and strain rate orient a particle, with other orientational biases responsible for the taxes such as gyrotaxis \citep[e.g.][]{Pedley1990}, phototaxis \citep[e.g.][]{Drescher2010a} and chemotaxis \citep[e.g.][]{Alt1980}.
If the suspension is dilute enough, the hydrodynamic stresses and forces from the particles to the flow and the hydrodynamic interactions between them are negligible and can be foregone. In that case, one may simulate them as (biased) active Brownian particles (ABPs): i.e. simulating an individual particle's positional and rotational trajectory through the Langevin equation while prescribing a flow field that is assumed to be not affected by the presence of other particles (Brownian dynamics simulation).

Alternatively, one can employ the statistical (continuum) equation which describes the probability density function of the ABPs governed by the Langevin equation \citep{Doi1988}. Some \cite[e.g.][]{Saintillan2018} refer to the equation as the Fokker-Planck equation, but in this work, this equation is referred to as the Smoluchowski equation so as not to be confused with the Fokker-Planck (FP) model introduced only for the particle orientation dynamics in earlier studies \citep{Pedley1990}.
While the Smoluchowski equation captures how the local flow field affects the trajectories of the particles, the hydrodynamic contribution of the particles can be later coupled with the flow equation through some averaged description of the bulk stresses and forces the particles exert on the flow \citep[e.g.][]{Batchelor1970,Hinch1972a,Hinch1972,Pedley1990}. This approach was detailed in \cite{Saintillan2015}. 

The continuum approach has several advantages over the individual-based approach. Firstly, the continuum approach is computationally more scalable as the cost of the individual-based method scales with the number of particles while the continuum method does not. Secondly, individual-based methods such as the Brownian dynamics simulation neglect the hydrodynamic contribution of the particles to the flow. For example, in bioconvection, buoyancy force from the particles is important even when the suspension is dilute, but it is difficult to capture the buoyancy force through individual-based simulations \citep[see][]{Bees2020}. While a particle-laden approach can be used to incorporate the hydrodynamic presence of the particles, the method comes with a significantly higher computational cost. Thirdly, it is easier to analyse the continuum solutions than their individual-based counterparts. For example, it is difficult to perform a more formal analysis for the instability generated by the pusher stresslets in an active suspension  without a continuum description \citep[see][]{Saintillan2008}.

Despite the descriptive merit of the Smoluchowski-based continuum model, there are only a handful of works that utilise a direct numerical simulation of this equation \citep[e.g.][]{Chen1999,Saintillan2008,Saintillan2010,Jiang2020}. Instead, many resort to the equivalent Brownian dynamics simulations. This is because the equation has a high number of dimensions (spatial, orientational and temporal), which renders the individual-based simulations more computationally attractive than the numerical simulations of the continuum equation.
However, as mentioned, the computational advantage of individual-based simulations is restricted to suspensions with a prescribed flow field. If the particles are exerting non-negligible forces or stresses on the flow, such as in the case of bioconvection, a continuum model is more attractive from both analytical and computational perspectives.

Several past works attempted to further reduce the computational cost of the Smoluchowski equation by modelling it with an approximated transport equation for motile particles.
For example, the Fokker-Planck (FP) model attempted to postulate the effective transport coefficients (i.e. drift and diffusivity) through the quasi-steady orientational distribution of particles at each location in the flow field \citep{Pedley1990,Pedley2010a}. Others attempted to employ the results from the generalised Taylor dispersion (GTD) theory in an unbounded homogeneous shear flow \citep{Frankel1991,Frankel1993,Hill2002,Manela2003,Bearon2003} to approximate the local behaviour of the particles suspended in a more general flow field \citep[e.g.][]{Bearon2011,Bearon2012,Croze2013,Croze2017,Fung2020a,Fung2020b}, although the original GTD theory is not meant to be used for such a purpose. These past works attempted to reduce the Smoluchowski equation by reducing the number of dimensions resolved simultaneously while keeping intact the dependency of transport properties on the flow field. The present work is formulated to achieve a similar goal by seeking a model equation for the transport of biased or non-biased ABPs. In particular, we propose a new transport equation that approximates the number density directly from the higher-dimensional Smoluchowski equation.

Recent works such as \citet{Croze2013,Croze2017} and \citet{Fung2020a} showed that the FP model of \cite{Pedley1990} is not as accurate as the GTD-based model utilising the local flow information in a unidirectional flow, especially at high shear rates. This is because the effective diffusion in the FP model is a phenomenological postulation with an \textit{ad hoc} constant for unknown diffusion time scale. Also, it is based on the Fokker-Planck equation for orientational distribution only, not the Smoluchowski equation like in the GTD-based model. Despite this merit, the application of the GTD theory in \cite{Hill2002} and \cite{Manela2003} was for unbounded homogeneous shear flows only.
When these models of were applied to a bounded inhomogeneous shear flow by \cite{Bearon2012}, they implicitly assumed a quasi-homogeneity of the given flow field. Therefore, this approach is more precisely a quasi-homogeneous approximation of the GTD theory developed for homogeneous shear flow. \cite{Jiang2019,Jiang2020} later demonstrated the rigorous applications of the GTD theory in both unbiased and biased ABP suspensions with inhomogeneous shear. In their works, the cross-stream direction was solved simultaneously with the orientation (local variables) while only the averaged streamwise dispersion was obtained in the streamwise direction (global variables). While \citeauthor{Jiang2019} had demonstrated the correct application of the GTD theory, their application remained specific to the parallel flow field of the problem. 

Alternatively, Brady and colleagues have established procedures to obtain the drift and dispersion of passive colloids \citep{Zia2010} and non-biased ABPs \citep{Takatori2014,Takatori2017,Peng2020} in the Fourier space. In the small-wavenumber limit, their works have demonstrated an alternative route to derive an advection-diffusion equation equivalent to that of GTD in the Fourier space (compare \cite{Takatori2017} with \cite{Hill2002} or \cite{Manela2003}).

This work aims to propose a new transport model for the biased or non-biased active Brownian particles that is independent of the nature of the global flow field. In particular,
we seek a model that can approximate the effective transport properties, such as the effective drift and dispersion, as a function of the local flow field. By writing the coefficients of the transport equation as a function of the local flow field, we aim to develop an approach that is generally applicable to any complex flow field, instead of re-evaluating such coefficients in each specific type of global flow field \citep[e.g.][]{Jiang2019}. We show that the Smoluchowski equation admits an exact transformation into a transport equation. Combining this transformation with the method of multiple scales, we propose a novel transport equation, in which the orientation dynamics is determined only by the local flow information in the physical space. 
Note that there is an important distinction between the GTD theory and the present model. In the GTD theory, one seeks effective drift and diffusion coefficients that are constant at macroscopic scale or in the homogeneous directions, while in the present method, we relax such requirement and seek effective drift and dispersion as a local variable that depends on the local flow field. More discussion on the difference between the present method and the GTD theory follows in \S\ref{subsec:GTD}.

This work is organised as follows. In \S\ref{sec:background}, the Smoluchowski equation is presented with the equations governing the motion of active (or swimming) Brownian particles. We also briefly introduce the GTD model and how it differs from the present work. In \S\ref{sec:exact_transform}, the exact transformation of the Smoluchowski equation into a transport equation is introduced. While the transformed equation cannot be directly used as a model, it sets up the mathematical platform for a further approximation. 
In \S\ref{sec:asymp}, the local approximation is presented for the development of a novel transport equation model. In \S\ref{sec:examples}, we present several examples using gyrotactic particle suspensions. We also demonstrate the accuracy of the newly introduced model by comparing the approximation with the full solution of the Smoluchowski equations in both one-dimensional shear flows and two-dimensional convective cells. In \S\ref{sec:Discussion}, we further dissect the physical interpretation of the transformation and discuss the physical origin of the many drifts and dispersions from the transformation. Lastly, in \S\ref{sec:conclusion}, we briefly outline the potential application of the local approximation and the remaining challenges in the proposed model. 

\section{Background\label{sec:background}}
\subsection{The Smoluchowski equation\label{sec:non_dim}}
We consider a dilute suspension of active Brownian particles (ABPs), where there is randomness in both the physical space $\mathbf{x}^*$ and orientational space $\mathbf{p}$. In this study, the term ABP is used to refer to a self-propelling particle (or microswimmer) subject to a translational and/or rotational random walk. Given the stochastic nature of the trajectory, 
we consider the probability density function $\Psi(\mathbf{x}^*,\mathbf{p},t^*)$ for particles located at $\mathbf{x}^*$ with orientation $\mathbf{p}$ at time $t^*$. The equation for $\Psi(\mathbf{x}^*,\mathbf{p},t^*)$ is governed by the Smoluchowski equation \citep{Doi1988}:
\begin{equation}
\frac{\partial \Psi}{\partial t^*}+\bnabla^*_\mathbf{x} \cdot \left[ \dot{\mathbf{x}}^* \Psi  - D_T^* \bnabla^*_\mathbf{x} \Psi \right] + \delp \cdot \left[ \dot{\mathbf{p}}^* \Psi - d_r^* \delp \Psi \right] =0,
\label{eq:smol_original_dim}
\end{equation}
where the deterministic motion for $\dot{\mathbf{x}}^*$ is governed by 
\begin{equation}
	\dot{\mathbf{x}}^*= \mathbf{u}^*+ V_c^*\mathbf{p}. \label{eq:xdot_dim}
\end{equation}
Here, the superscript ($\cdot)^*$ represents dimensional variables or parameters, $\mathbf{u}^*$ is the prescribed flow velocity and $V_c^*\mathbf{p}$ the velocity of particles by active motion (swimming/motility). 
Meanwhile, the deterministic form of orientational dynamics for $\dot{\mathbf{p}}^*$ is governed by
\begin{equation}
	\dot{\mathbf{p}}^*=\left(\frac{\boldsymbol{{\Omega}}^*(\mathbf{x})}{2} \wedge \mathbf{p}+\alpha_0 \mathbf{p} \bcdot \mathsfbi{E}^* \bcdot (\mathsfbi{I} - \mathbf{pp}) \right) + \frac{1}{2B^*} \left[ \mathbf{k} - (\mathbf{k} \cdot \mathbf{p})\mathbf{p} \right]. \label{eq:pdot_dim}
\end{equation}
Here, we assume that the particle is oriented by the local flow through the Jeffery's equation \citep{Jeffery1922,Bretherton1962}, 
where $\boldsymbol{{\Omega}}^* = \bnabla^*_\mathbf{x} \wedge \mathbf{u}^*$ is the vorticity, $\mathsfbi{E}^*=(\bnabla^*_\mathbf{x} \mathbf{u}^*+\bnabla^*_\mathbf{x} {\mathbf{u}^*}^T)/2$ the rate-of-strain tensor and $\alpha_0$ the Bretherton constant. In this work, we also consider gyrotaxis of the given particles \citep{Pedley1990}, as they will be used in the flow examples in \S\ref{sec:examples}. Gyrotaxis is given by the second term on the right-hand side of (\ref{eq:pdot_dim}), where $\mathbf{k}$ is the unit vector pointing upwards (against gravity) and $B^*$ the gyrotactic time scale.

In (\ref{eq:smol_original_dim}), we have also assumed that the random motions in $\mathbf{x}$- and $\mathbf{p}$-space can be modelled as translational diffusion with the corresponding diffusivity $D_T^*$ and rotational diffusion with the diffusivity $d_r^*$, respectively. The translational diffusion $D_T^*$ often originates from thermal fluctuation especially for small particles. However, it is often negligible for many microorganisms (e.g. microalgae), given their relatively large size. In this study, we keep this term without loss of generality, so that proposed framework can be extended to other types of particles. 

Equation (\ref{eq:smol_original_dim}) is subsequently non-dimensionalised with a suitable length scale and time scale. In this work, the characteristic length $h^*$ is chosen from the given flow field, and the inverse of rotational diffusivity $1/d_r^*$ is selected as the time scale. For convenience, we also use the characteristic speed $U^*$ of the flow for the non-dimensionalisation. Hence,
$$
\mathbf{x}=\frac{\mathbf{x}^*}{h^*}, \quad 
t={t^* d_r^*}, \mbox{and} \quad \mathbf{u}=\frac{\mathbf{u}^*}{U^*}.
$$
The dimensionless parameters for the speed of motility (swimming) $V_s^*$, the flow speed $U^*$, the translational diffusivity ${D}_T^*$ and the gyrotactic timescale $B^*$ are
$$
\Pe_s=\frac{V_s^* }{h^*d_r^*}, \quad
\Pe_f=\frac{U^* }{h^*d_r^*}, \quad
{D}_T=\frac{{D}_T^*}{(h^*)^{2} d_r^*}, \quad \mbox{and} \quad \beta=\frac{1}{2 d_r^* B^*},
 \label{eq:non_dimensionalise}
$$
respectively, in which $\Pe_f$ and $\Pe_s$ are the ambient flow and motility P\'{e}clet numbers. The dimensionless form of (\ref{eq:smol_original_dim}) is then given by
\begin{equation}
	\frac{\partial \Psi}{\partial t}+\nabla_\mathbf{x} \cdot \left[ (\Pe_f \mathbf{u}+\Pe_s \mathbf{p}) \Psi \right] + \mathcal{L}_p(\mathbf{x},t) \Psi=  {D}_T \nabla^2_x \Psi,
	\label{eq:smol}
\end{equation}
where 
\begin{equation}
	\mathcal{L}_p(\mathbf{x},t)\Psi =\delp \bcdot \left[ \left(\frac{\Pe_f}{2} \boldsymbol{\Omega} \wedge \mathbf{p}+\Pe_f \alpha_0 \mathbf{p} \bcdot \mathsfbi{E} \bcdot (\mathsfbi{I} - \mathbf{pp})  + \beta \left[ \mathbf{k} - (\mathbf{k} \cdot \mathbf{p})\mathbf{p} \right] \right) \Psi \right]-\lapp \Psi. \\ \label{eq:Lp_op}
\end{equation}

By the divergence theorem, the integration over $\mathbf{p}$-space of the operator $\mathcal{L}_p(\mathbf{x},t)$ acting on any arbitrary continuously differentiable function $a(\mathbf{p})$ satisfies
\begin{equation}
	\int_{S_p} \mathcal{L}_p(\mathbf{x},t) \, a(\mathbf{p}) \; d^2 \mathbf{p}=0, \label{eq:p-space-divergence}
\end{equation}
where $S_p$ is the unit sphere, i.e. the $\mathbf{p}$-space subject to $\|\mathbf{p}\|=1$. We note that this will be used as a solvability condition for inversion of the operator later (see eq. (\ref{eq:smol_Leinv})). 
Physically, it is related to the conservation of probability in $\mathbf{p}$-space. 
We also note that (\ref{eq:Lp_op}) may be modified to account for other taxes by including the relevant modelling terms, e.g. the run-and-tumble dynamics and chemotaxis \citep{Subramanian2009} or phototaxis \citep{Williams2011}. Therefore, we expect that many deterministic models for the orientation dynamics in $\mathbf{p}$-space would be given as a linear operator $\mathcal{L}_p(\mathbf{x},t)$ that satisfies (\ref{eq:p-space-divergence}). 
In the following sections, we use the linear operator $\mathcal{L}_p(\mathbf{x},t)$ to represent the orientation dynamics in $\mathbf{p}$-space to maintain this level of abstraction in the orientation dynamics. 

\subsection{Comparison with the generalised Taylor dispersion theory}\label{subsec:GTD}
The generalised Taylor dispersion theory was originally derived by \cite{Brenner1980} as a generalisation of the seminal work by \cite{Aris1956} on the axial dispersion of cross-sectional mean solute concentration in a flowing pipe. The same framework was later extended by \cite{Frankel1989,Frankel1991,Frankel1993} to homogeneous shear flow and more kinds of colloids. This framework approximates the underlying high-dimensional microscopic transport equation, such as (\ref{eq:smol_original_dim}), with a lower-dimensional macroscopic transport equation, such as an effective advection-diffusion equation for the active particles. The key step in the GTD theory is in the identification of the local and global variables, in which the macroscopic transport equation only spans the global space while the local space and its interaction with the global space are approximated through Aris' method of moments in the global space in a long-time limit. 

However, the GTD theory was developed under the notion that at the macroscopic scale, there is a homogeneity in the global space when the transport properties by local flow are averaged out. Hence, the theory only seeks constant macro-transport coefficients in the global space. By comparison, this work aims to relax the requirement of constant transport coefficients in the physical space $\mathbf{x}$. Instead, we seek drift and dispersion coefficients that can be written as a function of the local flow field, which may not be homogeneous in $\mathbf{x}$. Although both theories would yield some form of a transport equation for the ABPs in a dilute suspension, the scope of this work is different from that of the GTD theory. By providing spatially varying transport coefficients, the transport model from the present method is applicable to any scales at which the flow may be inhomogeneous, while the GTD theory only gives effective transport properties (or coefficients) at macroscopic scale after coarse-graining the transport by local flows at smaller scales. Moreover, by approximating the transport coefficients as a function of the local flow field (i.e. the local spatial derivatives of the flow velocity), the present approach offers a universal model independent of the global flow configuration for given ABPs. Therefore, the model derived from this work is applicable to any continuous flow fields. This is in contrast to the GTD theory, where the coefficients of macroscopic transport equation have to be re-derived whenever the flow configuration changes \citep[e.g.][]{Jiang2019}.

\section{Exact transformation into a transport equation \label{sec:exact_transform}}
The purpose of this work is to obtain a new transport model for biased or non-biased ABPs in a suspension only using the local flow information. In particular, the transport properties (drift and dispersion coefficients) are given at each position of $\mathbf{x}$, unlike in the previous work such as \cite[]{Jiang2019,Jiang2020} and \cite{Peng2020}. We start by seeking an exact mathematical transformation of (\ref{eq:smol}) into a transport equation that may well be subjected to a complex flow field. The resulting equation will remain exact, but dependent on the exact probability density function $\Psi(\mathbf{x},\mathbf{p},t)$ in both orientational and spatial spaces. Then, based on the same transformation, in \S\ref{sec:asymp} we further approximate the transport equation, so that the orientational distribution can be obtained using only the local flow information given at the spatial location $\mathbf{x}$. 

We define $n(\mathbf{x},t)$ and $f(\mathbf{x},\mathbf{p},t)$ as $\Psi(\mathbf{x},\mathbf{p},t)= n(\mathbf{x},t)f(\mathbf{x},\mathbf{p},t)$, so that $f(\mathbf{x},\mathbf{p},t)$ at each $\mathbf{x}$ becomes the probability density function in $\mathbf{p}$-space satisfying $\int_{S_p} f(\mathbf{p}) d^2 \mathbf{p}=1$. 
Now, from (\ref{eq:p-space-divergence}), integration of (\ref{eq:smol}) over $\mathbf{p}$-space gives the following equation in the $(\mathbf{x},t)$ space:
\begin{equation}
\partial_t n(\mathbf{x},t)+  \bnabla_{\mathbf{x}} \bcdot [(\Pe_s \pavg_f(\mathbf{x},t) + \Pe_f \mathbf{u}(\mathbf{x},t)) n(\mathbf{x},t)] = {D}_T \nabla^2_x n(\mathbf{x},t), \label{eq:smol_inte}
\end{equation}
where
\begin{equation}
\pavg_f(\mathbf{x},t) \equiv  \int_{S_p} \mathbf{p} f(\mathbf{x},\mathbf{p},t) d^2 \mathbf{p}.
\end{equation}
Here, we are effectively taking the zeroth-order moment of the Smoluchowski equation. Although (\ref{eq:smol_inte}) appears as a standard advection-diffusion equation, it is not solvable in the absence of the full information of $\Psi(\mathbf{x},\mathbf{p},t)$ because $\pavg_f(\mathbf{x},t)$ is still unknown. The term $\pavg_f(\mathbf{x},t)$ can be interpreted as the normalised polar-order parameter. Its time evolution can be obtained by taking the first-order moment of (\ref{eq:smol}). However, the equation again involves the second-order moment (or the nematic-order parameter) of the probability density function, which requires a closure model for the problem to be solved \citep{Saintillan2015}. This is a general limitation in this type of approach (the tensor-harmonic expansion). 

\begin{table}
	\begin{tabular}{c  l }
		Terms                                                                            & Physical meaning  \\
		                                                                                                     \hline
		$n \; \partial_t f$                                                              & Unsteadiness of $f$ in $\mathbf{p}$-space \\ [2pt]
		$\Pe_f n \mathbf{u} \bcdot \bnabla_{\mathbf{x}} f$ & Passive advection of $f$ in $\mathbf{x}$ by the ambient flow $\mathbf{u}$ \\ [2pt]
		$ - {D}_T (\nabla_x^2 f) n$                                                      & Translational diffusion of $f$ in $\mathbf{x}$ \\ [2pt]
		{$ - 2 {D}_T (\bnabla_{\mathbf{x}} f) \bcdot (\bnabla_{\mathbf{x}} n)$}          & Cross-translation diffusion in $\mathbf{x}$ between $n$ and $f$ \\ [2pt]
		\multirow{2}{*}{$\Pe_s (\mathbf{p}f - \pavg_f f) \bcdot \bnabla_{\mathbf{x}} n$} & Change in $f$ induced by motility and \\
		                                                                                 & gradient of number density in $\mathbf{x}$ \\	[2pt]
		\multirow{2}{*}{$\Pe_s n (\mathbf{p}\bcdot \bnabla_{\mathbf{x}} f - f \bnabla_{\mathbf{x}}\bcdot \pavg_f)$}  &  Change in $f$ induced by \\
																						& motility and inhomogeneity of $f$ in $\mathbf{x}$ 
	\end{tabular}
	\caption{Physical meaning of each term in equation (\ref{eq:smol_subtracted})\label{tab:phys_smol_sub}}
\end{table}

Instead of utilising the equations for higher-order moments, here we take a different approach based on an exact transformation of (\ref{eq:smol}) into a different form of transport equation. 
In the following, we transform the Smoluchowski equation by taking the inverse of operator $\mathcal{L}_p$. However, in order to ensure each term in the equation satisfies the solvability condition in (\ref{eq:p-space-divergence}), we must first 
multiply (\ref{eq:smol_inte}) by $f(\mathbf{x},\mathbf{p},t)$ and subtracting it from the Smoluchowski equation (\ref{eq:smol}), which gives
\begin{eqnarray}
  n \; \partial_t f
&+& (\Pe_f \mathbf{u} \bcdot \bnabla_{\mathbf{x}} f - {D}_T \nabla_x^2 f) n - 2 {D}_T (\bnabla_{\mathbf{x}} f) \bcdot (\bnabla_{\mathbf{x}} n) \nonumber \\
&+& \Pe_s (\mathbf{p}f - \pavg_f f) \bcdot \bnabla_{\mathbf{x}} n
+\Pe_s n (\mathbf{p}\bcdot \bnabla_{\mathbf{x}} f - f \bnabla_{\mathbf{x}}\bcdot \pavg_f) \nonumber \\
&+& n  \mathcal{L}_p(\mathbf{x},t) f  =0, \label{eq:smol_subtracted}
\end{eqnarray}
Now, each term in (\ref{eq:smol_subtracted}) satisfies the solvability condition (\ref{eq:p-space-divergence}), and may be interpreted physically as described in table \ref{tab:phys_smol_sub}. Note that (\ref{eq:smol_subtracted}) is the real-space equivalent of equation (13) in \mbox{\cite{Peng2020}} before they applied the small-wavenumber approximation. Up to this point, the Fourier-based method of \mbox{\cite{Peng2020}} is somewhat similar to the transformation in this study. However, the two methods diverge from this point onwards.

Next, we introduce a new set of variables $f_\star$ and $\mathbf{b}_\star$, with $(\cdot)_\star$ indicating any subscript below, by the following set of linear equations:
\begin{subequations} \label{eq:f_b} 
\begin{eqnarray}
 \mathcal{L}_p(\mathbf{x},t) f_u(\mathbf{x},\mathbf{p},t) & = & \Pe_f \mathbf{u} \bcdot \bnabla_{\mathbf{x}} f , \label{eq:fu}\\
  \mathcal{L}_p(\mathbf{x},t) f_{D_T}(\mathbf{x},\mathbf{p},t) & = & - {D}_T \nabla_x^2 f, \label{eq:fDT}\\
 \mathcal{L}_p(\mathbf{x},t) \mathbf{b}_{D_T}(\mathbf{x},\mathbf{p},t) & = & - 2 {D}_T \bnabla_{\mathbf{x}} f, \label{eq:bDT}\\
 \mathcal{L}_p(\mathbf{x},t) \mathbf{b}_c(\mathbf{x},\mathbf{p},t) & = & \Pe_s(\mathbf{p} - \pavg_f) f, \label{eq:bc}\\
 \mathcal{L}_p(\mathbf{x},t) f_c(\mathbf{x},\mathbf{p},t)  & = & \Pe_s ( \mathbf{p} \bcdot \bnabla_{\mathbf{x}} f - f \bnabla_{\mathbf{x}} \bcdot \pavg_f), \label{eq:fc} \\
 \mathcal{L}_p(\mathbf{x},t) f_{\partial t}(\mathbf{x},\mathbf{p},t)  & = & \partial_t f \label{eq:fdt}.
\end{eqnarray} 
\end{subequations}
Each new variable $f_\star$ and $\mathbf{b}_\star$ corresponds to a term in (\ref{eq:smol_subtracted}) or table \ref{tab:phys_smol_sub}.
Note that the introduced variables are still functions of both $\mathbf{x}$ and $t$, because $\mathcal{L}_p$ can have coefficients varying in $\mathbf{x}$ and $f$ depends on both $\mathbf{x}$ and $t$.
With the introduced variables, (\ref{eq:smol_subtracted}) can be rewritten as
\begin{equation}
 \left[\mathcal{L}_p (\mathbf{b}_{D_T} + \mathbf{b}_c) \right] \bcdot \bnabla_{\mathbf{x}} n(\mathbf{x},t)
+ \left[\mathcal{L}_p (f_u+f_{D_T}+f_c+f_{\partial t} + f) \right] n(\mathbf{x},t) =0. \label{eq:smol_substituted}
\end{equation}
Now, we are ready to take the inverse of the operator $\mathcal{L}_p$.
This leads to
\begin{equation}
	\left[\mathbf{b}_{D_T} + \mathbf{b}_c \right] \bcdot \bnabla_{\mathbf{x}} n(\mathbf{x},t)
   + \left[f_u+f_{D_T}+f_c+f_{\partial t} + f \right] n(\mathbf{x},t) =n(\mathbf{x},t) \, g, \label{eq:smol_Leinv}
\end{equation}
where $f_\star$ and $\mathbf{b}_\star$ are defined such that they are subject to the integral condition $\int_{S_p} f_\star d^2  \mathbf{p}=0$ and $\int_{S_p} \mathbf{b}_\star d^2 \mathbf{p}=\mathbf{0}$, while the homogeneous solution $g(\mathbf{x},t;\mathbf{p})$, defined by 
\begin{subequations}\label{eq:FK_g}
	\begin{eqnarray}
	\mathcal{L}_p(\mathbf{x},t) g(\mathbf{x},t;\mathbf{p}) &=& 0 \quad \mbox{subject to}\\
	\int_{S_p} g(\mathbf{x},t;\mathbf{p}) d^2 \mathbf{p} &=& 1,
\end{eqnarray}
\end{subequations}
is added to the equation after being multiplied by $n(\mathbf{x},t)$. The multiplication factor $n(\mathbf{x},t)$ can be obtained by integrating (\ref{eq:smol_Leinv}) over $\mathbf{p}$-space. 

\begin{table}
	\begin{tabular}{c  l }
		Terms & Physical meaning  \\ \hline
		$\pavg_g$ & Averaged motility of individual particle from the homogeneous solution of $\mathcal{L}_p$ \\ [2pt]
		\multirow{2}{*}{$\mathbf{V}_{\partial t}$} & Drift due to interaction between particles' orientational dynamics \\ & and the unsteadiness of $f$ in $\mathbf{p}$-space \\ [2pt]
		\multirow{2}{*}{$\mathbf{V}_{u}$} & Drift due to interaction between particles' orientational dynamics \\ & and passive advection of $f$ in $\mathbf{x}$ by the flow field $\mathbf{u}$ \\ [2pt]
		 \multirow{2}{*}{$\mathbf{V}_{c}$}  & Drift due to interaction between particles' motility \\ & and the inhomogeneity of particles' orientational dynamics in $\mathbf{x}$ \\ [2pt]
		 \multirow{2}{*}{$\mathbf{V}_{D_T}$} &  Drift due to interaction between particles' orientational dynamics \\ & and translational diffusion of $f$ in $\mathbf{x}$ \\ [2pt]
		\multirow{2}{*}{$\mathsfbi{D}_{D_T}$} & Dispersion from interaction between particles' orientational dynamics \\ 
		&  and the dispersion of $n$ and $f$ due to translational diffusion of $f$ and $n$ \\ [2pt]
		{$\mathsfbi{D}_{c}$} & Dispersion due to interaction between particles' motility and orientational dynamics
	\end{tabular}
	\caption{Physical meaning of each derived term in equation (\ref{eq:smol_coarse_final})\label{tab:phys_smol_coarse}}
\end{table}

Multiplying $\mathbf{p}$ by (\ref{eq:smol_Leinv}) and integrating in $\mathbf{p}$-space then yields 
\begin{equation}
	(\mathsfbi{D}_{D_T}+ \mathsfbi{D}_c) \bcdot \bnabla_{\mathbf{x}} n
	+ \left[ \mathbf{V}_{u}+ \mathbf{V}_{D_T}+\mathbf{V}_c + \mathbf{V}_{\partial t} + \pavg_{f} \right] n =n \pavg_{g}, \label{eq:smol_full_coarseG}
\end{equation}
where
\begin{equation}
	\pavg_g(\mathbf{x},t) \equiv  \int_{S_p} \mathbf{p} g(\mathbf{x},t;\mathbf{p}) d^2 \mathbf{p},
\end{equation}
\begin{equation}
\mathbf{V}_{\star}(\mathbf{x},t)  =  \int_{S_p} \mathbf{p} f_{\star}(\mathbf{x},\mathbf{p},t) d^2 \mathbf{p}, \label{eq:V_ast}
\end{equation}
\begin{equation}
\mathsfbi{D}_{\star} (\mathbf{x},t)  =  \int_{S_p} \mathbf{p} \mathbf{b}_{\star}(\mathbf{x},\mathbf{p},t) d^2 \mathbf{p} \label{eq:D_ast}
\end{equation}
with $(\cdot)_\star$ indicating any of the subscripts used in (\ref{eq:f_b}). Here, this step is equivalent to expressing $n \langle \mathbf{p} \rangle_g$ in terms of $n \langle \mathbf{p} \rangle_f$ and the other related terms, in which $n \langle \mathbf{p} \rangle_g$ and $n \langle \mathbf{p} \rangle_f$ may be interpreted as the first-order Lagrangian and Eulerian moments, respectively (see \S\ref{sec:Discussion_trans} for a further discussion of their physical interpretations). However, it should be distinguished from the one with the typical tensor-harmonic expansion \cite[e.g.][]{Saintillan2015}, in which the Smoluchowski equation (\ref{eq:smol}) was used to derive the time-evolving equation for the polar or nematic order. Indeed, in such an approach, the equation for each moment has a closure problem, as it requires the information involving higher-order moments. On the contrary, an equation such as (\ref{eq:smol_Leinv}) only defines the relation between $n \langle \mathbf{p} \rangle_g$ and $n \langle \mathbf{p} \rangle_f$, and it is not a time-evolution equation that can be solved only with the information from higher-order moments.

Replacing $n \pavg_{f}$ in (\ref{eq:smol_inte}) with that of (\ref{eq:smol_full_coarseG}) leads to the following transport equation:
\begin{eqnarray}
 & & \partial_t  n+  \bnabla_{\mathbf{x}} \bcdot 
 \left[ (\Pe_f \mathbf{u}  + \Pe_s (\pavg_{g}-\mathbf{V}_u-\mathbf{V}_{D_T}- \mathbf{V}_{c} -\mathbf{V}_{\partial t} ) ) n \right] \nonumber \\ 
 & = & D_T \nabla_x^2 n + \Pe_s \bnabla_{\mathbf{x}} \bcdot (\mathsfbi{D}_{D_T}+ \mathsfbi{D}_c) \bcdot \bnabla_{\mathbf{x}} n. \label{eq:smol_coarse_final}
\end{eqnarray}
This is an exact transport equation directly obtained from (\ref{eq:smol}) without making any approximations. Here, we note that previous studies, including the FP model \citep{Pedley1990}, the GTD model \citep{Hill2002,Manela2003} and the Fourier method for ABPs \citep{Takatori2017,Peng2020} used $\pavg_g$ to describe the average drift from the particle motility. However, it is now seen that the average motility of individual particles $\pavg_g$ does not necessarily constitute all the drifts (a further discussion of this issue follows in \S\ref{sec:Discussion_trans}).

Lastly, it should be mentioned that (\ref{eq:smol_coarse_final}) is not the only transport equation one can obtain from (\ref{eq:smol}) -- indeed, we have already retrieved a different form of transport equation from (\ref{eq:smol}), that is, (\ref{eq:smol_inte}).This is essentially the consequence of reducing the dimensions of the given system (\ref{eq:smol}) from the $(\mathbf{x},\mathbf{p})$-space to just $\mathbf{x}$-space. However, expression (\ref{eq:smol_Leinv}) directly derived from (\ref{eq:smol}) enables us to decompose $\pavg_f$ in (\ref{eq:smol_inte}) into the averaged motility of individual particles $\langle\mathbf{p}\rangle_g$ and the other terms from (\ref{eq:smol}) in a mathematically exact manner. Hence, each term in (\ref{eq:smol_coarse_final}) would admit a physical interpretation, as listed in table \ref{tab:phys_smol_coarse}. Finally, it is important to note that $\mathsfbi{D}_{D_T}$ and $\mathsfbi{D}_c$ in (\ref{eq:smol_coarse_final}) do not necessarily describe a diffusion process, as they are not guaranteed to be either symmetric or positive definite. Therefore, one should be careful in understanding and interpreting their actual roles, and, in this sense, (\ref{eq:smol_coarse_final}) may not precisely be referred to as an advection-diffusion equation. 

\section{Local approximation of the transformed transport equation \label{sec:asymp}}
While the transport equation in (\ref{eq:smol_coarse_final}) is obtained without applying any approximation to (\ref{eq:smol}), the formulae for $\mathbf{V}_\star$ and $\mathsfbi{D}_\star$ given in (\ref{eq:f_b}) are based on $f=\Psi/n$, requiring the full knowledge of $\Psi$ (i.e. the solution to (\ref{eq:smol})). 
Therefore, the transformation discussed in \S\ref{sec:exact_transform} does not alleviate the difficulty related to the computational cost of the full Smoluchowski equation (\ref{eq:smol}). To resolve this issue, in this section, we combine the transformation technique leading to (\ref{eq:smol_full_coarseG}) with a multiple-time-scale asymptotic analysis. This results in an approximated form of (\ref{eq:smol_coarse_final}) utilising only the local flow information (i.e. a local approximation). 

First, we assume $\Pe_s (\equiv \epsilon) \ll 1$,  $\Pe_f \lesssim O(\epsilon)$ and $D_T \lesssim O(\epsilon)$, and define $\tilde{\Pe}_f=\Pe_f / \epsilon$ and $\tilde{D}_T=D_T / \epsilon$. Physically, these assumptions imply that the timescale in the orientational $\mathbf{p}$-space is much faster than that in $\mathbf{x}$-space (i.e. a quasi-steady assumption). 
They would be valid for most of the phenomena of interest such as gyrotactic focusing and bioconvection, where the timescale for swimmers to swim across the container width or characteristic flow length $h^*/V_c^*$ is longer than the orientation timescale $1/d_r^*$, and the flow speed is comparable to the swimming speed from $\Pe_s \sim \Pe_f \sim O(\epsilon)$.
An alternative but equivalent interpretation of the assumptions is that the run length of the ABPs, $V_h^*/d_r^*$, is much smaller than the characteristic flow length.
Under these assumptions, the orientational component of $\Psi$ (i.e. $f(\mathbf{x},\mathbf{p},t)$) will first relax to a quasi-equilibrium in $\mathbf{p}$-space while the $\mathbf{x}$-dependency of $\Psi$ is still evolving slowly. This then enables us to introduce a slowly varying timescale $T=\epsilon t $ for the dynamics of $\Psi$ in $\mathbf{x}$-space. 
The standard multiple-scale asymptotic analysis is subsequently applied by expanding $\Psi=\Psi^{(0)}(t,T,\mathbf{x},\mathbf{e})+ \epsilon \Psi^{(1)}(t,T,\mathbf{x},\mathbf{e}) + \epsilon^2 \Psi^{(2)}(t,T,\mathbf{x},\mathbf{e})+O(\epsilon^3)$. Following a transformation similar to that in \S\ref{sec:exact_transform} and retaining the terms up to $O(\epsilon^2)$ (see appendix \ref{app:asymp_details} for further details), we derive an approximated transport equation given by
\begin{align}
 \partial_t n & + \bnabla_{\mathbf{x}} \bcdot \left[ \Pe_s \left(\pavg_{g}+\tilde{\Pe}_f \mathbf{u} \right) n-\Pe_s^2 \left(\mathbf{V}_{g,u}+\mathbf{V}_{g,D_T}+\mathbf{V}_{g,c}+\mathbf{V}_{g,\partial T} \right) n \right]  \nonumber \\ 
 & \approx  \Pe_s \tilde{D}_T \nabla^2_x n + \Pe_s^2 \bnabla_{\mathbf{x}} \bcdot  \left[ (\mathsfbi{D}_{g,c}+\mathsfbi{D}_{g,D_T}) \bcdot \bnabla_{\mathbf{x}}  n \right] \label{eq:asymp_coarse_final}
\end{align}
for the transport of $n(\mathbf{x},t)$, where the drifts and dispersion coefficients are defined by (\ref{eq:V_ast}-\ref{eq:D_ast}) with 
\begin{subequations}	\label{eq:f_b_g} 
	\begin{eqnarray}
	\mathcal{L}_p(\mathbf{x},T) f_{g,u}(\mathbf{x},T;\mathbf{p}) & = & \tilde{\Pe}_f \mathbf{u} \bcdot \bnabla_{\mathbf{x}} g , \label{eq:fu_g}\\
	\mathcal{L}_p(\mathbf{x},T) f_{g,D_T}(\mathbf{x},T;\mathbf{p}) & = & - \tilde{D}_T \nabla_x^2 g, \label{eq:fDT_g}\\
	\mathcal{L}_p(\mathbf{x},T) \mathbf{b}_{g,D_T}(\mathbf{x},T;\mathbf{p}) & = & - 2 \tilde{D}_T \bnabla_{\mathbf{x}} g, \label{eq:bDT_g}\\
	\mathcal{L}_p(\mathbf{x},T) \mathbf{b}_{g,c}(\mathbf{x},T;\mathbf{p}) & = & (\mathbf{p} - \pavg_g) g, \label{eq:bc_g}\\
	\mathcal{L}_p(\mathbf{x},T) f_{g,c}(\mathbf{x},T;\mathbf{p})  & = &  ( \mathbf{p} \bcdot \bnabla_{\mathbf{x}} g - g \bnabla_{\mathbf{x}} \bcdot  \pavg_g). \label{eq:fc_g} \\
	\mathcal{L}_p(\mathbf{x},T) f_{g,\partial T}(\mathbf{x},T;\mathbf{p})  & = & \partial_T g, \label{eq:fdt_g}
	\end{eqnarray}
\end{subequations}
where all $f_{g,\star}$ and $\mathbf{b}_{g,\star}$ are subjected to the integral condition $\int_{S_p} d^2 \mathbf{p}=0$. The approximated transport equation (\ref{eq:asymp_coarse_final}) is identical to (\ref{eq:smol_coarse_final}), except that their coefficients in (\ref{eq:f_b_g}) are now obtained by replacing $f$ in (\ref{eq:f_b}) with $g$ in (\ref{eq:FK_g}). This is a crucial advantage of (\ref{eq:asymp_coarse_final}) over (\ref{eq:smol_coarse_final}) because $g$ in (\ref{eq:FK_g}) can be solved pointwise at each $\mathbf{x}$ if the local flow information (i.e. $\boldsymbol{\Omega}$ and $\mathsfbi{E}$) is known. This is effectively a `local' approximation of $f$ using $g$, in which (\ref{eq:asymp_coarse_final}) no longer requires the full solution to (\ref{eq:smol}).  

Here, the derivation above is similar to that of \cite{Bearon2015} and \cite{Vennamneni2020}. However, in deriving (\ref{eq:asymp_coarse_final}), we have assumed $T=\epsilon t$. This time-scale separation is different from $T= \epsilon^2 t$ of \cite{Bearon2015} and \cite{Vennamneni2020}. We note that the $\mathbf{V}_{g,\star}$ and $\mathsfbi{D}_{g,\star}$ terms in (\ref{eq:asymp_coarse_final}) scale with $\Pe_s^2$, while the rest of the equation scales with $\Pe_s$. 
Therefore, the effect of these terms appears only at $O(\epsilon^2)$, while the rest of the terms are still non-zero at $O(\epsilon)$. This contrasts with the APB suspensions considered in \cite{Bearon2015} and \cite{Vennamneni2020}. In their cases, the translational diffusion was negligible ($D_T=0$), the averaged orientation of individual particles was not biased ($\pavg_g=\mathbf{0}$) and the flow was parallel such that $\mathbf{u} \bcdot \bnabla_{\mathbf{x}}=0$. Hence, if $T=\epsilon t$ was assumed, the equation at $O(\epsilon)$ would simply be degenerate. However, in general, there is no reason that the leading-order equation has to have such a trivial solution especially in the presence of taxes, translational diffusion or a non-parallel flow field. Therefore, these leading-order effects require us to retain the scaling $T=\epsilon t$. 

\section{Flow examples \label{sec:examples}}
Now, we test the accuracy of the local approximation model proposed in \cref{sec:asymp}. To this end, we numerically solve the transport equation for the number density from the local approximation model and compare it with the full solution to the Smoluchowski equation (\ref{eq:smol}). In \S\ref{sec:example_upright} and \S\ref{sec:example_HS}, we consider the suspension of bottom-heavy motile (i.e. gyrotactic) microorganisms in a one-dimensional parallel shear flow, where the number density is assumed to be uniform in the streamwise direction. This example was also considered in the recent work based on the GTD theory of \cite{Jiang2019,Jiang2020} and the Fourier method in \cite{Peng2020}. However, it is important to mention that the scope of the local approximation model in this study is different from theirs. 
In their work, the cross-stream direction and the orientation were solved simultaneously, while they seek a global effective drift and dispersion of a number density that were \emph{averaged} in the cross-stream direction. By contrast, the local approximation model in the present study solves the orientational space separately from the cross-streamwise direction, and the effective drifts and dispersions are defined at every cross-streamwise location from the exact transformation. It is therefore expected that the results from \cite{Jiang2019,Jiang2020} and \cite{Peng2020} must have a better accuracy than those from our local approximation model, as their approach is equivalent to computing the full solution of the Smoluchowski equation on the cross-streamwise plane. In essence, they are working on a coarse-graining level different from ours, and their approaches are computationally more costly.

Finally, in \S\ref{sec:example_2D}, we consider a two-dimensional Taylor-Green-type vortical flow in a periodic box. Since the flow is complex and inhomogeneous, the advantage of the local approximation will be demonstrated as it provides a good prediction for the steady number density without directly solving the Smoluchowski equation. 

\begin{figure}
	\centering{}
	\sidesubfloat[]{\includegraphics[width = 0.45 \columnwidth]{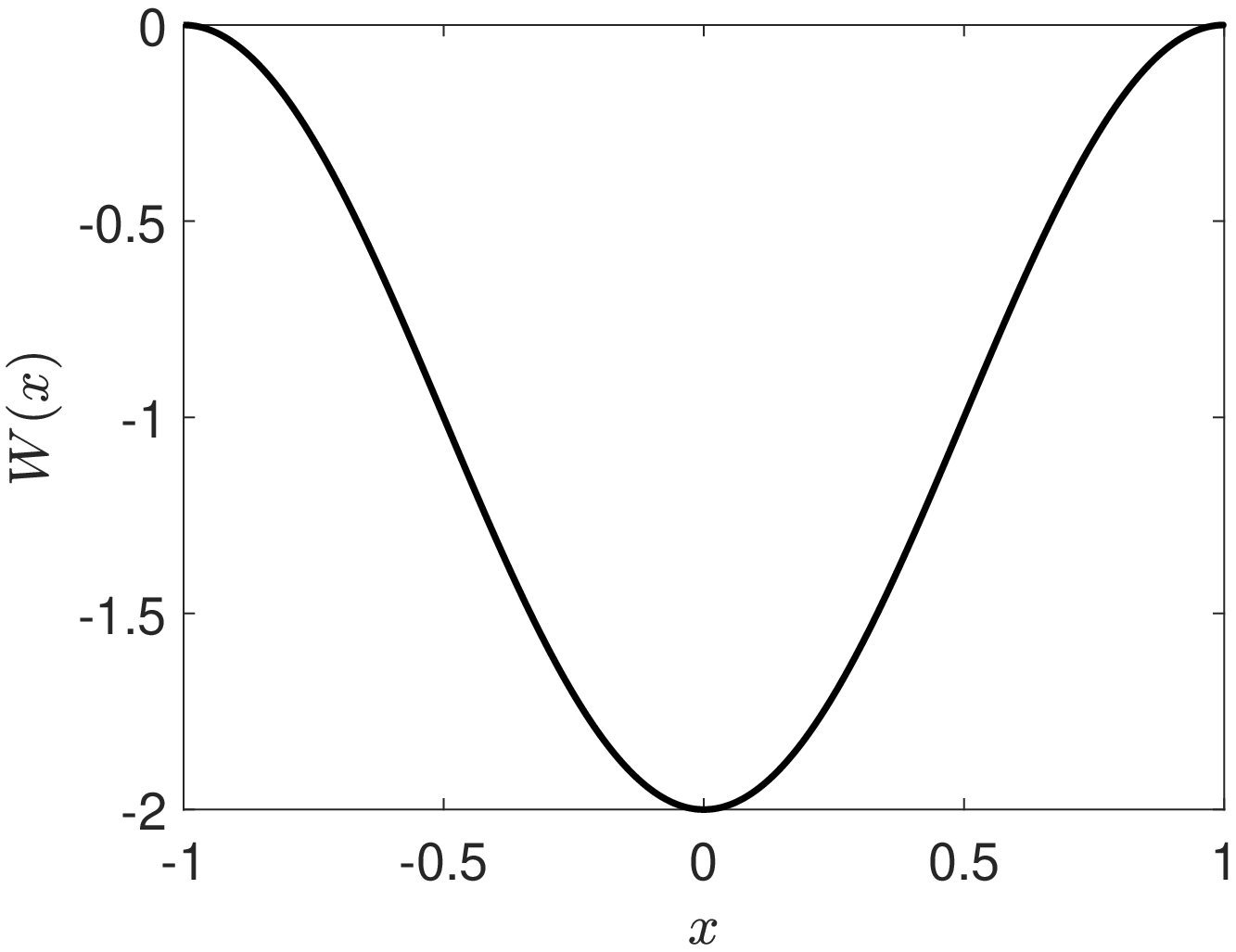}\label{fig:vertical_profile}}
	\sidesubfloat[]{\includegraphics[width = 0.45 \columnwidth]{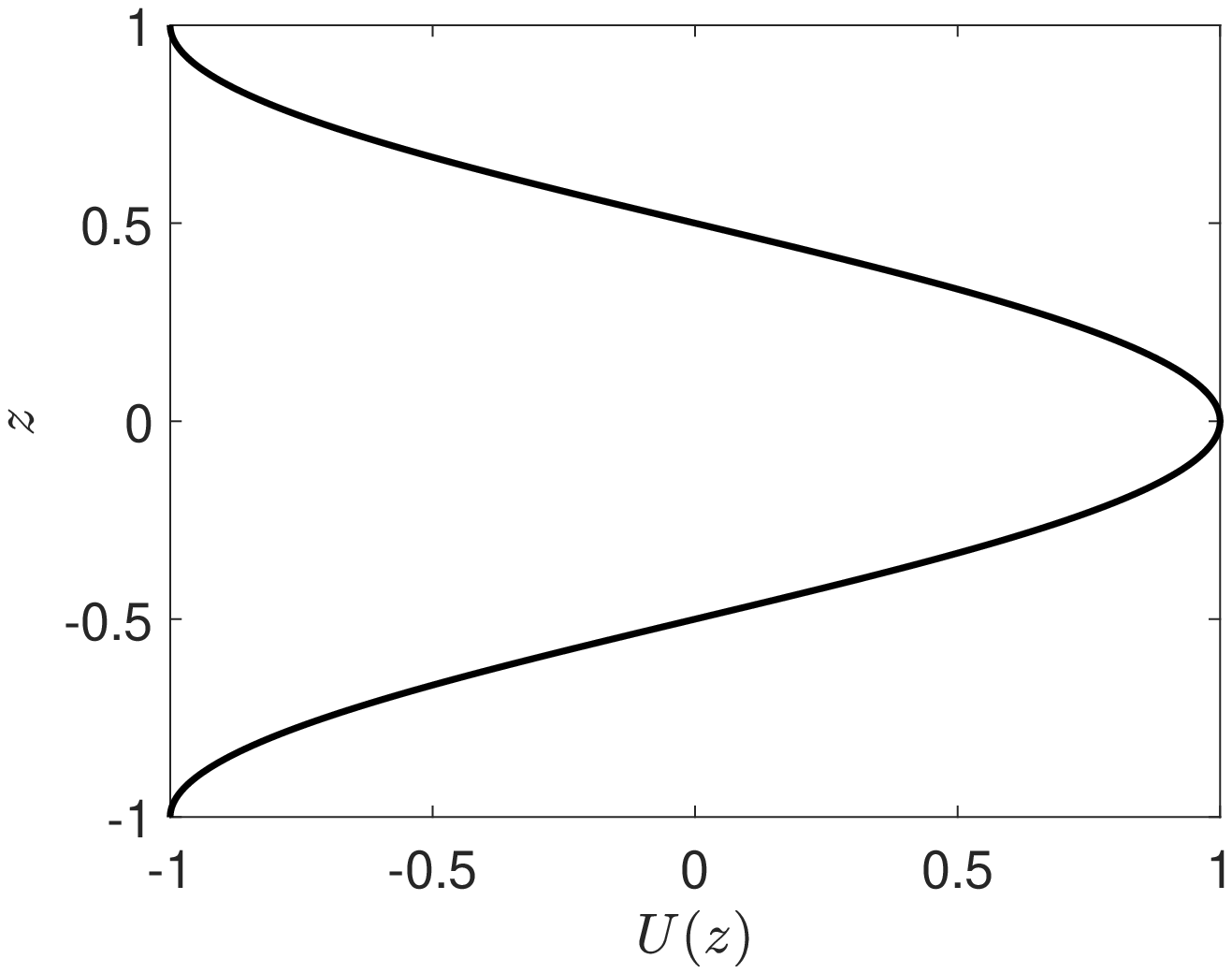}\label{fig:horizontal_profile}}
	\caption{The prescribed flow profile for the examples in $(a)$ \S\ref{sec:example_upright} and $(b)$ \S\ref{sec:example_HS}. In $(a)$, the vertical flow is $W(x)=-\cos(\pi x)-1$. In $(b)$, the horizontal flow is $U(z)=\cos(\pi z)$. \label{fig:flow_profile}}
\end{figure}

\subsection{Numerical method}
Our numerical method is loosely based on the Spherefun package \citep{Townsend2016}, which utilises the double Fourier sphere (DFS) method to represent the spherical space $\mathbf{p}$. The method transforms the longitude and latitude coordinates $(\phi,\theta) \in [-\pi,\pi] \times [0,\pi]$ into two independent Fourier space variables. Here, we follow the definition of \citet[][p. C405]{Townsend2016} and define $\phi$ and $\theta$ such that each component of $\mathbf{p}=[p_x,p_y,p_z]^T$ can be written as
\begin{equation}
	p_x=\cos{\phi}\sin{\theta}, \qquad p_y=\sin{\phi}\sin{\theta}, \qquad p_z=\cos{\theta}.
\end{equation}
Periodicity in the spherical space was maintained by enforcing the reflectional symmetry in its transformed coefficients \citep[see][p. C406]{Townsend2016}. The $\delp \cdot \left[ \dot{\mathbf{p}} \Psi \right]$ operation and the $\mathbf{p}$-dependent part of the $\delx \cdot \left[ \dot{\mathbf{x}} \Psi \right]$ operation in (\ref{eq:smol}) were completely implemented in the spectral space. Meanwhile, based on the parallel assumption in the physical space $\mathbf{x}$, we have discretised the cross-stream direction ($x$ or $z$, depending on the prescribed flow field) by a 6th order central difference scheme with an equispaced grid. Time integration was conducted semi-implicitly, in which the $\lapp$ term was advanced with the second-order Crank-Nicolson method while the rest are marched with a third-order low-storage Runge-Kutta method. The matrix inversion arising in the Crank-Nicolson method was solved using the algorithm for the Helmholtz equation in the Spherefun package. For simplicity, we have implemented a periodic boundary condition in the cross-stream direction. The method was validated by comparing the $\mathbf{p}$-space results with the previous solver used in \cite{Hwang2014a} and with the analytical solution of the following example. 

Since the numerical solution of the Smoluchowski equation is compared with the steady results from the local approximation model, we have also computed the effective drifts and dispersions of the model (\ref{eq:f_b_g}) by directly inverting the linear $\mathcal{L}_\mathbf{p}$ operator in the spectral space. The resulting effective drifts and dispersions are then used in the transport equation to solve for the steady solution of $n$ (see (\ref{eq:ns_exact}-\ref{eq:ns_asymp})). The method was also validated with the previous solver used to compute the drifts and diffusivity from the FP and GTD model \citep{Fung2020a}. 

\subsection{A suspension of gyrotactic active particles in a prescribed vertical flow \label{sec:example_upright}}

\begin{figure}
	\centering{}
	\sidesubfloat[]{\includegraphics[width = 0.45 \columnwidth]{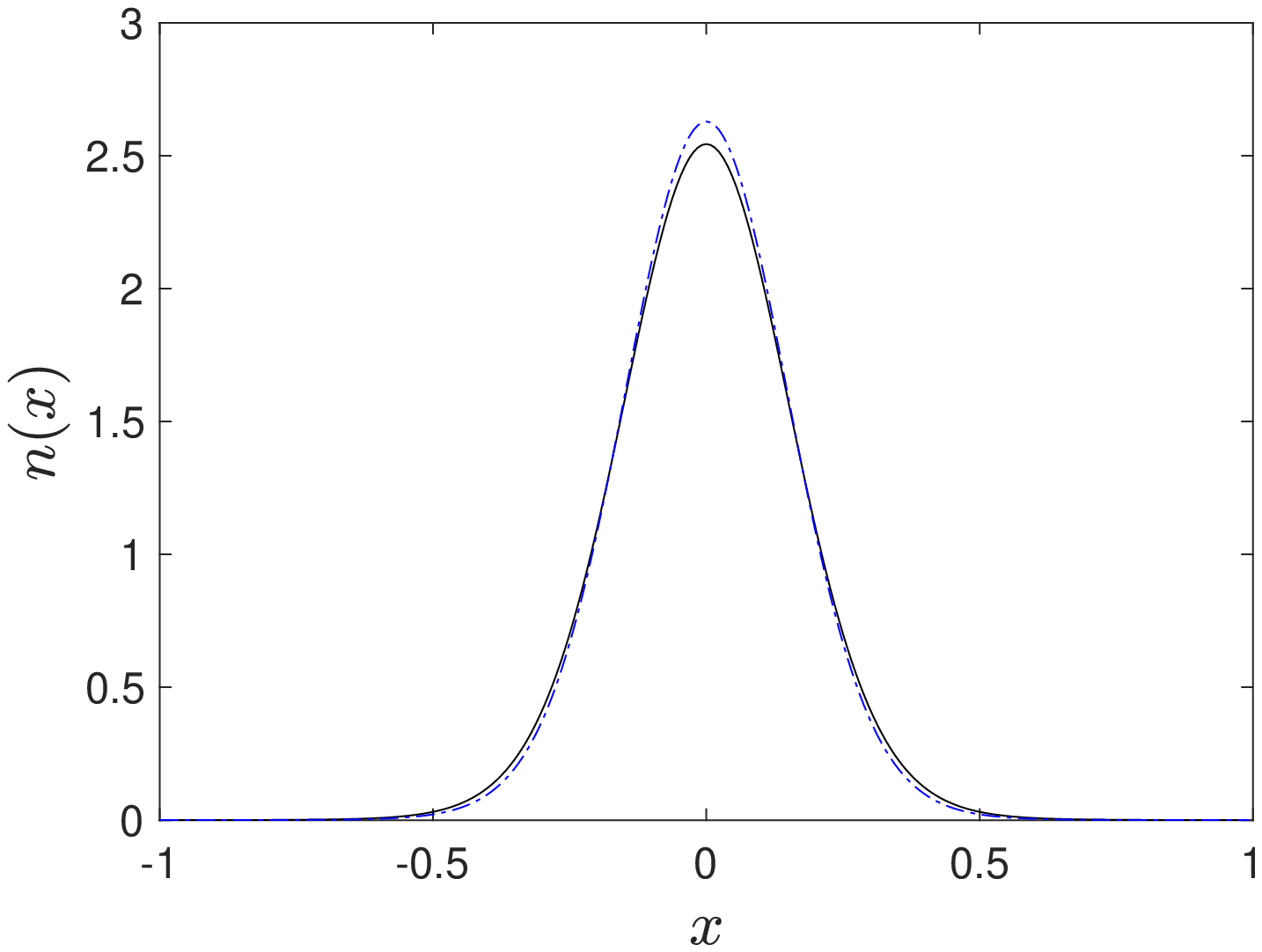}\label{fig:nx_sphere}}
	\sidesubfloat[]{\includegraphics[width = 0.45 \columnwidth]{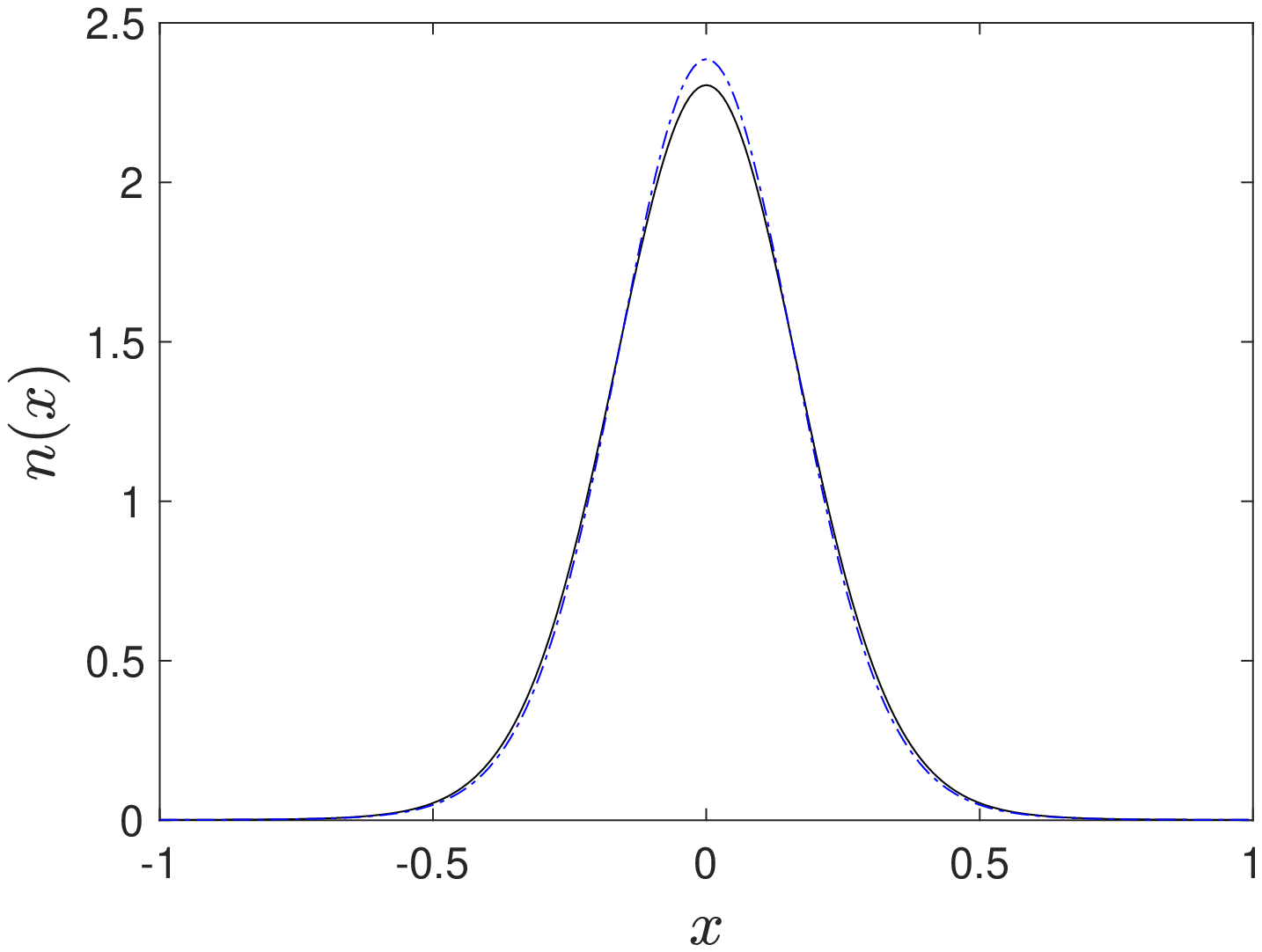}\label{fig:nx_strong_Pef1}}\\
	\sidesubfloat[]{\includegraphics[width = 0.45 \columnwidth]{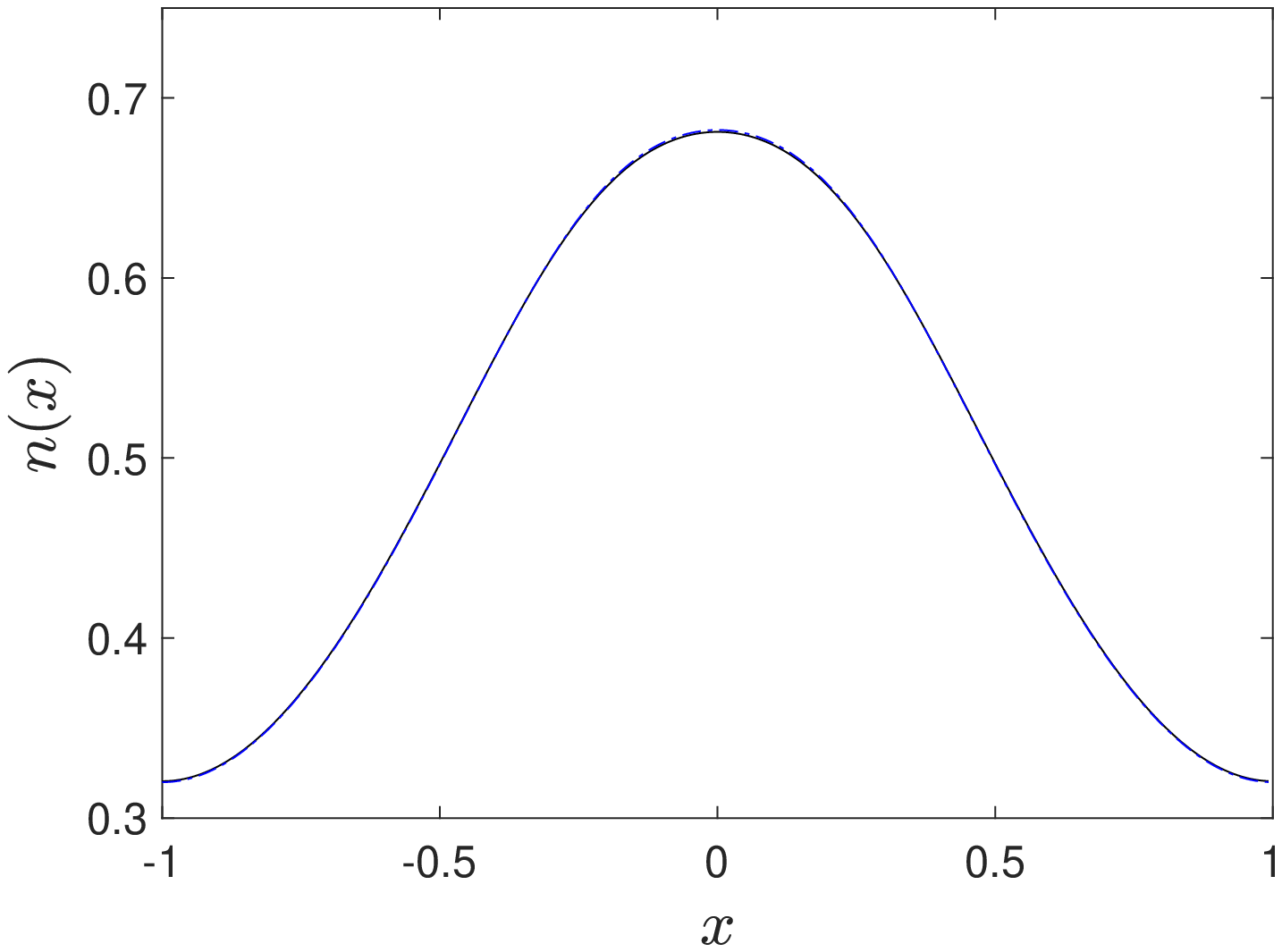}\label{fig:nx_weak_Pef1}}
	\sidesubfloat[]{\includegraphics[width = 0.45 \columnwidth]{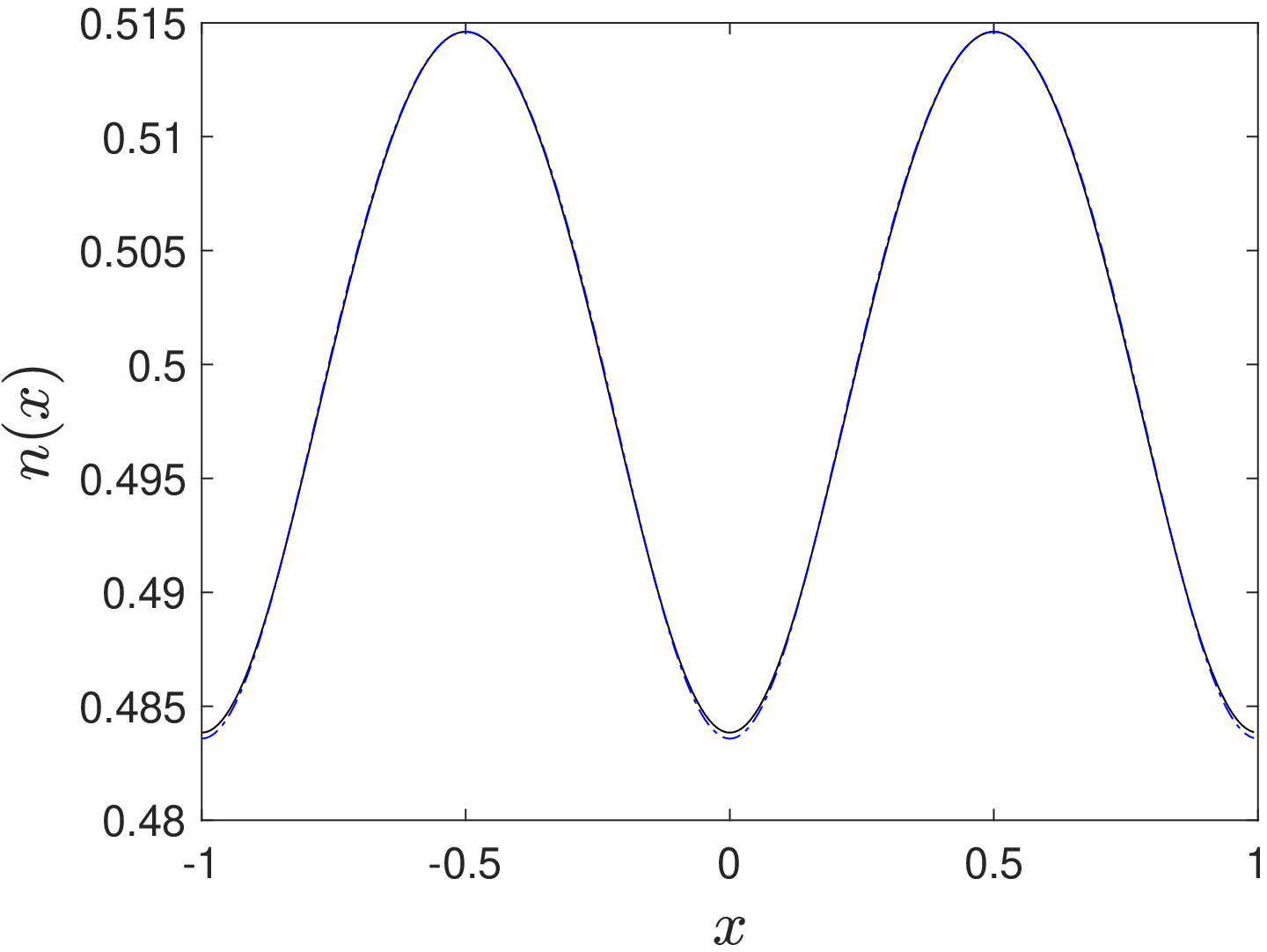}\label{fig:nx_nongyro}}
	\caption{Comparison of the steady-state number density given by the direct integration of (\ref{eq:smol}) (black solid line, $n_{f,s}$) and the local approximation model of \S\ref{sec:asymp} (blue dot-dashed line, $n_{g,s}$) of suspensions of $(a)$ spherical and strongly gyrotactic ($\beta=2.2$, $\alpha_0=0$), $(b)$ non-spherical and strongly gyrotactic ($\beta=2.2$, $\alpha_0=0.31$), $(c)$ non-spherical and weakly gyrotactic ($\beta=0.21$, $\alpha_0=0.31$) and $(d)$ non-spherical and non-gyrotactic ($\beta=0$, $\alpha_0=0.31$) particles. The suspensions are subjected to a vertical flow $W(x)=-\cos(\pi x)-1$ with $\Pe_s=0.25$ and $\Pe_f=1$. Note that the vertical scale for $n(x)$ in $(c,d)$ is much smaller than that in $(a,b)$. \label{fig:nx_steady}}
\end{figure}

In this example, we revisit the classical problem of the formation of a gyrotactic plume by bottom-heavy motile microorganisms/particles \citep[][]{Kessler1986a,Hwang2014,Jiang2020,Fung2020a,Fung2020b}. For simplicity, we do not take into account how the particles may influence the flow via buoyancy or hydrodynamic interactions. Instead, we apply a prescribed parallel shear flow to the suspension $\mathbf{u}(\mathbf{x})=[0,0,W(x)]^T$, in which $x$ is the horizontal direction and $z$ is the vertical direction pointing upwards (i.e. the same direction as $\mathbf{k}$).

Four types of idealised motile microorganisms are considered: a strongly gyrotactic and spherical particle ($\beta=2.2$, $\alpha_0=0$), a strongly gyrotactic and non-spherical particle ($\beta=2.2$, $\alpha_0=0.31$), a weakly gyrotactic non-spherical particle ($\beta=0.21$, $\alpha_0=0.31$) and a non-gyrotactic and non-spherical particle ($\beta=0$, $\alpha_0=0.31$). The parameters $\beta=2.2$ and $\alpha_0=0.31$ for the strongly gyrotactic particle are based on \textit{Chlamydomonas augustae} \citep{Pedley1990,Croze2010}, while the gyrotactic parameter $\beta=0.21$ for the weakly gyrotactic particle is based on \textit{Dunaliella salina} \citep{Croze2017}. Since we cannot find any experimental value of $\alpha_0$ for \textit{D. salina}, we assume the weakly gyrotactic particle shares the same value of $\alpha_0=0.31$ for comparisons. Lastly, we also consider a suspension of non-spherical and non-gyrotactic particles for completeness.

In \S\ref{sec:V_compare} and \ref{sec:transient}, we first assume that the gyrotactic particle undergoes no translational diffusion and that the dilute suspension is well described by (\ref{eq:smol}) with $D_T=0$. Later in \S\ref{sec:example_DT}, we add translational diffusion (i.e. finite $D_T$) to the particles to show the extra drift and dispersion that may arise from it. Also, to avoid the additional complication that may arise due to the boundary conditions in the physical space \citep[e.g. wall accumulation of][]{Ezhilan2015}, we assume a periodicity of $2 h^*$ in the $x$-direction. Therefore, the shear flow profile $W(x)$ is periodic in $x \in [-1,1]$. For convenience, we also define the shear profile $S(x)=-(\Pe_f/2) \partial_x W(x)$, with  $W(x)=-\cos(\pi x)-1$. The flow profile is plotted in \cref{fig:vertical_profile}. The initial condition of the suspension is given to be uniform in both $(\mathbf{x},\mathbf{p})$-space.

\subsubsection{Steady solution and shear trapping \label{sec:V_compare}}

In this subsection, we first compare the converged steady state with the prediction from \S\ref{sec:asymp}.
\Cref{fig:nx_steady} shows the number density at converged steady state $n_{f,s}$ after the numerical integration of the Smoluchowski equation for the suspensions of the idealised particles. Here, a non-negligibly large $Pe_s(\equiv 0.25)$ is deliberately chosen to highlight the deviation of the prediction by the local approximation model from the exact solution to the Smoluchowski equation. 
\begin{figure}
	\centering{}
	\sidesubfloat[]{\includegraphics[width = 0.45 \columnwidth]{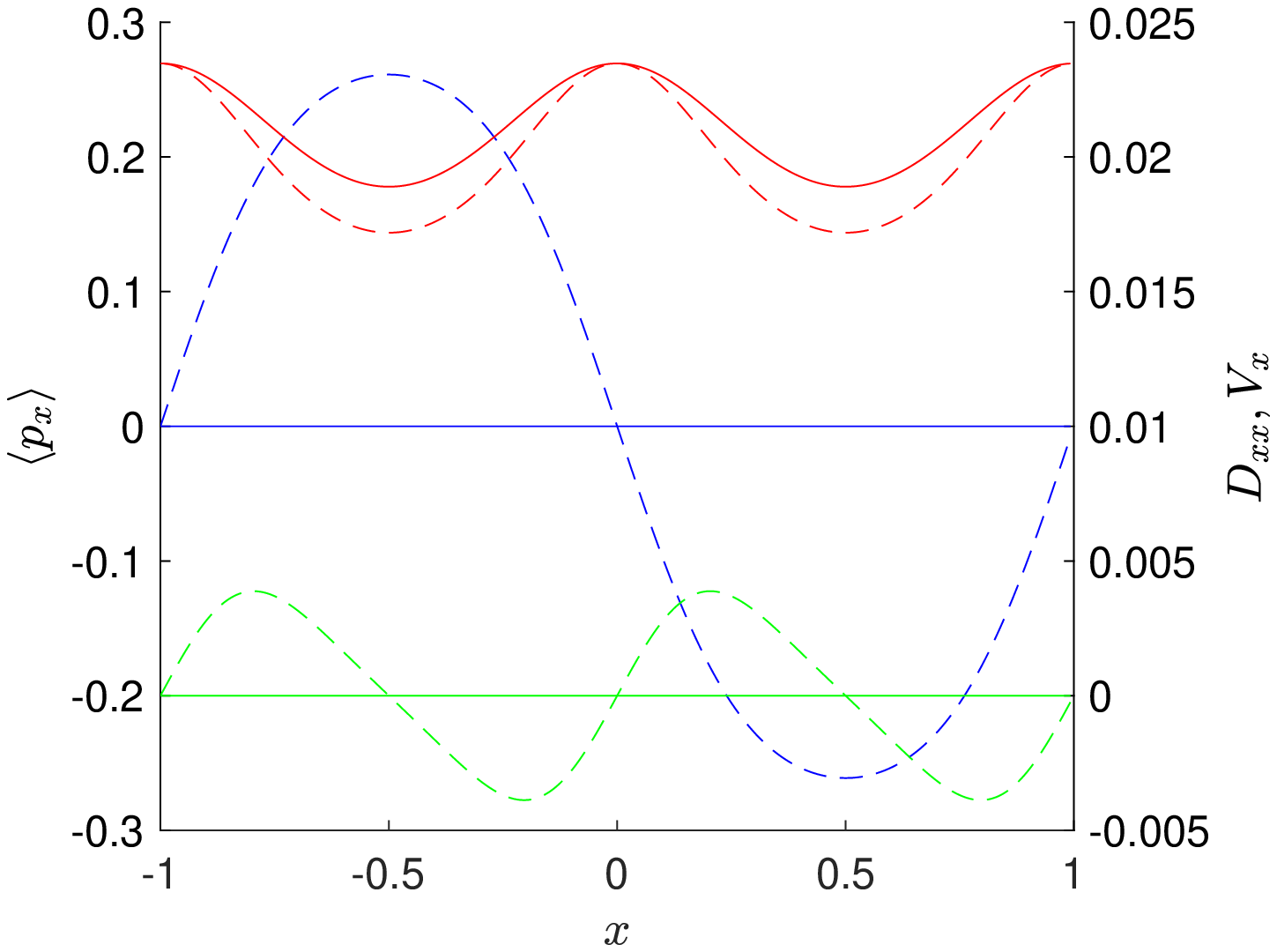}\label{fig:DVpavg_sphere}}
	\sidesubfloat[]{\includegraphics[width = 0.45 \columnwidth]{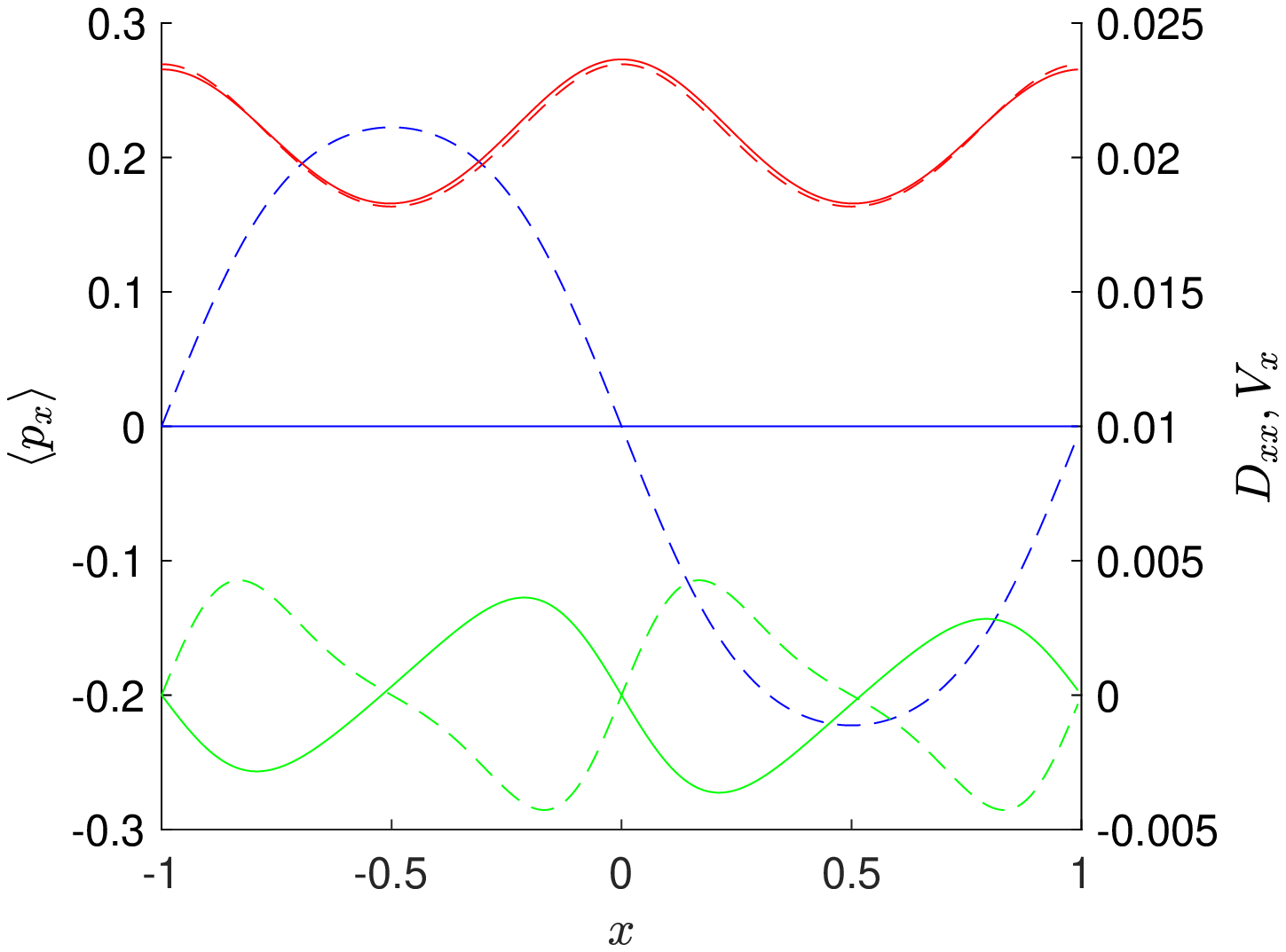}\label{fig:DVpavg_strong_Pef1}}\\
	\sidesubfloat[]{\includegraphics[width = 0.45 \columnwidth]{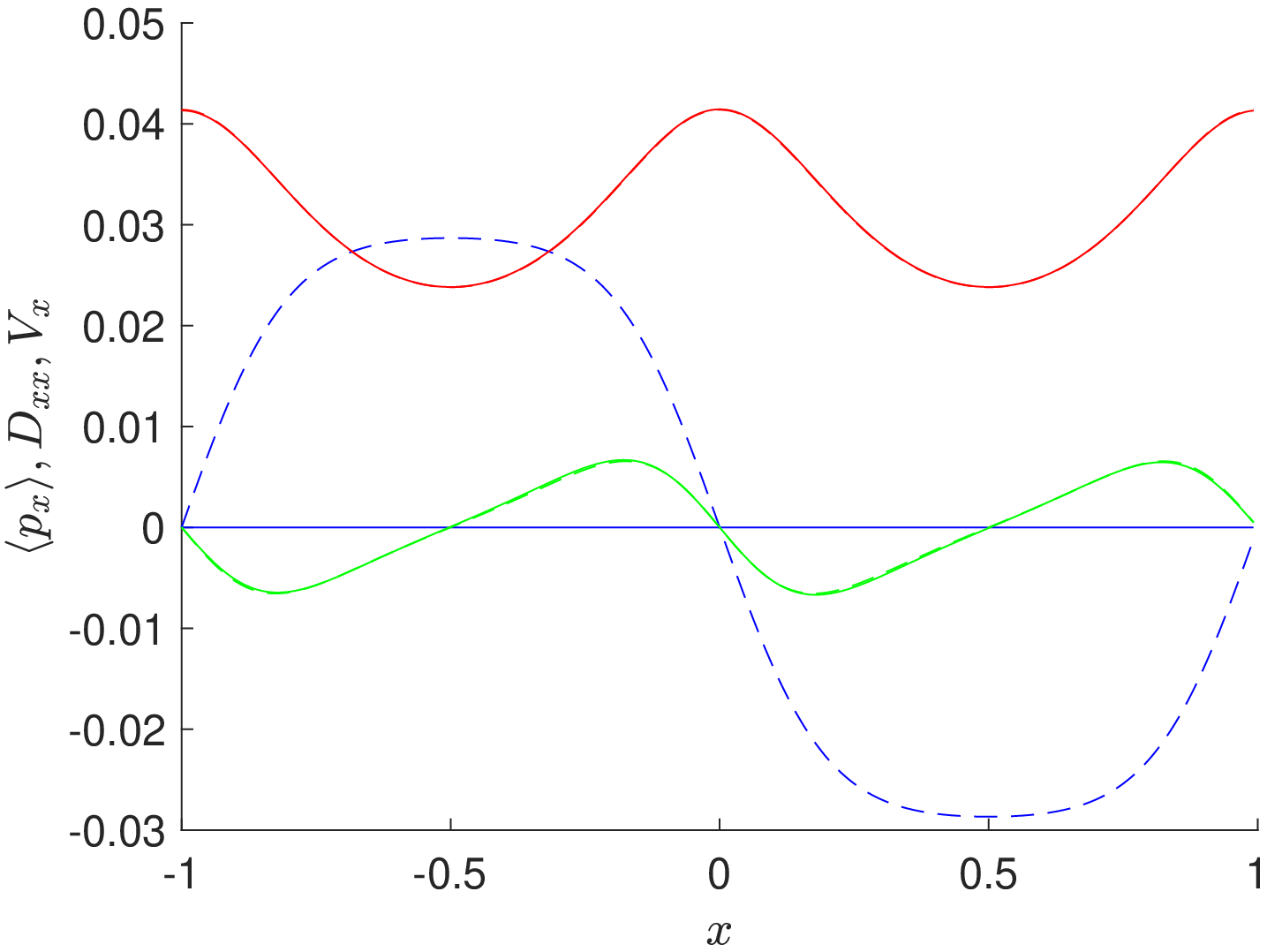}\label{fig:DVpavg_weak_Pef1}}
	\sidesubfloat[]{\includegraphics[width = 0.45 \columnwidth]{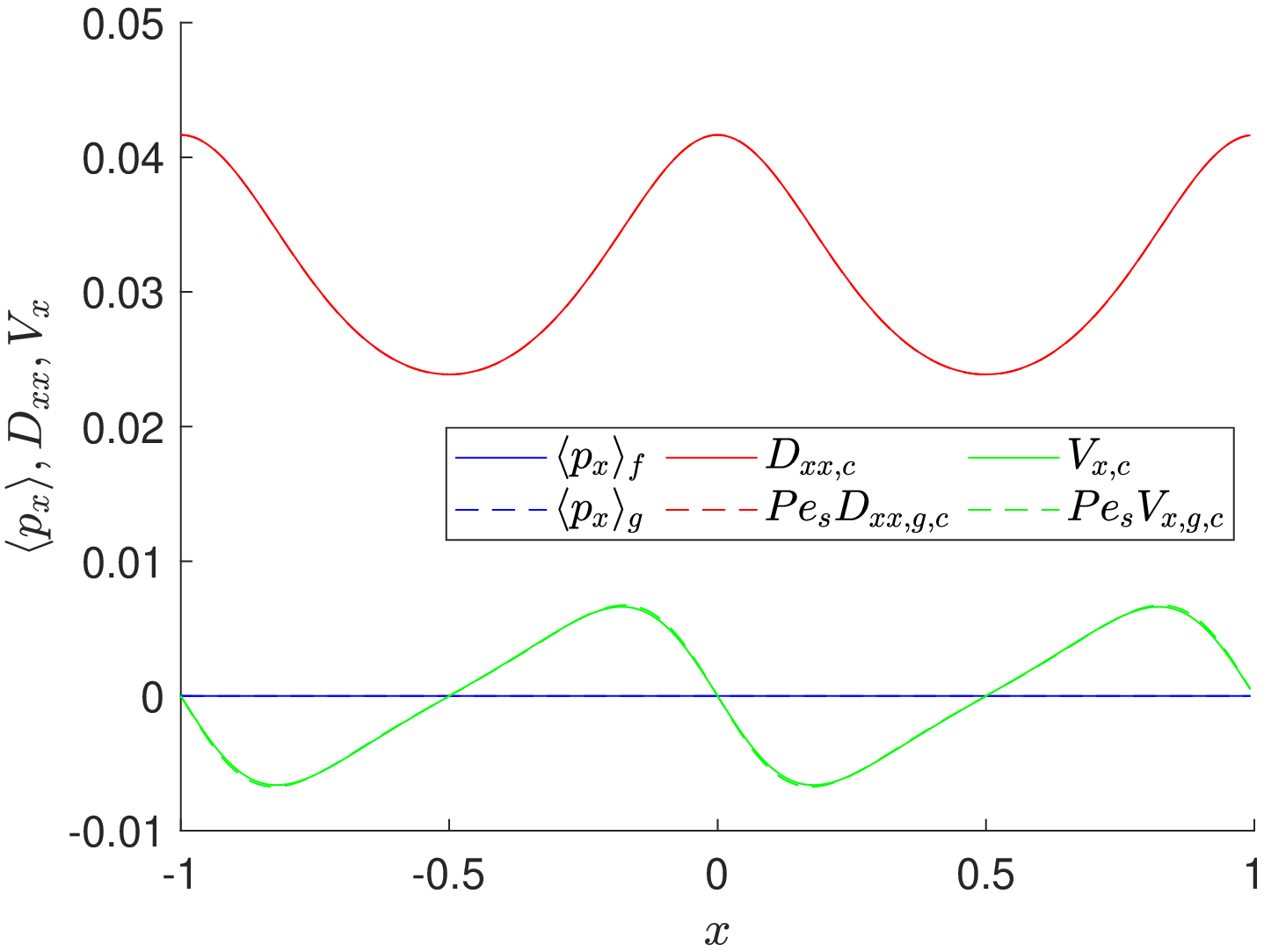}\label{fig:DVpavg_nongyro}}	
	\caption{Comparison of the drifts and dispersion terms in (\ref{eq:ns_exact}-\ref{eq:ns_asymp},\ref{eq:px_f}) at the steady state. The plots show the values of $\psth{x}_f$ (blue, solid), $\psth{x}_g$ (blue, dashed), $\mathsfi{D}_{xx,c}$ (red, solid), $\Pe_s \mathsfi{D}_{xx,g,c}$ (red, dashed), $V_{x,c}$, (green, solid) and $\Pe_s V_{x,g,c}$ (green, dashed) in suspensions of $(a)$ spherical and strongly gyrotactic ($\beta=2.2$, $\alpha_0=0$), $(b)$ non-spherical and strongly gyrotactic ($\beta=2.2$, $\alpha_0=0.31$), $(c)$ non-spherical and weakly gyrotactic ($\beta=0.21$, $\alpha_0=0.31$) and $(d)$ non-spherical and non-gyrotactic ($\beta=0$, $\alpha_0=0.31$) particles. The suspensions are subjected to a vertical flow $W(x)=-\cos(\pi x)-1$ with $\Pe_s=0.25$ and $\Pe_f=1$.\label{fig:DVpavg}}
\end{figure}
In the case of spherical gyrotactic particle suspension \pcref{fig:nx_sphere}, an analytical solution (\ref{eq:sphere_analytical}) has been found for the steady state of spherical gyrotactic particle suspension in a vertical flow (appendix \ref{app:analytical_sol}). The numerical solution, in this case, agrees very well with the analytical solution. In \cref{fig:nx_steady}, we have also plotted the steady-state number density given by the local approximation model in \S\ref{sec:asymp} ($n_{g,s}$). For strongly gyrotactic particles \pcreftwo{fig:nx_sphere}{fig:nx_strong_Pef1}, the model gives predictions close to the full solution. For weakly gyrotactic particles \pcref{fig:nx_weak_Pef1} and non-gyrotactic particles \pcref{fig:nx_nongyro}, the local approximation makes predictions almost identical to the exact result from the Smoluchowski equation. In spite of the small variation in $n(x)$ in the non-gyrotactic case (note the small scale variation of $n(x)$ in the ordinate of \cref{fig:DVpavg_nongyro}), the shear trapping mechanism that causes such aggregation is the dominant effect in this case, and the aggregation can be observed in experiments \citep{Rusconi2014}. Here, we note that if the homogeneous approximation model of the GTD theory is applied here in the manner of \cite{Bearon2012} and \cite{Croze2017}, it cannot predict the aggregation of particles due to shear trapping, resulting in a uniform distribution instead. However, if the $x$ direction is included as a local coordinate in the more rigorous application of GTD \citep{Jiang2019,Jiang2020}, the resulting number density will predict the shear trapping as it is equivalent to solving the full Smoluchowski equation in \cref{fig:DVpavg_nongyro}. 

Now, we investigate the performance of the local approximation model in terms of the coefficients of the transport equation given by each model. For a suspension of gyrotactic particles with $D_T=0$ in a prescribed parallel shear flow, the exact steady solution for the number density $n_{f,s}=n(x,\infty)$ is given from (\ref{eq:smol_coarse_final}) by 
\begin{equation}
	\partial_x [(\Pe_s \psth{x}_g - \Pe_s {V}_{x,c}) n_{f,s}] = \Pe_s \partial_x  [ \mathsfi{D}_{xx,c} \partial_x n_{f,s} ]. \label{eq:ns_exact}
\end{equation}
Similarly, the steady solution to the local approximation model in (\ref{eq:asymp_coarse_final}), denoted by $n_{g,s}(x)$, is given by 
\begin{equation}
	\partial_x [(\Pe_s \psth{x}_g - \Pe_s^2 {V}_{x,g,c}) n_{g,s}] = \Pe_s^2 \partial_x [\mathsfi{D}_{xx,g,c} \partial_x n_{g,s} ]. \label{eq:ns_asymp}
\end{equation}

\Cref{fig:DVpavg} shows the $x$ components of the effective drift and dispersion coefficients. First, we compare the effective dispersion from the local approximation with those from the exact transformation (see the right-hand side of (\ref{eq:ns_exact}-\ref{eq:ns_asymp})). In particular, when particles are spherical, $\mathsfi{D}_{xx,c}$ can be directly extracted as a function of the local vertical shear rate $S$ using the analytic solution of (\ref{eq:smol}) given in appendix \ref{app:analytical_sol}. In \cref{fig:Dallm}, $\mathsfi{D}_{xx,c}$ and $\mathsfi{D}_{xx,g,c}$ are plotted as a function of the vertical shear rate $S$. It is found that $\Pe_s \mathsfi{D}_{xx,g,c}$ approximates $\mathsfi{D}_{xx,c}$ quite well for all the range of $S$ considered. In general, $\Pe_s \mathsfi{D}_{xx,g,c}$ remains a good approximation for $\mathsfi{D}_{xx,c}$ for all the four cases considered at all the horizontal locations $x$ \pcref{fig:DVpavg}. 

\begin{figure}
	\centering
	\includegraphics[width = 0.5 \columnwidth]{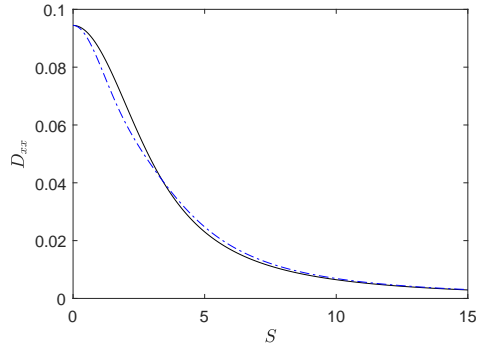}
	\caption{Comparison of the $xx$ component of $\mathsfbi{D}_c/\Pe_s$ (black line) and $\mathsfbi{D}_{g,c}$ (blue dot-dashed line) as a function of the local vertical shear $S(x)$ for spherical gyrotactic particles ($\beta=2.2$,$\alpha_0=0$), in which $\mathsfbi{D}_c/\Pe_s$ is computed from $f_s(\mathbf{p})$ of Appendix \ref{app:analytical_sol}.
	}\label{fig:Dallm} 
\end{figure}

As for the left-hand sides of (\ref{eq:ns_exact}) and (\ref{eq:ns_asymp}), the exact transformed equation and approximated model share the same $\psth{x}_g$ term. However, as shown in \cref{fig:DVpavg}, ${V}_{x,g,c}$ only follows ${V}_{x,c}$ closely in the weakly gyrotactic cases \pcreftwo{fig:DVpavg_weak_Pef1}{fig:DVpavg_nongyro} but poorly in the strongly gyrotactic cases \pcreftwo{fig:DVpavg_sphere}{fig:DVpavg_strong_Pef1}. Nevertheless, in the strongly gyrotactic cases, the left-hand sides of (\ref{eq:ns_exact}) and (\ref{eq:ns_asymp}) are dominated by $\psth{x}_g$, so the poor estimation of ${V}_{x,c}$ does not strongly affect the overall performance of the local approximation model \pcreftwo{fig:nx_weak_Pef1}{fig:nx_nongyro}.

Given the observation in the weakly gyrotactic cases \pcreftwo{fig:DVpavg_weak_Pef1}{fig:DVpavg_nongyro}, the drift $\mathbf{V}_{g,c}$ seems to be an important term in (\ref{eq:asymp_coarse_final}). Here, we further discuss the importance of this term from a physical perspective. The term $\mathbf{V}_{g,c}$ arises from the inhomogeneity of the local flow field (i.e. shear $S(x)$ in this example). Given its assumption, the quasi-homogeneous approximation model of the GTD theory in \cite{Bearon2012} and \cite{Croze2017} cannot obviously capture this effect of inhomogeneity in the shear $S(x)$. Hence, there is no equivalent of $\mathbf{V}_{g,c}$ in their model. On the other hand, a more rigorous application of the GTD theory will result in the equivalent of solving the Smoluchowski equation directly in this context \citep[e.g.~][]{Jiang2019,Jiang2020}. The form of (\ref{eq:fc_g}) for $\mathbf{V}_{g,c}$ suggests that there are two physical mechanisms at play that contribute to $\mathbf{V}_{g,c}$. One is the net flux caused by different levels of gyrotactic drift at different levels of shear at the adjacent location. The flux mainly manifests in the $-g \delx \bcdot \pavg_g$ term in (\ref{eq:fc_g}), which diminishes in the absence of gyrotaxis. The other is the shear trapping mechanism of \cite{Bearon2015} and \cite{Vennamneni2020}, which arises from the `eccentric shape' of the particles and their alignment with the flow. In the presence of inhomogeneous shear, the non-spherical shape leads to some inhomogeneity of $g$ in the $\mathbf{x}$-space \cite[for the detailed mechanism, see][]{Vennamneni2020}. Therefore, having a non-uniform shear in $\mathbf{x}$-space can lead to non-zero $\delx g$, even if the particle does not exhibit biased motility (i.e. $\pavg_g=0$). This behaviour would primarily manifest in the $\mathbf{p} \bcdot \delx g$ term in (\ref{eq:fc_g}). 

The importance of the drift term with $\mathbf{V}_{g,c}$ can further be understood by examining the scaling of the four cases in figures \ref{fig:nx_steady} and \ref{fig:DVpavg}. In the first case where the particles are spherical and strongly gyrotactic ($\alpha_0=0$, $\beta \sim O(1)$), the form of (\ref{eq:asymp_coarse_final}) implies $Pe_s^2 {V}_{x,g,c} \sim O(Pe_s^2)$, an order-of-magnitude smaller than $\Pe_s \psth{x}_g$: i.e. $\pavg_g \gg \Pe_s {V}_{x,g,c}$. This behaviour remains the same in the second case, where the particles are non-spherical and strongly gyrotactic ($\alpha_0 \ne 0$, $\beta \sim O(1)$).
However, in the third case where the particles are spheroidal and weakly gyrotactic ($\alpha_0 \sim \beta \sim O(Pe_s)$), $\psth{x}_g \sim Pe_s {V}_{x,g,c}$ due to $\psth{x}_g \sim O(Pe_s)$ from $\beta \sim O(Pe_s)$. Hence, if the particles are weakly gyrotactic, ${V}_{x,g,c}$ is of significance. Lastly, for the spheroidal and non-gyrotactic particles ($\alpha_0 \neq  0$, $\beta=0$), ${V}_{x,g,c}$ becomes dominant while $\psth{x}_g=0$. In this case, ${V}_{x,g,c}$ is purely from the shear trapping mechanism, as explained by \cite{Bearon2015} and \cite{Vennamneni2020}. 

\subsubsection{Transient dynamics\label{sec:transient}}

In this subsection, we investigate the transient dynamics from the perspective of the exact transformed equation. Rewriting (\ref{eq:smol_inte}) for this example, we have
\begin{equation}
	\partial_t n + \Pe_s \partial_x \left[ \psth{x}_f n \right] = 0, \label{eq:smol_inte_x}
\end{equation}
in which $\psth{x}_f$ can be expanded through (\ref{eq:smol_full_coarseG}) into 
\begin{equation}
	\psth{x}_f= \psth{x}_g - V_{x,c} - V_{x,\partial t} - \mathsfi{D}_{xx,c} \frac{\partial_x n}{n}. \label{eq:px_f}
\end{equation}
Substituting (\ref{eq:px_f}) into (\ref{eq:smol_inte_x}) yields the transport equation
\begin{equation}
	\partial_t n + \Pe_s \partial_x  \left[(\psth{x}_g-V_{x,c}- V_{x,\partial t}) n \right] = \partial_x \mathsfi{D}_{xx,c} \partial_x n. \label{eq:transport_x}
\end{equation}

Movies 1-4 show how the balance in (\ref{eq:px_f}) evolves in time from a uniform suspension. In the beginning, all terms were zeros, except for $\psth{x}_g$ and the unsteadiness in $f$ which balance out each other.  Note that the unsteadiness in $f$ was transformed into a drift $V_{x,\partial t}$ in the transport equation (see (\ref{eq:fdt_g})). As the suspension starts to evolve, the $\mathbf{p}$-space evolves first in the time scale of order unity (i.e. the fast time scale in \S\ref{sec:asymp}) -- note that the time scale in the $\mathbf{p}$-space is $1/d_r^*$ (see \S\ref{sec:background}). The fast-changing $f$ drives the drift $V_{x,\partial t}$ away from $\psth{x}_g$ in the beginning, resulting in non-zero $\psth{x}_f$ in (\ref{eq:px_f}), which in turn generates the unsteadiness in $n$ in (\ref{eq:smol_inte_x}). Therefore, $n(x,t)$ does not start evolving until $V_{x,\partial t}$ has become significantly different from $\psth{x}_g$. For $t \sim \mathcal{O}(1)$, $V_{x,\partial t}$ is close to zero, indicating that $f$ has reached the quasi-steady regime, justifying the assumption of \S\ref{sec:asymp}. It is also in this time interval where $V_{x,c} \approx V_{x,g,c}$ and $\mathsfi{D}_{xx,c} \approx \mathsfi{D}_{xx,g,c}$, implying that the local approximation in \S\ref{sec:asymp} would be valid after this short initial transient. 

For $t\gtrsim \mathcal{O}(1)$, $n(x,t)$ evolves slowly, while $\psth{x}_f$ diminishes towards zero, mainly due to the increasing magnitude of $(\partial_x n/n)$ to balance $\psth{x}_g$ in (\ref{eq:px_f}). As $\psth{x}_f$ vanishes, $n(x,t)$ reaches a steady equilibrium. During this slow transient period, $f$ also evolves but slowly enough such that $V_{x,\partial t}$ remains insignificant. Note that, in this example, the prescribed flow field was steady, such that $V_{x,g,\partial T}$ vanishes. If the prescribed flow were unsteady in the long timescale $T$, we would also expect $V_{x,\partial t}$ to be significant and to be well approximated by $V_{x,g,\partial T}$.
In all the examples considered, $\mathsfi{D}_{xx,c}$ remains close to the approximation $\mathsfi{D}_{xx,g,c}$. In weakly and non-gyrotactic suspensions, $V_{x,c}$ does not evolve far from $V_{x,g,c}$ either, but in strongly gyrotactic suspension, $V_{x,c}$ is found to change direction as $t\rightarrow \infty$. As mentioned in \S\ref{sec:V_compare}, $V_{x,c}$ is considerably small compared to $\psth{x}_g$ in this case. Therefore, regardless of the fact that $V_{x,g,c}$ differs from $V_{x,c}$, the local approximation model still performs well.

\subsubsection{Translational diffusion\label{sec:example_DT}}
\begin{figure}
	\centering{}
	\sidesubfloat[]{\includegraphics[width = 0.45 \columnwidth]{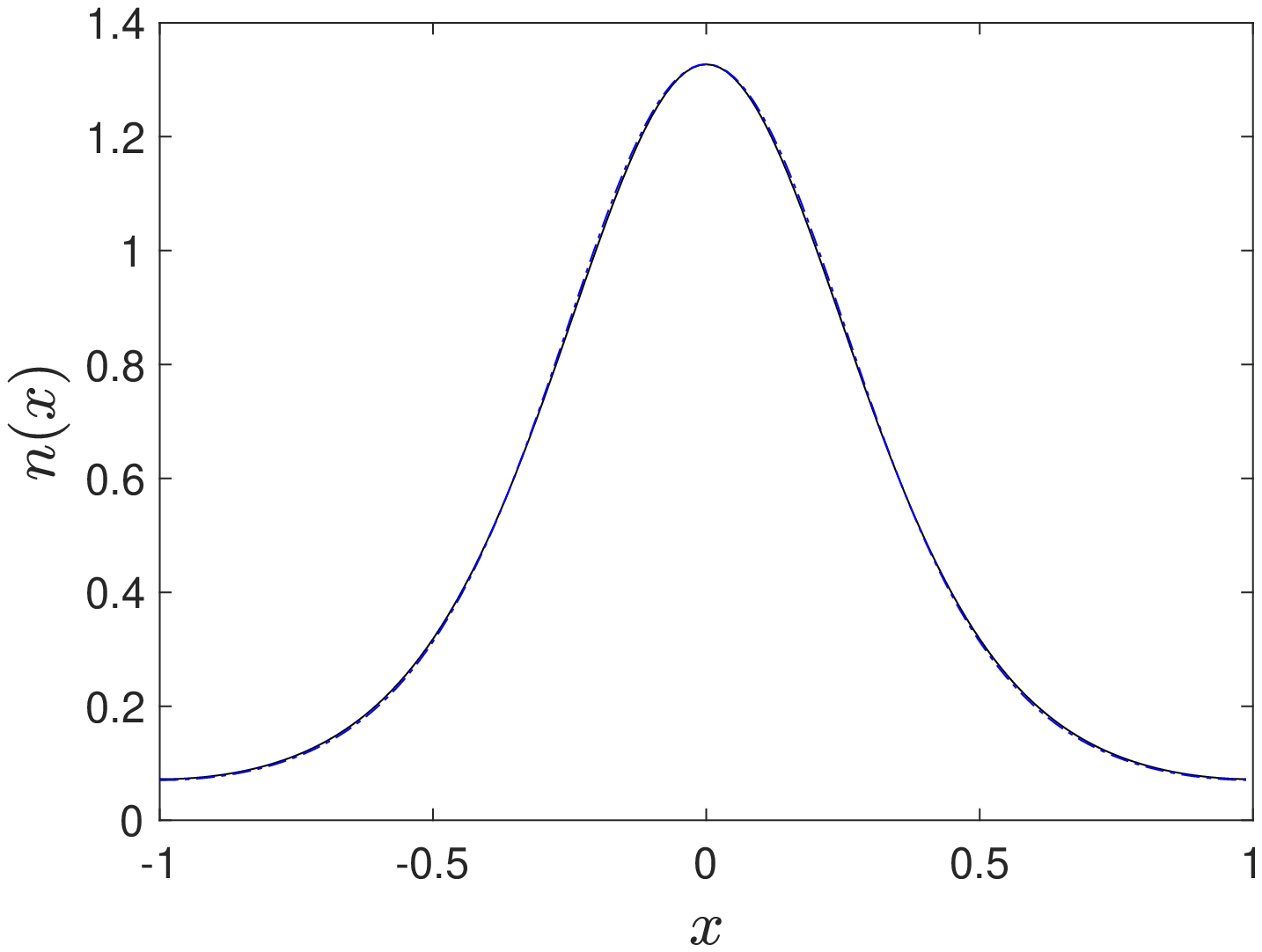}\label{fig:nx_sphere_DTmid}}
	\sidesubfloat[]{\includegraphics[width = 0.45 \columnwidth]{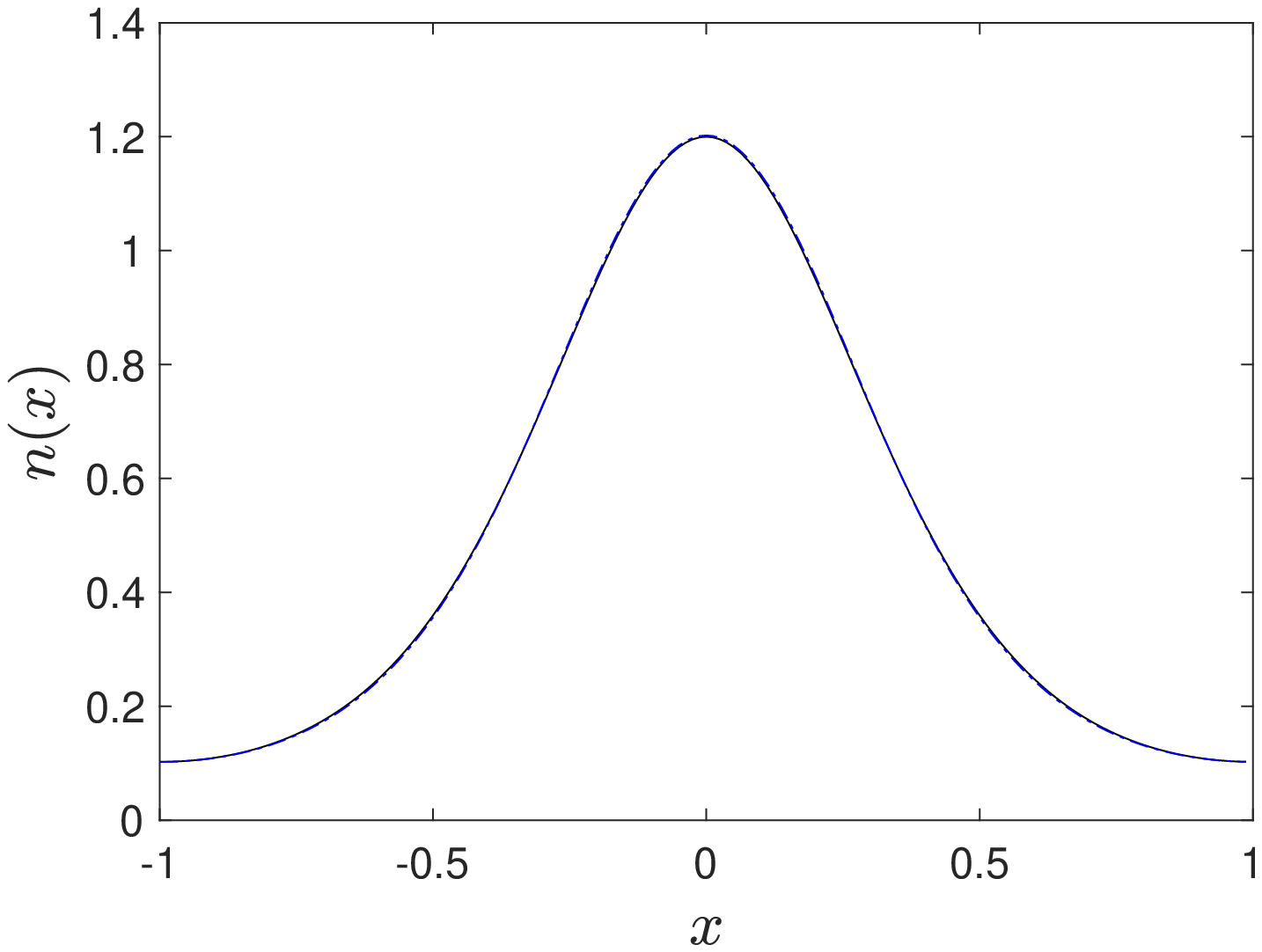}\label{fig:nx_strong_Pef1_DTmid}}\\
	\sidesubfloat[]{\includegraphics[width = 0.45 \columnwidth]{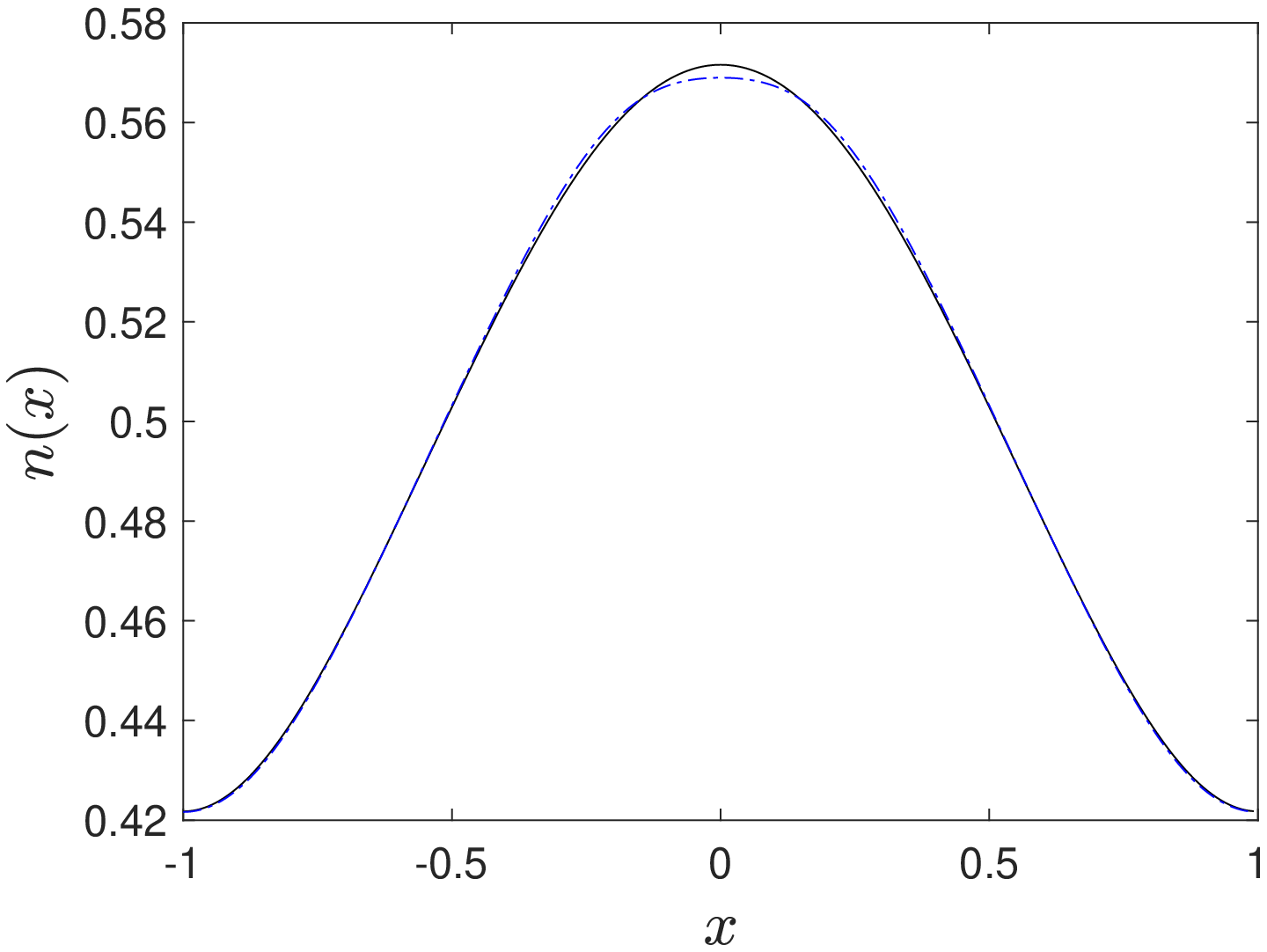}\label{fig:nx_weak_Pef1_DTmid}}
	\sidesubfloat[]{\includegraphics[width = 0.45 \columnwidth]{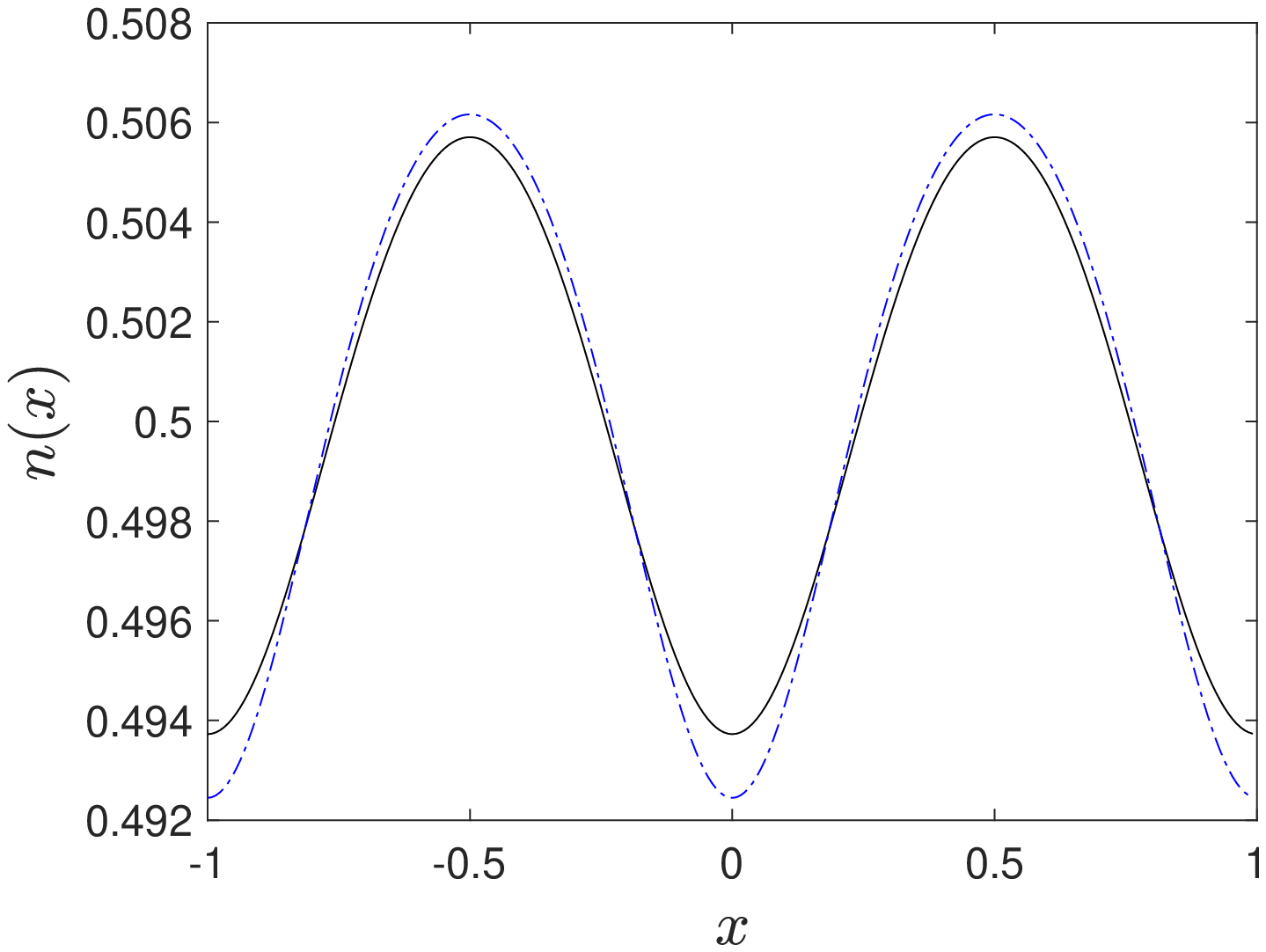}\label{fig:nx_nongyro_DTmid}}
	\caption{Comparison of the steady-state number density given by the direct integration of (\ref{eq:smol}) (black solid line, $n_{f,s}$) and the local approximation of \S\ref{sec:asymp} (blue dot-dashed line, $n_{g,s}$) of suspensions of $(a)$ spherical and strongly gyrotactic ($\beta=2.2$, $\alpha_0=0$), $(b)$ non-spherical and strongly gyrotactic ($\beta=2.2$, $\alpha_0=0.31$), $(c)$ non-spherical and weakly gyrotactic ($\beta=0.21$, $\alpha_0=0.31$) and $(d)$ non-spherical and non-gyrotactic ($\beta=0$, $\alpha_0=0.31$) particles. The particles are diffusive such that $D_T=0.01$. The suspensions are subjected to a vertical flow $W(x)=-\cos(\pi x)-1$ with $\Pe_s=0.25$ and $\Pe_f=1$.  Note that the vertical scale for $n(x)$ in $(c,d)$ is much smaller than that in $(a,b)$.  \label{fig:nx_steady_DTmid}}
\end{figure}

Lastly, we consider a non-zero translational diffusion for the previous examples. Microalgae such as \textit{Chlamydomos} and \textit{Dunaliella} are often considered to have negligible thermal diffusion given their relatively large sizes \citep[see reviews by][]{Pedley1992,Saintillan2018,Bees2020}.
Their random walk is often modelled only through rotational diffusion by assuming that the intracellular biochemical noise only affects the rotational motion. However, in theory, non-zero $D_T$ might also provide a mechanism to model some part of the randomness. The swimming mechanisms often involve sophisticated noisy beating dynamics of cilia and flagella \citep[e.g.][]{Wan2014}, which might result in random translation in addition to random rotation. Given the ambiguity in choosing a biologically relevant value for $D_T$, here we simply consider some values of $D_T$ to demonstrate the role of translational diffusion in the transport equation, i.e. $\mathbf{V}_{D_T}$ and $\mathsfbi{D}_{D_T}$.

We consider the steady-state number density at an arbitrary value of $D_T=0.01$, which is chosen to be of magnitude similar to that of $\Pe_s \mathsfbi{D}_c$. This arbitrary choice was made to highlight the potential role of translational diffusion. Also, for biological microparticles, any $D_T$ value larger than $\Pe_s \mathsfbi{D}_c$ would be physically unrealistic (c.f. experimental measurements of \cite{Croze2017}).  We have also computed the steady state at $D_T=0.002$, but since the results are qualitatively the same, we only present the $D_T=0.01$ case here.

The exact steady-state number density $n_{f,s}({x})$ from the Smoluchowski equation (\ref{eq:smol}) is given by 
\begin{equation}
	\partial_x [(\Pe_s \psth{x}_g - \Pe_s ({V}_{x,c} + {V}_{x,D_T})) n_{f,s}] = \partial_x [ (D_T+ \Pe_s (\mathsfi{D}_{xx,c}+\mathsfi{D}_{xx,D_T})) \partial_x n_{f,s} ], \label{eq:ns_exact_DT}
\end{equation}
and the number density from the local approximation $n_{g,s}$ is given by 
\begin{equation}
	\partial_x [(\Pe_s \psth{x}_g - \Pe_s^2 ({V}_{x,g,c}+{V}_{x,g,D_T})) n_{g,s}] =  \partial_x [(D_T+\Pe_s^2(\mathsfi{D}_{xx,g,c}+\mathsfi{D}_{xx,g,D_T})) \partial_x  n_{g,s} ].  \label{eq:ns_asymp_DT}
\end{equation}
Note that $\Pe_s^2 {V}_{x,g,D_T}$ and $\Pe_s^2 \mathsfi{D}_{xx,g,D_T}$ scale with $\Pe_s D_T$ from (\ref{eq:fDT_g}) and (\ref{eq:bDT_g}). 

\begin{figure}
	\centering{}
	\sidesubfloat[]{\includegraphics[width = 0.45 \columnwidth]{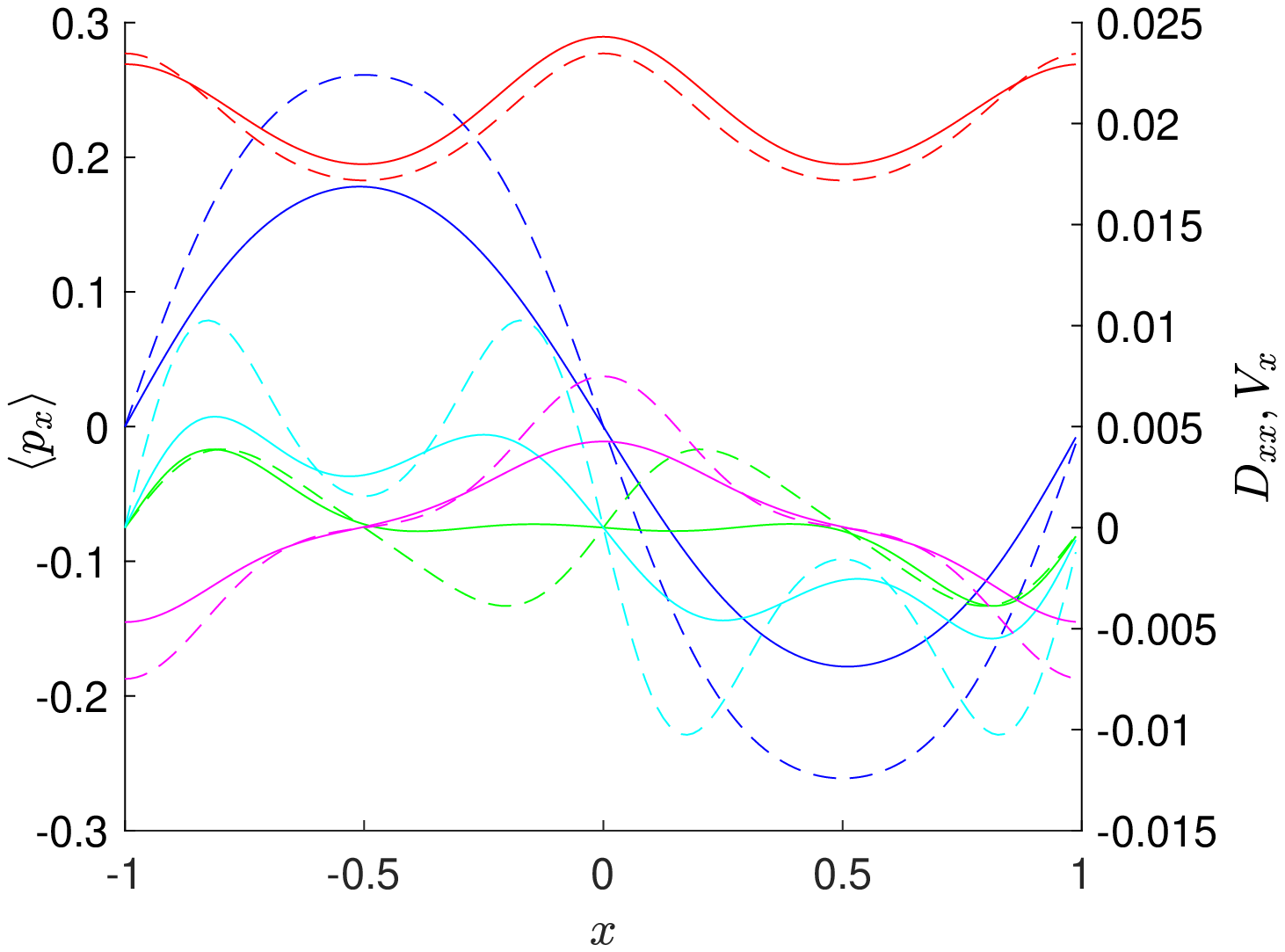}\label{fig:DVpavg_sphere_DTmid}}
	\sidesubfloat[]{\includegraphics[width = 0.45 \columnwidth]{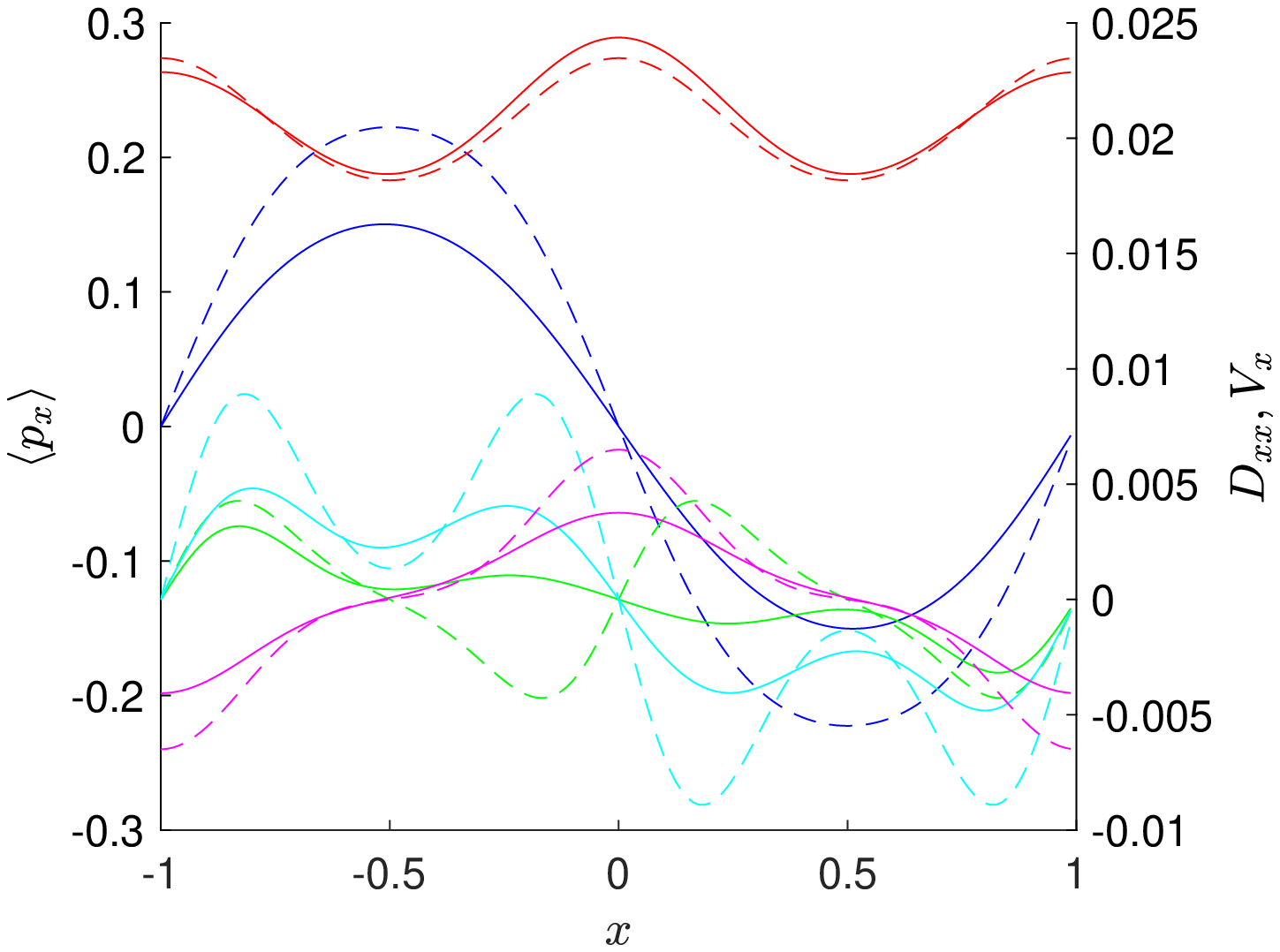}\label{fig:DVpavg_strong_Pef1_DTmid}}\\
	\sidesubfloat[]{\includegraphics[width = 0.45 \columnwidth]{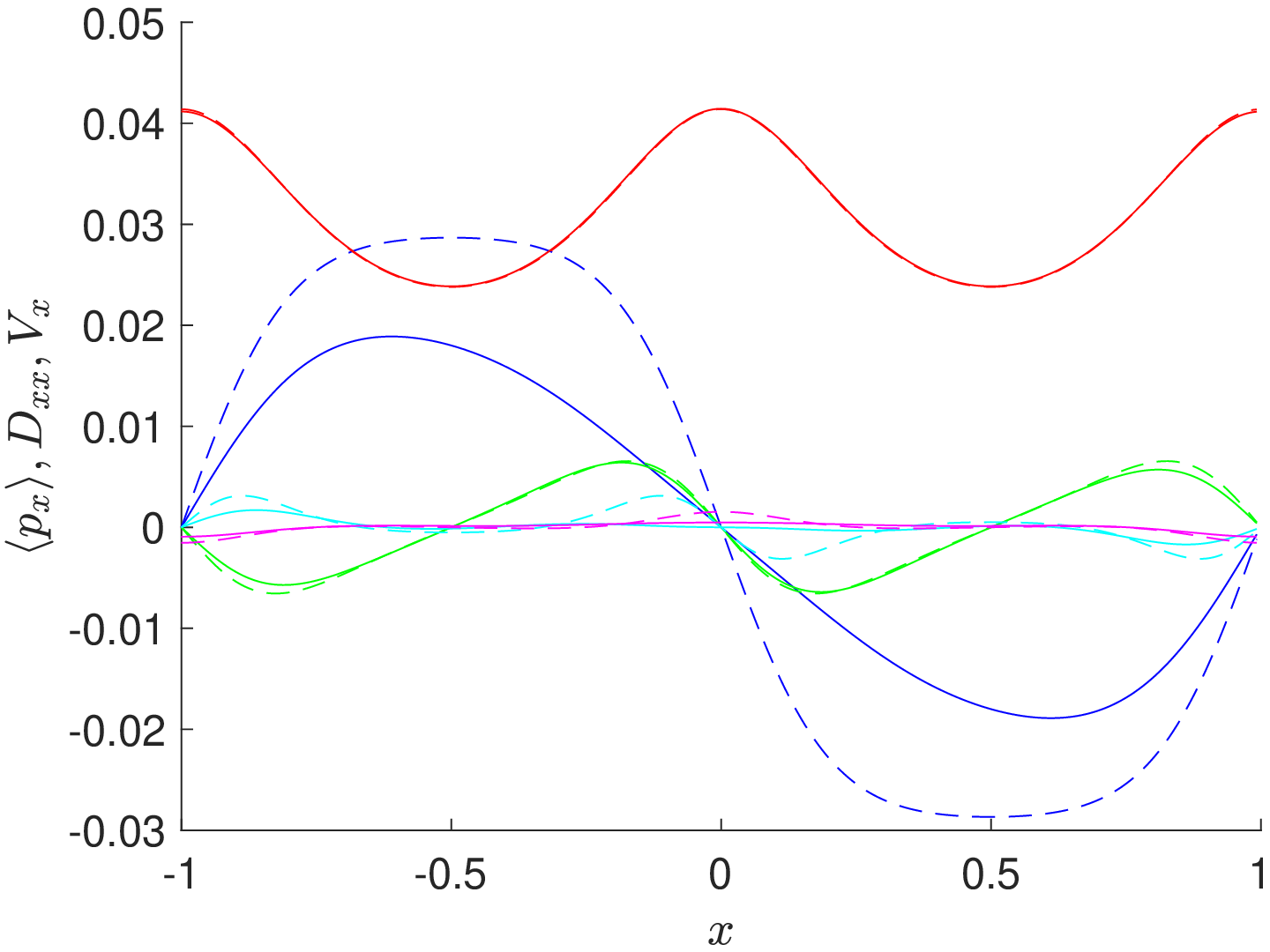}\label{fig:DVpavg_weak_Pef1_DTmid}}
	\sidesubfloat[]{\includegraphics[width = 0.45 \columnwidth]{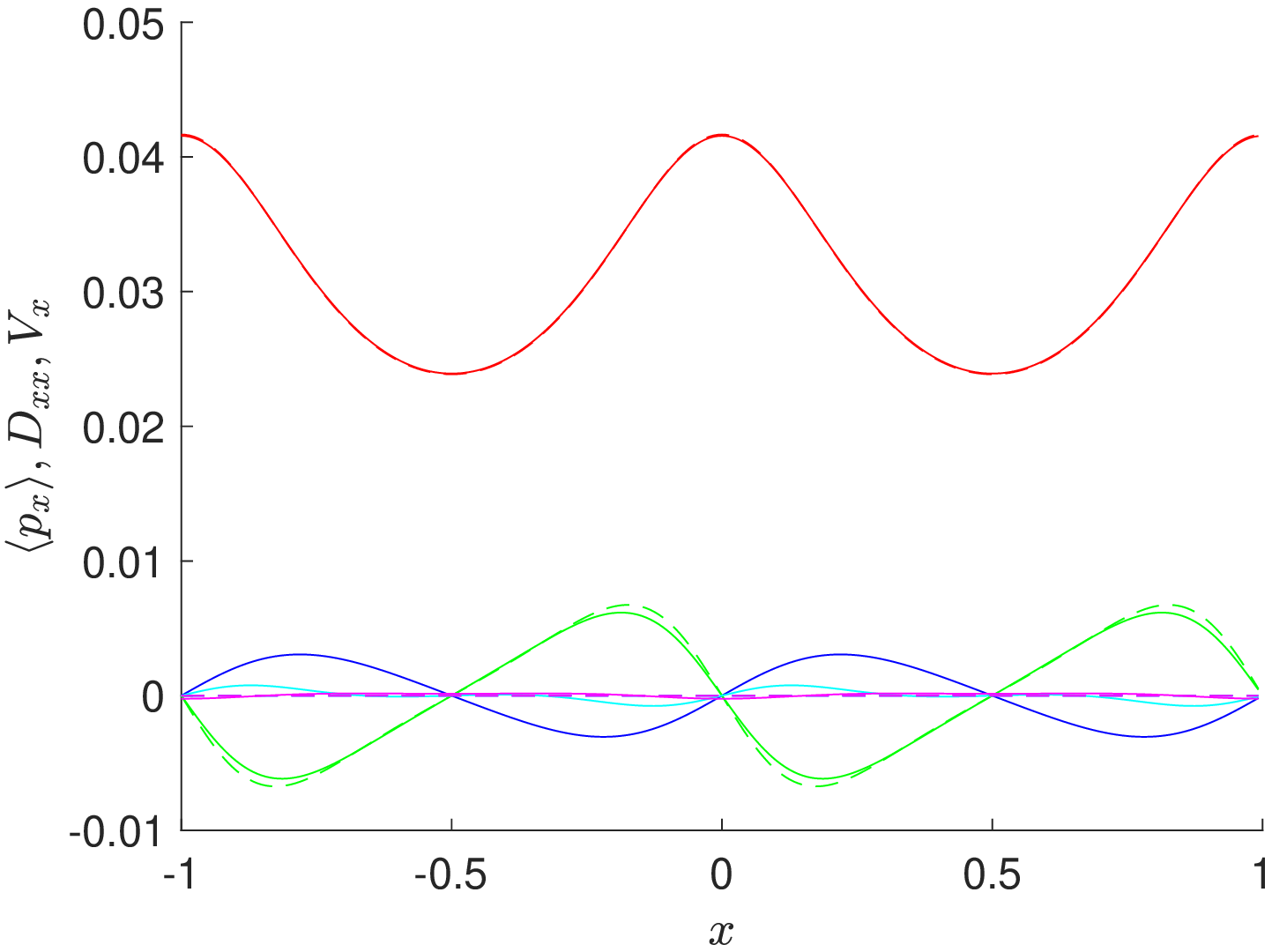}\label{fig:DVpavg_nongyro_DTmid}}	
	\caption{Comparison of the drifts and dispersion terms in (\ref{eq:ns_exact_DT}-\ref{eq:ns_asymp_DT}) at the steady state. The plots show the values of $\psth{x}_f$ and $\psth{x}_g$ (blue), $\mathsfi{D}_{xx,c}$ and $\Pe_s \mathsfi{D}_{xx,g,c}$ (red), $V_{x,c}$ and $\Pe_s V_{x,g,c}$ (green),  $\mathsfi{D}_{xx,D_T}$ and $\Pe_s \mathsfi{D}_{xx,g,D_T}$ (magneta), $V_{x,D_T}$ and $\Pe_s V_{x,g,D_T}$ (cyan) calculated using the steady-state $f(\mathbf{x},\mathbf{p},\infty)$ (solid lines) and $g(\mathbf{x},\infty;\mathbf{p})$ (dashed lines) of suspensions of $(a)$ spherical and strongly gyrotactic ($\beta=2.2$, $\alpha_0=0$), $(b)$ non-spherical and strongly gyrotactic ($\beta=2.2$, $\alpha_0=0.31$), $(c)$ non-spherical and weakly gyrotactic ($\beta=0.21$, $\alpha_0=0.31$) and $(d)$ non-spherical and non-gyrotactic ($\beta=0$, $\alpha_0=0.31$) particles. The particles are diffusive such that $D_T=0.01$. The suspensions are subjected to a vertical flow $W(x)=-\cos(\pi x)-1$ with $\Pe_s=0.25$ and $\Pe_f=1$. \label{fig:DVpavg_DTmid}}
\end{figure}

Figure \ref{fig:nx_steady_DTmid} shows the steady-state number density with $D_T=0.01$ for the same parameters considered in figure \ref{fig:nx_steady}. One can see that the introduction of non-zero $D_T$ has further smoothed out the number density in all cases considered by comparing figures \ref{fig:nx_steady} and \ref{fig:nx_steady_DTmid}. However, $D_T$ does not seem to have significantly altered most of the conclusions drawn in \S\ref{sec:V_compare}. 

The effective drift and dispersion terms in this case are also shown in figure \ref{fig:DVpavg_DTmid}. The introduction of non-zero $D_T$ also gives rise to an extra drift ${V}_{x,D_T}$ and an extra dispersion $\mathsfi{D}_{xx,D_T}$. Their approximations ${V}_{x,g,D_T}$ and $\mathsfi{D}_{xx,D_T}$ loosely follow the trend of their exact counterpart, such that the overall approximation $n_{g,s}$ remains an accurate prediction of the full solution.
Comparing the strongly gyrotactic case (figure \ref{fig:DVpavg_strong_Pef1_DTmid}) with the weakly gyrotactic case (figure \ref{fig:DVpavg_weak_Pef1_DTmid}), one can also conclude that the effects of ${V}_{x,D_T}$ and $\mathsfi{D}_{xx,D_T}$ are much stronger in strongly gyrotactic suspensions. Since ${V}_{x,D_T}$ and $\mathsfi{D}_{xx,D_T}$ are driven by $\delx f$ and $\nabla^2_\mathbf{x} f$ according to (\ref{eq:fDT}) and (\ref{eq:bDT}), the large ${V}_{x,D_T}$ and $\mathsfi{D}_{xx,D_T}$ are likely driven by the larger variation of $f$ in $x$ induced by the stronger gyrotaxis. 
Lastly, it is worth noting that $\mathsfi{D}_{xx,D_T}$ and $\mathsfi{D}_{xx,g,D_T}$ can be negative for some domain in $x$. As mentioned in \S\ref{sec:exact_transform}, the terms with $\mathsfi{D}_{xx,D_T}$ and $\mathsfi{D}_{xx,g,D_T}$ do not necessarily represent diffusion -- they depict the dispersive behaviour introduced by translational diffusion. 
Therefore, negative diagonal values in $\mathsfbi{D}_{D_T}$ are allowed. 

\subsection{A suspension of gyrotactic active particles subjected to a prescribed horizontal flow\label{sec:example_HS}}

\begin{figure}
	\centering{}
	\sidesubfloat[]{\includegraphics[width = 0.46 \columnwidth]{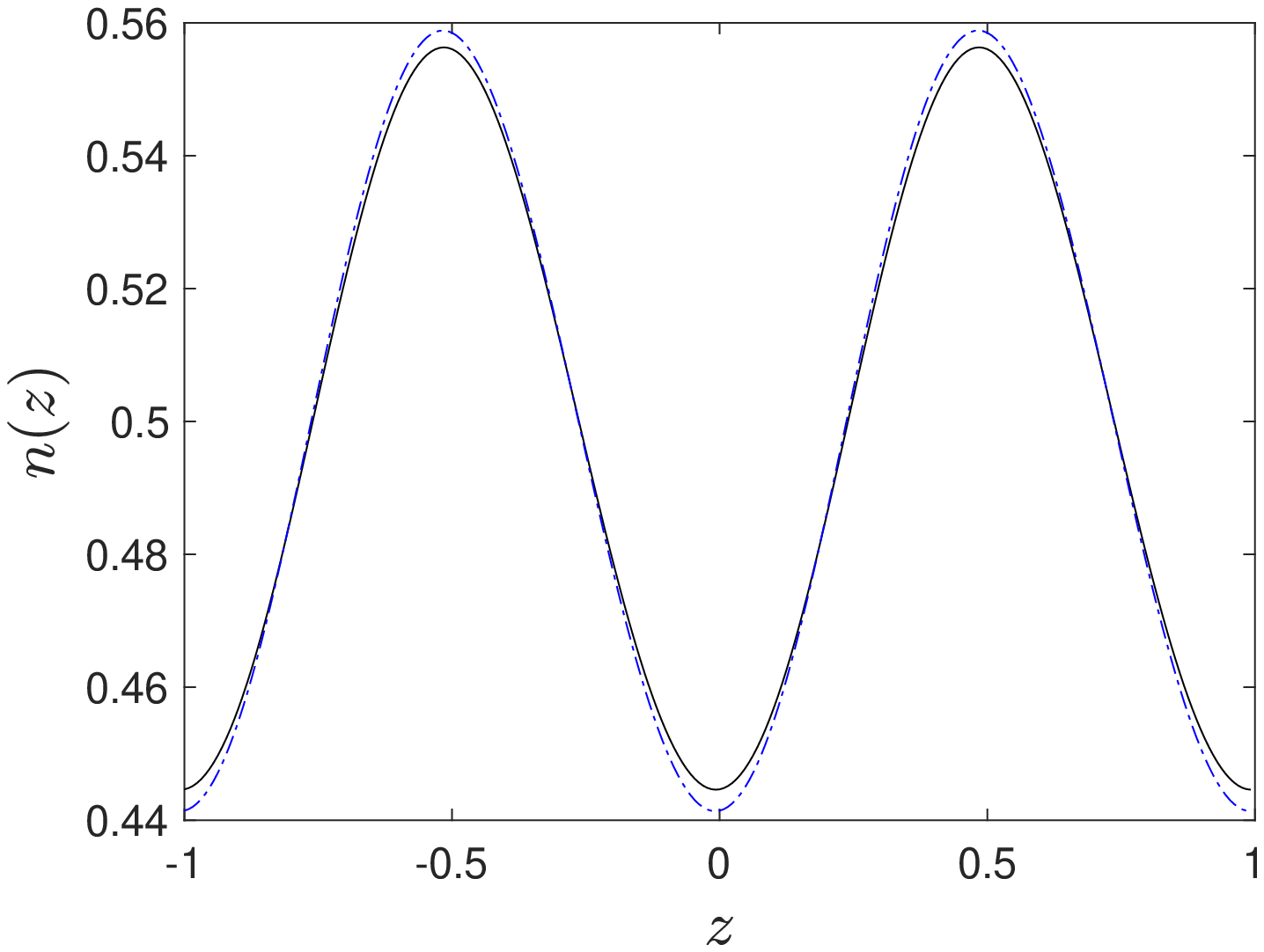}\label{fig:nz_steady_sphere}}
	\sidesubfloat[]{\includegraphics[width = 0.46 \columnwidth]{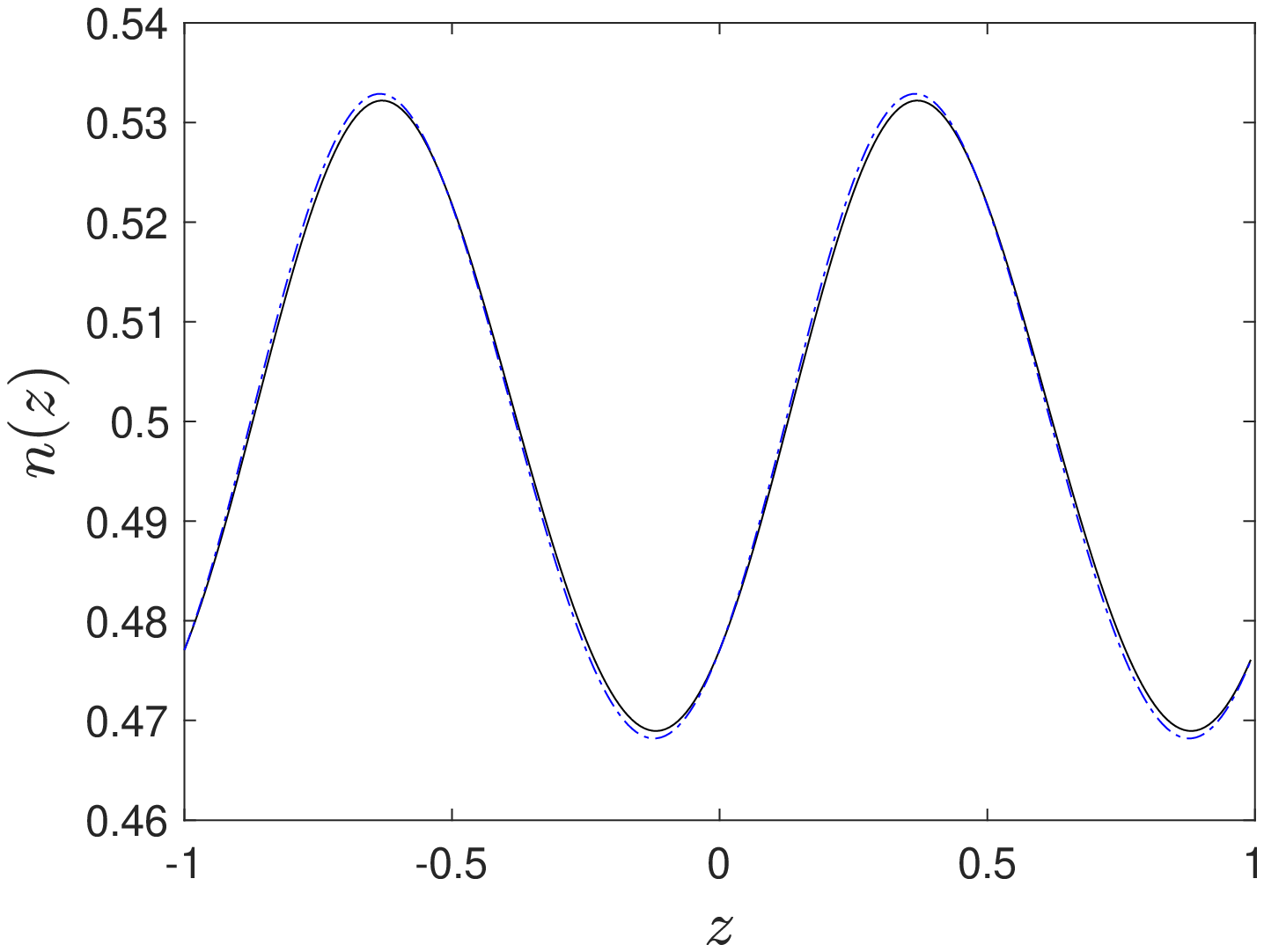}\label{fig:nz_steady_weak}}
	\caption{Comparison of the steady-state number density given by the direct integration of (\ref{eq:smol}) (black solid line, $n_{f,s}$) and the local approximation of \S\ref{sec:asymp} (blue dot-dashed line, $n_{g,s}$) of suspensions of $(a)$ strongly gyrotactic particles ($\beta=2.2$, $\alpha_0=0.31$) and $(b)$ weakly gyrotactic particles ($\beta=0.21$, $\alpha_0=0.31$). The suspensions are subjected to horizontal shear flow $U(z)=\cos (\pi z)$ with $\Pe_s=0.25$ and $\Pe_f=1$. \label{fig:HS_n}}
\end{figure}

\begin{figure}
	\centering{}
	\sidesubfloat[]{\includegraphics[width = 0.46 \columnwidth]{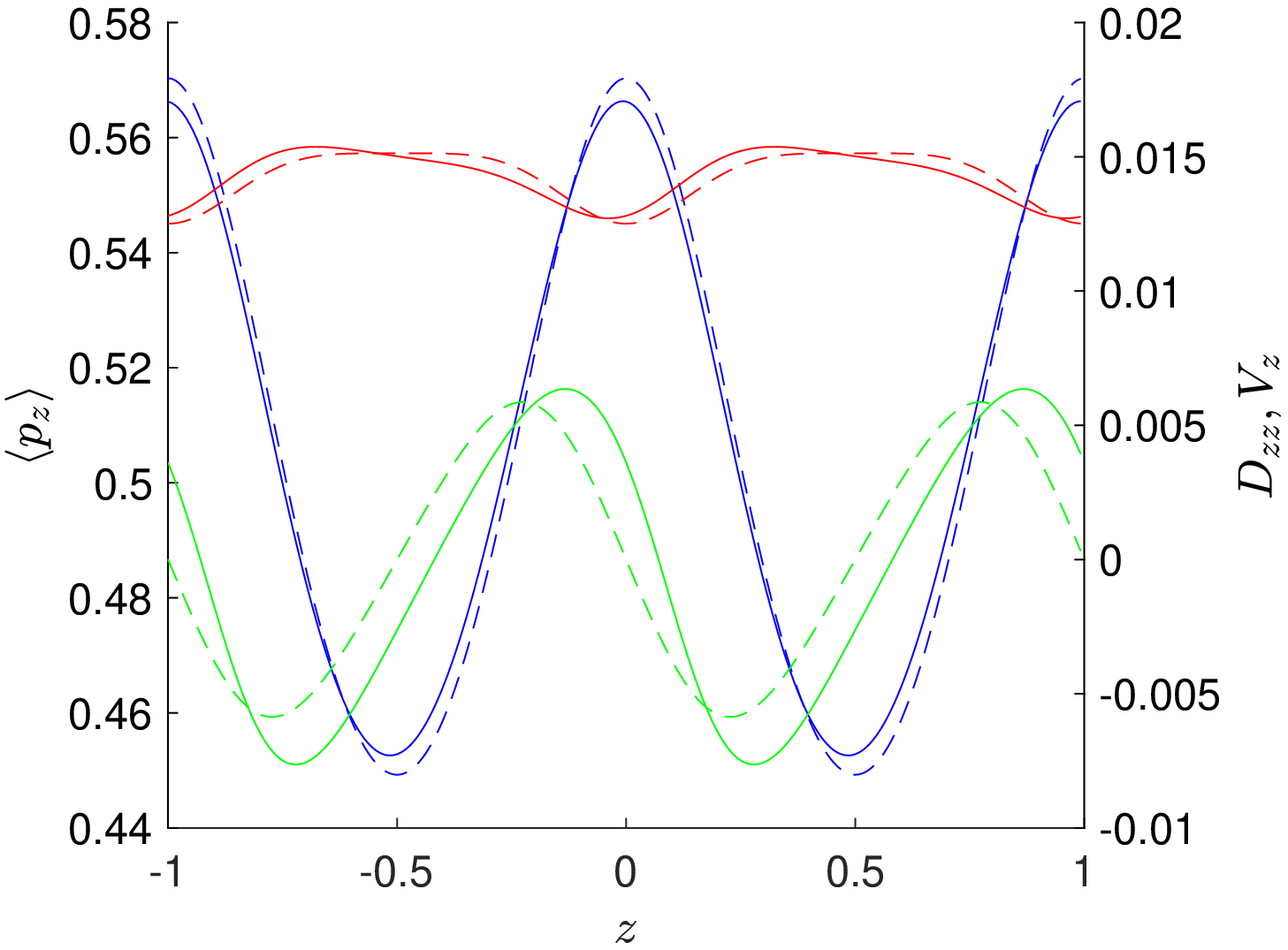}\label{fig:DVpavg_strong_Pef1_HS}}
	\sidesubfloat[]{\includegraphics[width = 0.46 \columnwidth]{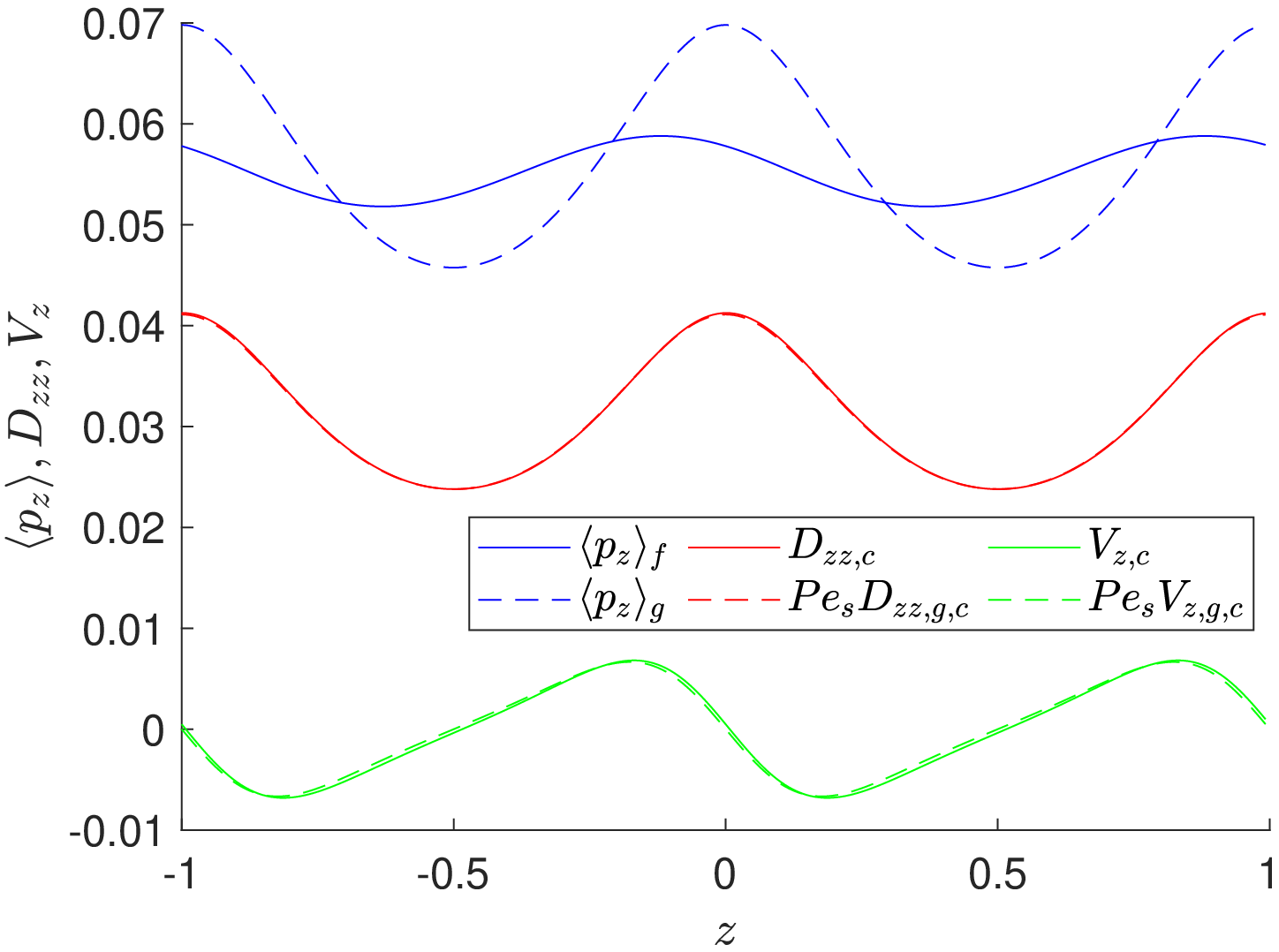}\label{fig:DVpavg_weak_Pef1_HS}}\\
	\caption{Comparison of the vertical drifts and dispersion terms at the steady state. The plots show the values of $\psth{z}_f$ (blue, solid), $\psth{z}_g$ (blue, dashed), $\mathsfi{D}_{zz,c}$ (red, solid), $\Pe_s D_{zz,g,c}$ (red, dashed), $V_{z,c}$, (green, solid) and $\Pe_s V_{z,g,c}$ (green, dashed) at the steady state of suspensions of $(a)$ strongly gyrotactic particles ($\beta=2.2$, $\alpha_0=0.31$) and $(b)$ weakly gyrotactic particles ($\beta=0.21$, $\alpha_0=0.31$). The suspensions are subjected to a horizontal shear flow $U(z)=\cos (\pi z)$ with $\Pe_s=0.25$ and $\Pe_f=1$.\label{fig:DVpavg_HS}} 
\end{figure}

In this section, we consider a horizontal shear flow $\mathbf{u}=[U(z),0,0]^T$ in the gyrotactic suspension instead of a vertical shear flow. Similar to \S\ref{sec:example_upright}, we first assume an infinite $\mathbf{x}$-domain with a periodicity in $z$ and no translational diffusion. The horizontal shear flow is prescribed as $U(z)=\cos{(\pi z)}$, as shown in \cref{fig:horizontal_profile}. 
Figure \subref{fig:nz_steady_sphere} shows that the steady-state number density profiles $n(z)$ of the strongly gyrotactic suspension ($\beta=2.2$,$\alpha_0=0.31$) computed from the local approximation model, which follows the full solution closely.
Similar to the case in \S\ref{sec:V_compare}, the small differences mainly come from the difference between $V_{z,g,c}$ and $V_{z,c}$, which is relatively small when compared with $\pavg_g$ \pcref{fig:DVpavg_strong_Pef1_HS}. As for the weakly gyrotactic non-spherical case ($\beta=0.21$,$\alpha_0=0.31$), the prediction from the local approximation is virtually the same as the full solution to the Smoluchowski equation (\ref{eq:smol}).

The transient dynamics is also investigated for the horizontal flow.
As shown in movies 5-6, the simulation initially shows the dominant balance between $V_{z,\partial t}$ and $\psth{z}_g$. At the time scale of order unity, $V_{z,\partial t}$ diminishes quickly, driven by the fast-changing $f$. At $t \gtrsim \mathcal{O}(1)$, $V_{z,\partial t}$ becomes insignificant, indicating that $f$ has reached the quasi-steady regime. Meanwhile, the local approximation accurately predicts $V_{z,c} \approx Pe_s V_{z,g,c}$ and $\mathsfi{D}_{zz,c} \approx Pe_s D_{zz,g,c}$, similar to $V_{x,c} \approx Pe_s V_{x,g,c}$ and $\mathsfi{D}_{zz,c} \approx Pe_s \mathsfi{D}_{xx,g,c}$ in \S\ref{sec:transient}. However, unlike the vertical flow cases, movies 5-6 show that $\psth{z}_f$ does not tend to zero as $t \rightarrow \infty$ in these horizontal flow cases. Instead, figure \ref{fig:DVpavg_HS} shows that they stay in roughly the same order as $\psth{z}_g$ at steady equilibrium.
Moreover, both $Pe_s V_{z,g,c}$ and $Pe_s D_{zz,g,c}$ remain good approximations to $V_{z,c}$ and $\mathsfi{D}_{zz,c}$, respectively, for a long time, even when particles are strongly gyrotactic. 

\subsection{A suspension of gyrotactic active particles in periodic convective cells \label{sec:example_2D}}

\begin{figure}
	\centering{}
	\includegraphics[width = 0.3\columnwidth]{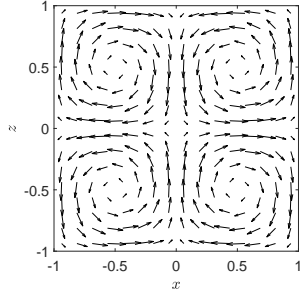}
	\caption{Vector plot of the two-dimensional periodic convective flow field (\ref{eq:conv_cells}). \label{fig:conv_vec}}
\end{figure}

\begin{figure}
	\centering{}
	\sidesubfloat[]{\includegraphics[width = 0.3 \columnwidth]{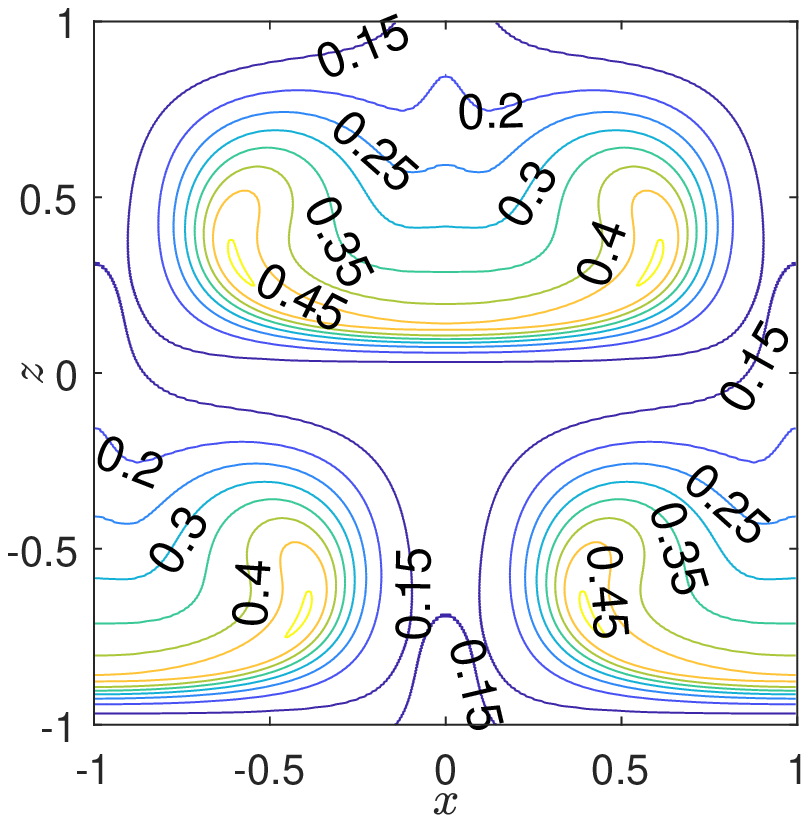}\label{fig:2D_beta2.2_B0.31_nfs}}
	\sidesubfloat[]{\includegraphics[width = 0.3 \columnwidth]{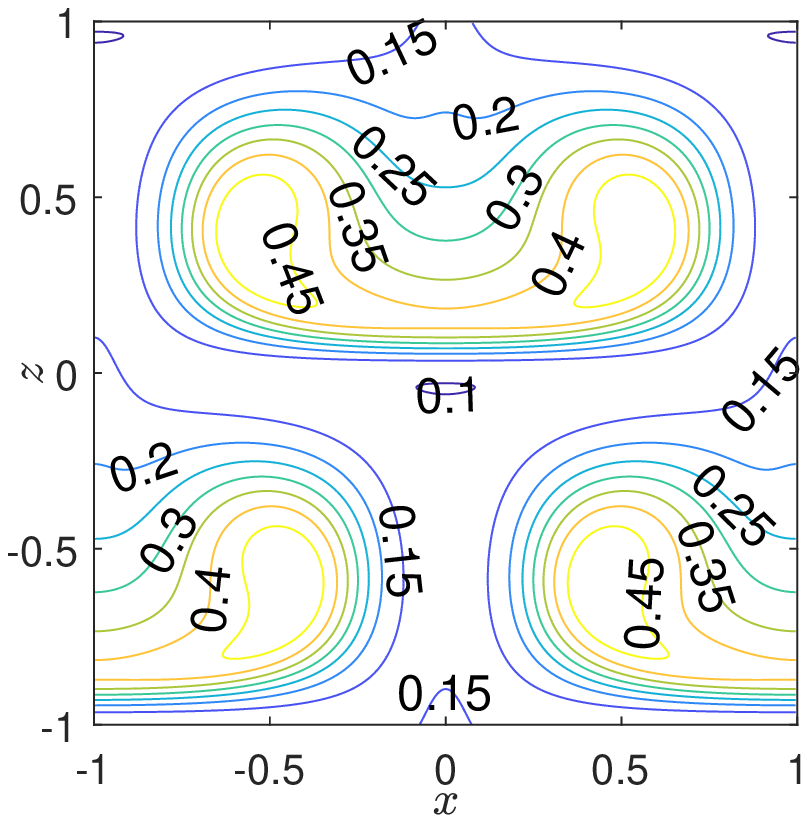}\label{fig:2D_beta2.2_B0.31_ngs}}\\
	\sidesubfloat[]{\includegraphics[width = 0.3 \columnwidth]{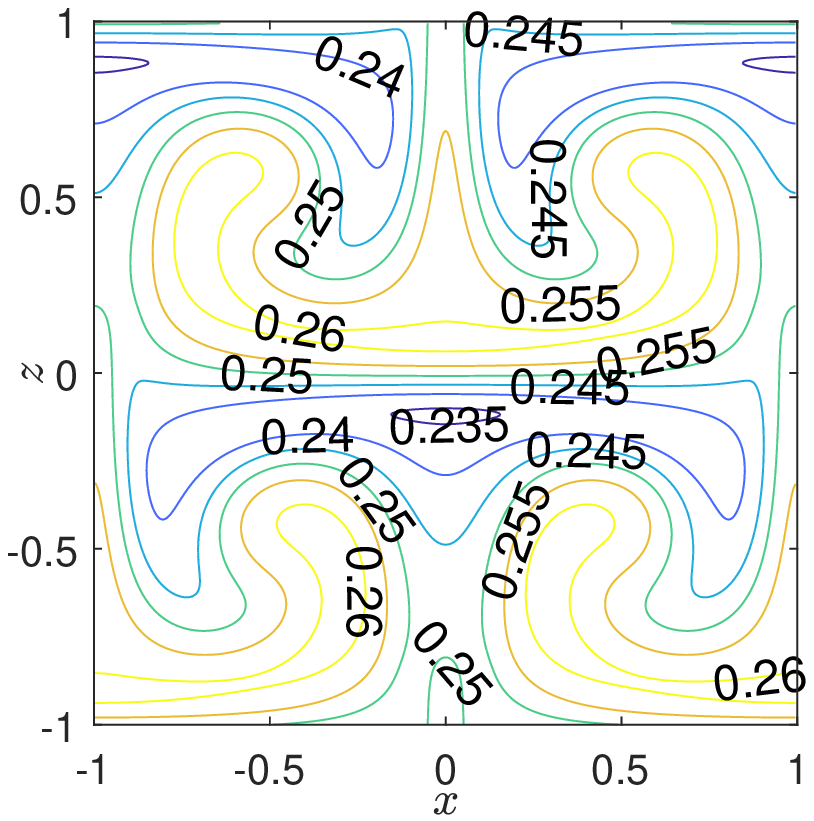}\label{fig:2D_beta0.21_B0.31_nfs}}
	\sidesubfloat[]{\includegraphics[width = 0.3 \columnwidth]{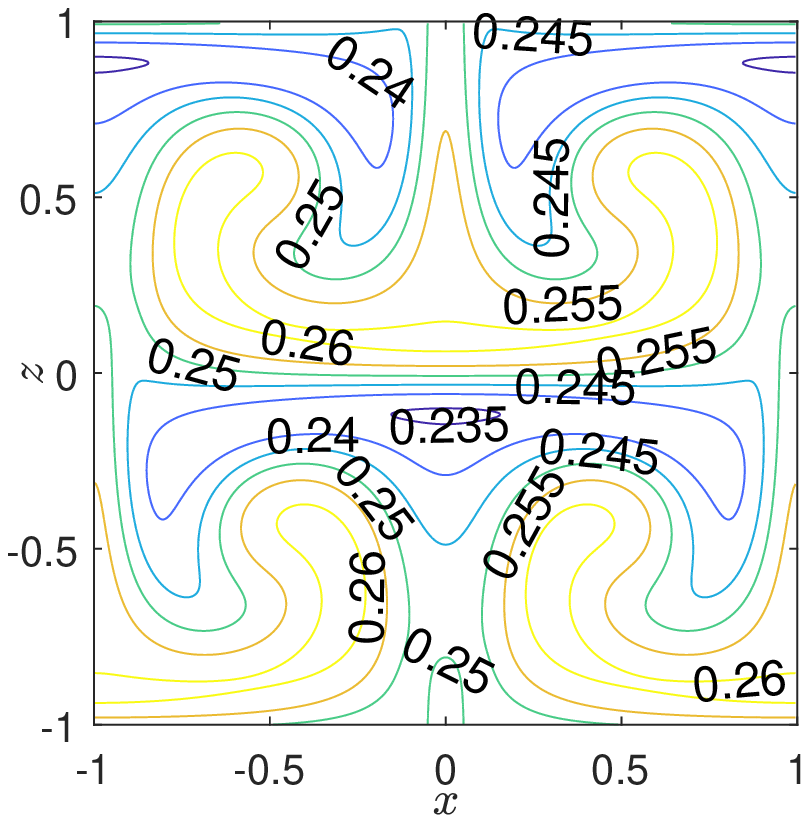}\label{fig:2D_beta0.21_B0.31_ngs}}
	\caption{Comparison of the steady-state number density of $(a,b)$ strongly and $(c,d)$ weakly gyrotactic suspensions in the convective flow (\ref{eq:conv_cells}) given by $(a,c)$ direct integration of (\ref{eq:smol}) and $(b,d)$ the local approximation. \label{fig:2D_ns}}
\end{figure}

In this example, we consider a steady two-dimensional Taylor-Green-type vortex flow in a periodic box  $(x,z) \in [-1,1] \times [-1,1]$ with the velocity field
\begin{equation}
	\mathbf{u}=(\sin{(\pi x)} \cos{(\pi z)},0,-\cos{(\pi x)}\sin{(\pi z)})^T. \label{eq:conv_cells}
\end{equation} 
The flow is also shown in figure \ref{fig:conv_vec}. The flow is introduced as a simple model to study mixing by a vortical flow and has been used to study gyrotactic motility in vortical flow in past works \citep[e.g.][]{Bearon2011,Durham2011}. We note that the exact Taylor-Green vortex flow will decay over time under the Navier-Stokes equation. However, it maintains its structure while it decays and satisfies the continuity equation. For simplicity, here we consider the steady vortex rather than the decaying one.

We note that in this inhomogeneous flow, both the GTD theory and the Fourier method can also provide the macrotransport properties over many periodic boxes if one treats the unit box as the local space (or the microscale) and solves the Smoluchowski equation within the box. However, neither the GTD theory nor the Fourier method can approximate to the transport of ABP number density within the box without fully solving the Smoluchowski equation. Hence, this example demonstrates the advantage of the local approximation over the previous method in providing an approximated equation of transport within the box (the microscale).

\begin{figure}
	\centering{}
	\sidesubfloat[]{\includegraphics[width = 0.3 \columnwidth]{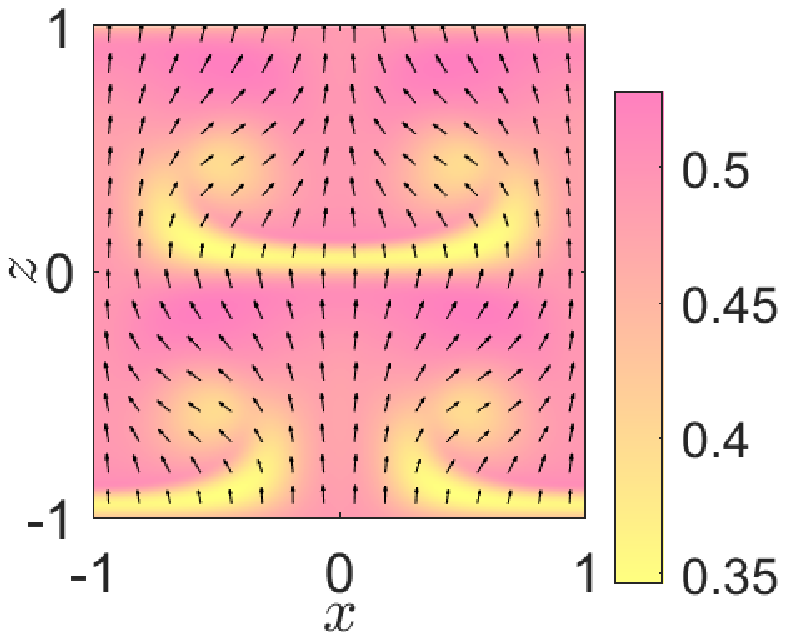}\label{fig:2D_beta2.2_B0.31_ef}}
	\sidesubfloat[]{\includegraphics[width = 0.3 \columnwidth]{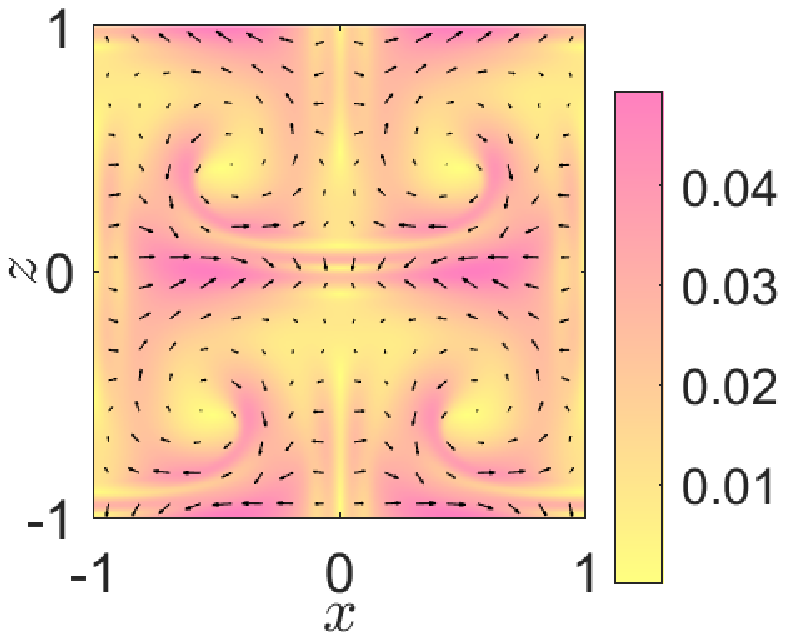}\label{fig:2D_beta2.2_B0.31_Vif}}
	\sidesubfloat[]{\includegraphics[width = 0.3 \columnwidth]{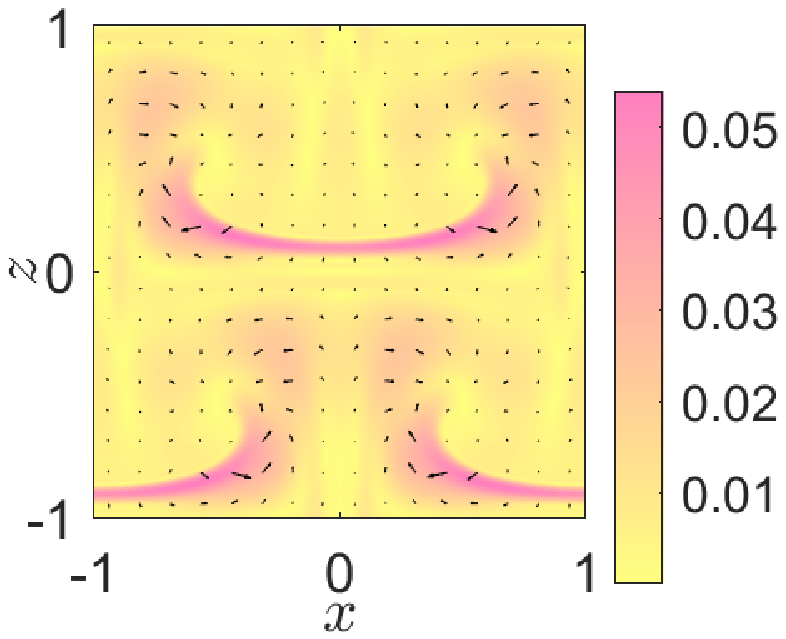}\label{fig:2D_beta2.2_B0.31_Vuf}}\\
	\sidesubfloat[]{\includegraphics[width = 0.3 \columnwidth]{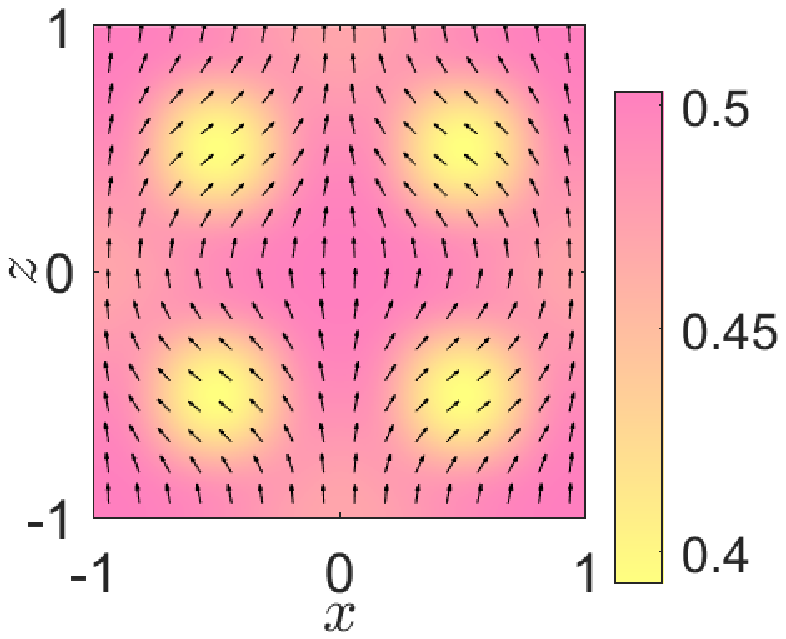}\label{fig:2D_beta2.2_B0.31_eg}}
	\sidesubfloat[]{\includegraphics[width = 0.3 \columnwidth]{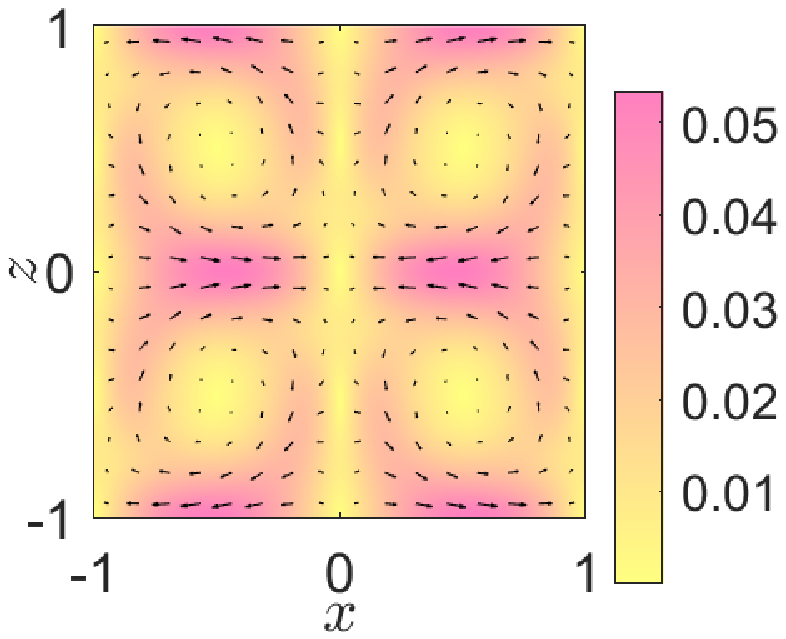}\label{fig:2D_beta2.2_B0.31_Vig}}
	\sidesubfloat[]{\includegraphics[width = 0.3 \columnwidth]{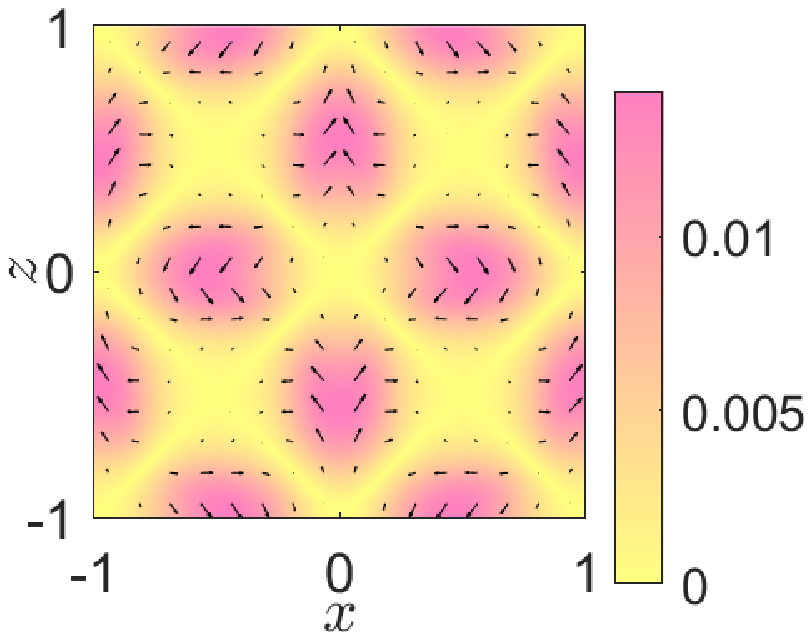}\label{fig:2D_beta2.2_B0.31_Vug}}
	\caption{Comparison of the drifts at the steady state. The plots show the values of $(a)$ $\pavg_f$, $(b)$ $\mathbf{V}_c$, $(c)$ $\mathbf{V}_\mathbf{u}$, $(d)$ $Pe_s\pavg_g$, $(e)$ $Pe_s\mathbf{V}_{g,c}$ and $(f)$ $Pe_s\mathbf{V}_{g,\mathbf{u}}$  calculated using the steady-state $f(\mathbf{x},\mathbf{p},\infty)$ and $g(\mathbf{x},\infty;\mathbf{p})$ of a suspension of strongly gyrotactic particles ($\beta=2.2,\alpha_0=0.31$). The suspension is subjected to the convective flow (\ref{eq:conv_cells}) with $\Pe_s=0.25$ and $\Pe_f=0.5$. \label{fig:2D_beta2.2_V}}
\end{figure}
\begin{figure}
	\centering{}
	\sidesubfloat[]{\includegraphics[width = 0.22 \columnwidth]{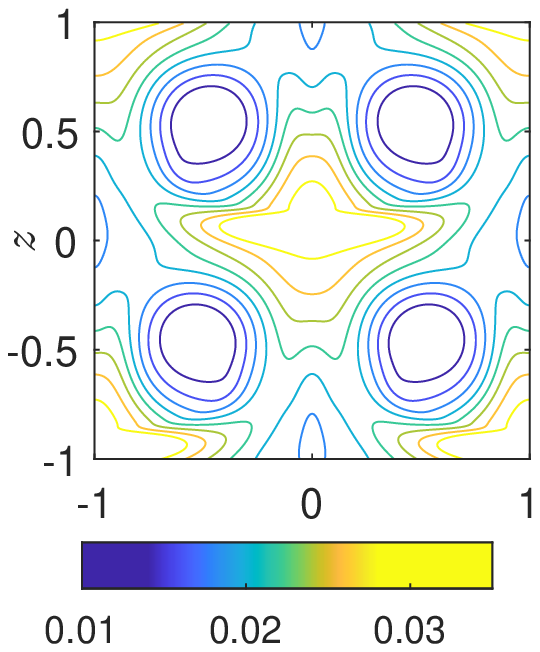}\label{fig:2D2.2beta_0.31B_Dxxf}}
	\sidesubfloat[]{\includegraphics[width = 0.22 \columnwidth]{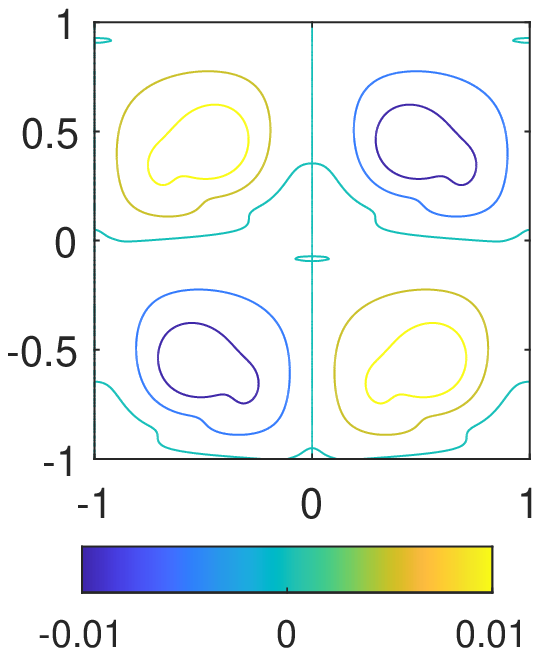}\label{fig:2D2.2beta_0.31B_Dxzf}}
	\sidesubfloat[]{\includegraphics[width = 0.22 \columnwidth]{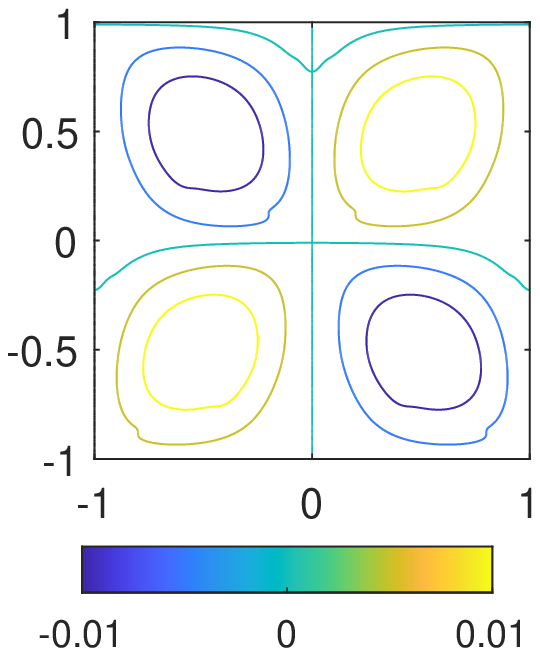}\label{fig:2D2.2beta_0.31B_Dzxf}}
	\sidesubfloat[]{\includegraphics[width = 0.22 \columnwidth]{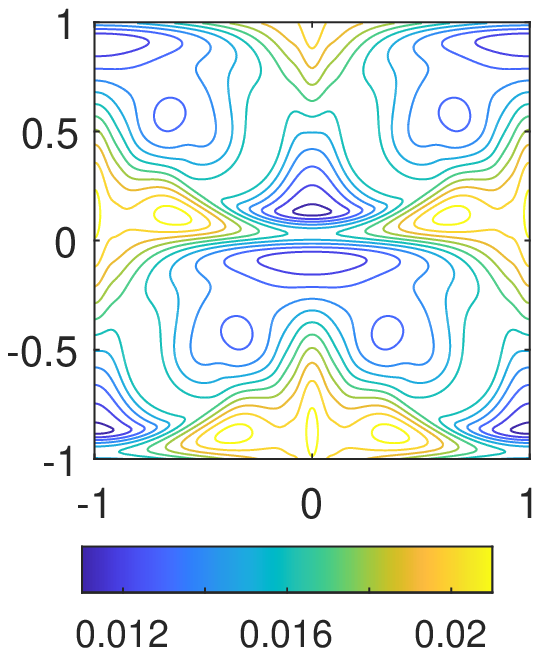}\label{fig:2D2.2beta_0.31B_Dzzf}}\\
	\sidesubfloat[]{\includegraphics[width = 0.22 \columnwidth]{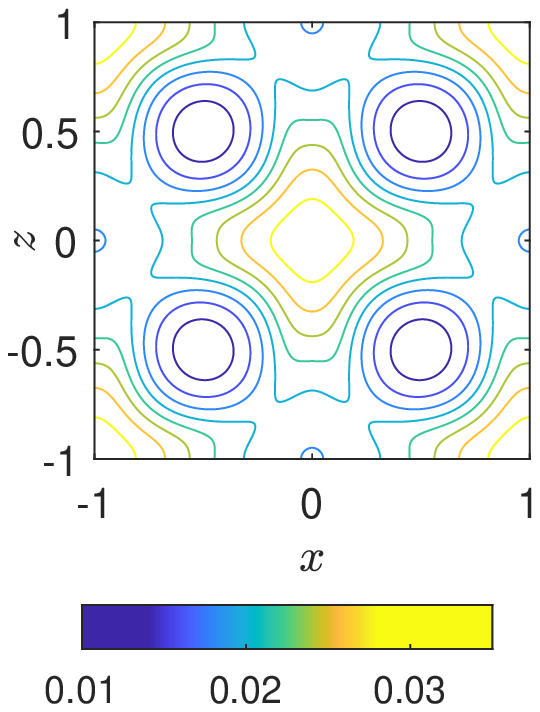}\label{fig:2D2.2beta_0.31B_Dxxg}}
	\sidesubfloat[]{\includegraphics[width = 0.22 \columnwidth]{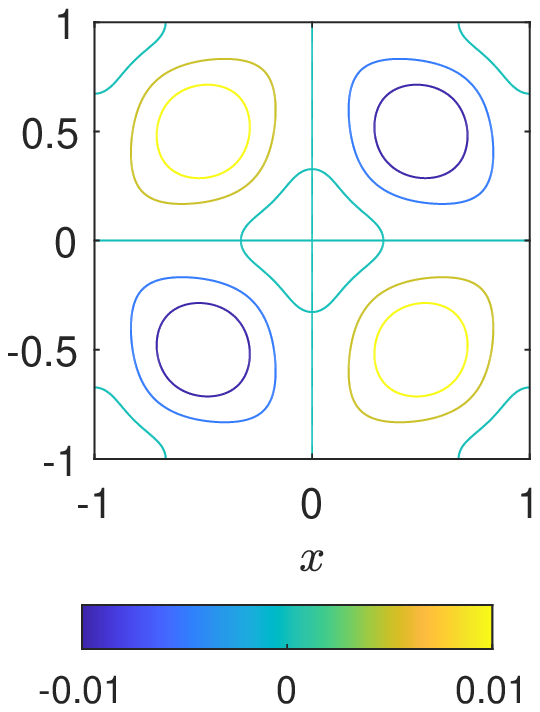}\label{fig:2D2.2beta_0.31B_Dxzg}}
	\sidesubfloat[]{\includegraphics[width = 0.22 \columnwidth]{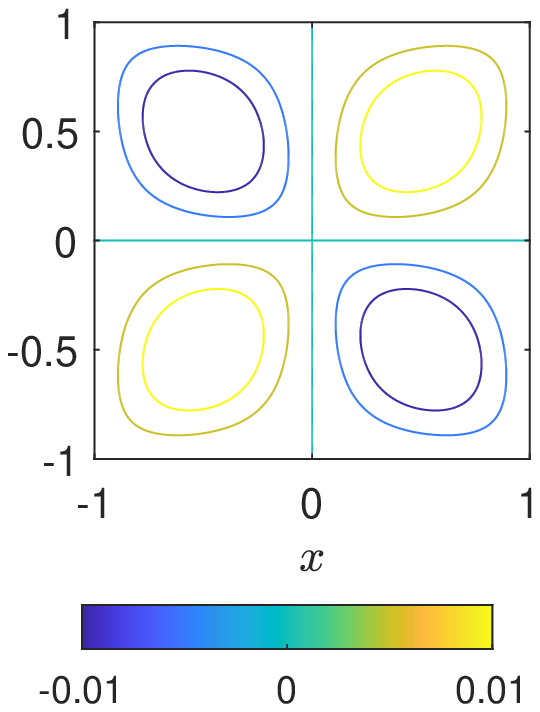}\label{fig:2D2.2beta_0.31B_Dzxg}}
	\sidesubfloat[]{\includegraphics[width = 0.22 \columnwidth]{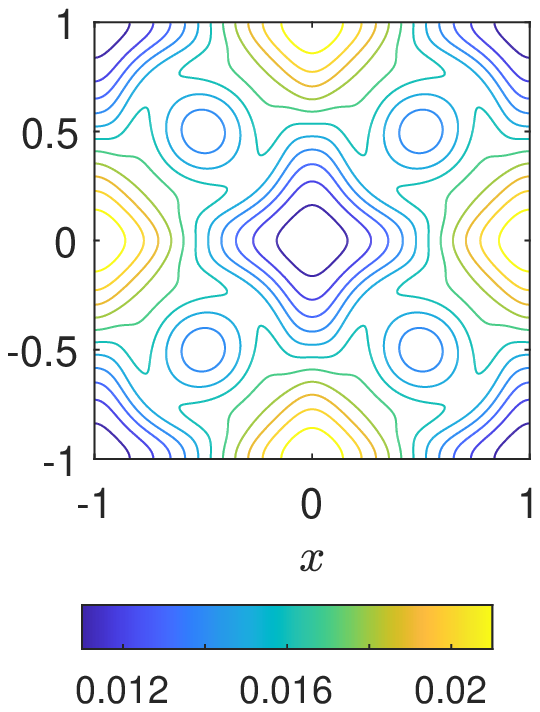}\label{fig:2D2.2beta_0.31B_Dzzg}}	
	\caption{Comparison of $(a-d)$ the dispersion $\mathsfbi{D}_c$ and $(e-h)$ its approximation $Pe_s\mathsfbi{D}_{g,c}$ at the steady state. The plots show the $(a,e)$ $xx$, $(b,f)$ $xz$, $(c,g)$ $zx$, $(d,h)$ $zz$ components of dispersion tensors. The suspension is strongly gyrotactic ($\beta=2.2,\alpha_0=0.31$) and subjected to the convective flow (\ref{eq:conv_cells}) with $\Pe_s=0.25$ and $\Pe_f=0.5$.\label{fig:2D_beta2.2_D}}
\end{figure}

\Cref{fig:2D_ns} shows the comparisons between the number density from the full solution of the Smoluchowski equation and the local approximation in suspensions of strongly and weakly gyrotactic non-spherical particles. We note that $\mathbf{V}_{u}$ is no longer vanishing like in the previous examples, as the prescribed flow field is no longer parallel. In all cases, the predictions by the local approximation model appear to be good compare with the full solution to the Smoluchowski equation. 

Nevertheless, the local approximation model performs a little more poorly for the strongly gyrotactic suspension (figures \ref{fig:2D_ns}$a,b$) than for the weakly gyrotactic suspension (figures \ref{fig:2D_ns}$c,d$),  similar to the examples in \S\ref{sec:example_upright} and \S\ref{sec:example_HS}. In particular, the approximations for $\mathbf{V}_{u}$, $\mathbf{V}_{c}$, $\mathsfi{D}_{xx,c}$ and $\mathsfi{D}_{zz,c}$ near $z=0$ and $z=\pm 1$ exhibit such a behaviour (see \cref{fig:2D_beta2.2_V,fig:2D_beta2.2_D}). 
Nevertheless, as shown in \cref{fig:2D_beta2.2_V}, the drift remains dominated by $\pavg_g$, while the dispersion $\mathsfbi{D}_c$ is well approximated by $\Pe_s \mathsfbi{D}_{g,c}$ in most of the domain (\cref{fig:2D_beta2.2_D}). Therefore, the approximation remains qualitatively close to the exact result, despite the local inaccuracy. 

\begin{figure}
	\centering{}
	\sidesubfloat[]{\includegraphics[width = 0.28 \columnwidth]{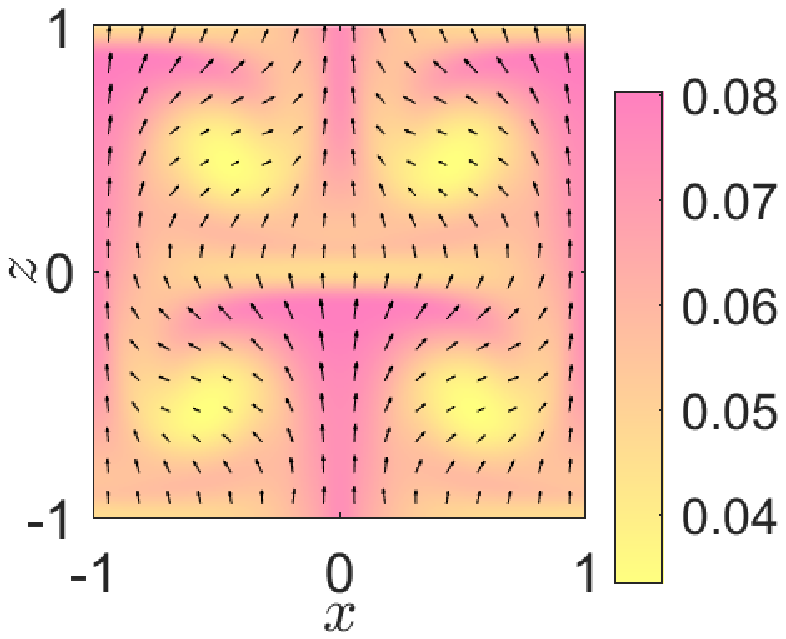}\label{fig:2D_beta0.21_B0.31_ef}}
	\sidesubfloat[]{\includegraphics[width = 0.28 \columnwidth]{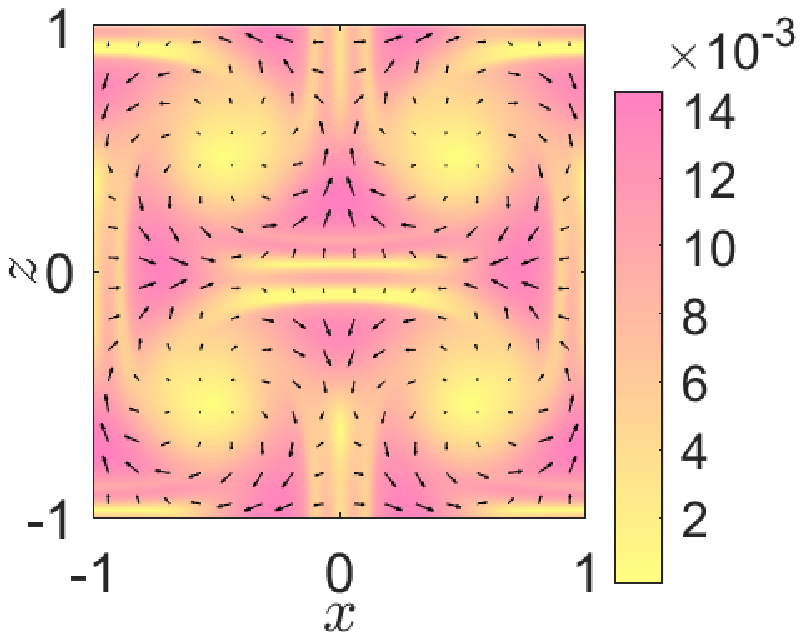}\label{fig:2D_beta0.21_B0.31_Vif}}
	\sidesubfloat[]{\includegraphics[width = 0.28 \columnwidth]{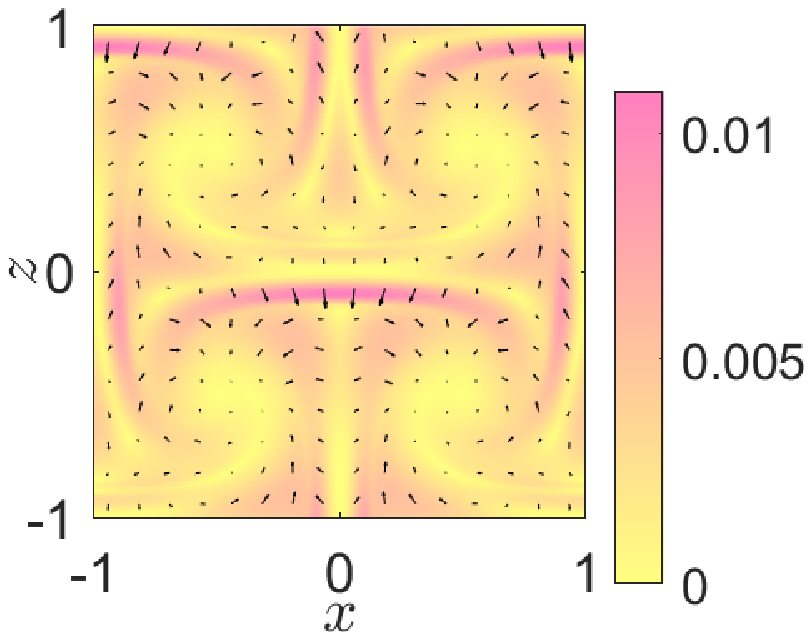}\label{fig:2D_beta0.21_B0.31_Vuf}}\\
	\sidesubfloat[]{\includegraphics[width = 0.28 \columnwidth]{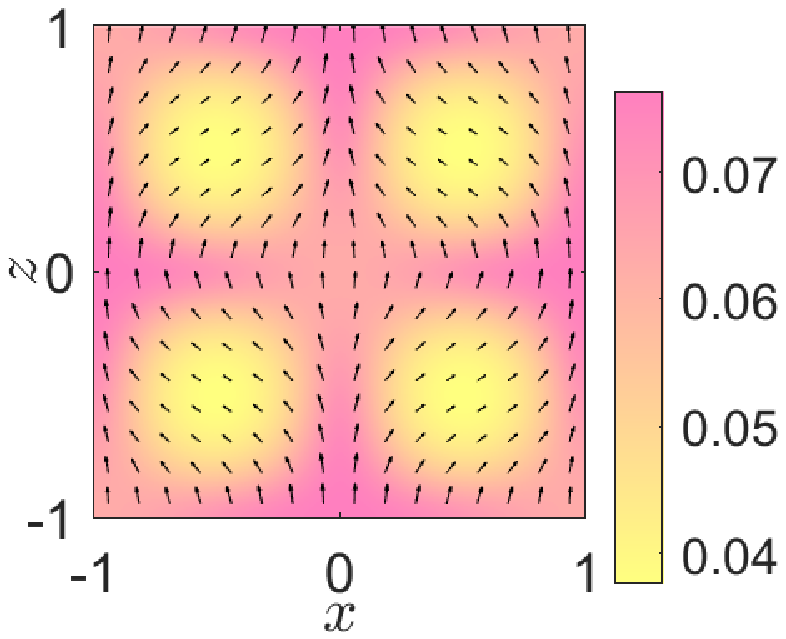}\label{fig:2D_beta0.21_B0.31_eg}}
	\sidesubfloat[]{\includegraphics[width = 0.28 \columnwidth]{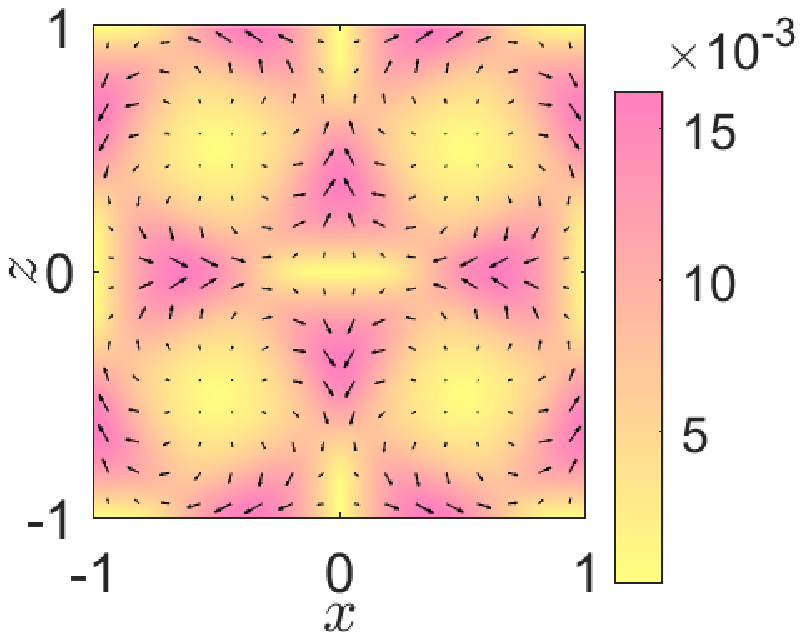}\label{fig:2D_beta0.21_B0.31_Vig}}
	\sidesubfloat[]{\includegraphics[width = 0.28 \columnwidth]{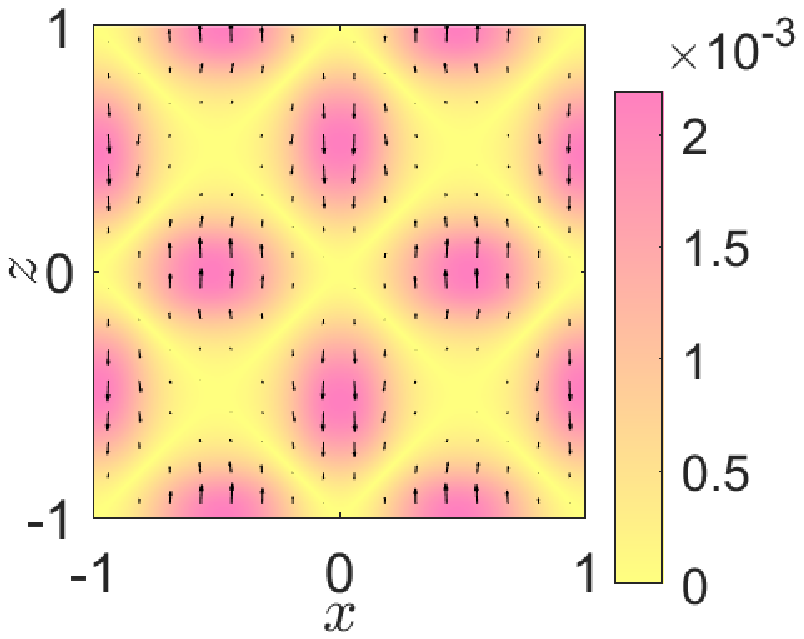}\label{fig:2D_beta0.21_B0.31_Vug}}
	\caption{Comparison of the drifts at the steady state. The plots show the values of $(a)$ $\pavg_f$, $(b)$ $\mathbf{V}_c$, $(c)$ $\mathbf{V}_\mathbf{u}$, $(d)$ $Pe_s\pavg_g$, $(e)$ $Pe_s\mathbf{V}_{g,c}$ and $(f)$ $Pe_s\mathbf{V}_\mathbf{g,u}$ calculated using the steady-state $f(\mathbf{x},\mathbf{p},\infty)$ and $g(\mathbf{x},\infty;\mathbf{p})$ of a suspension of weakly gyrotactic particles ($\beta=0.21,\alpha_0=0.31$). The suspension is subjected to the convective flow (\ref{eq:conv_cells}) with $\Pe_s=0.25$ and $\Pe_f=0.5$. \label{fig:2D_beta0.21_V}}
\end{figure}

\begin{figure}
	\centering{}
	\sidesubfloat[]{\includegraphics[width = 0.22 \columnwidth]{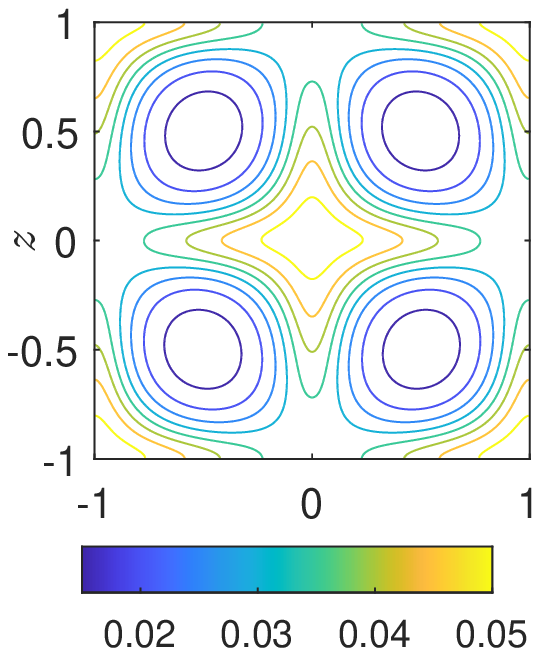}\label{fig:2D0.21beta_0.31B_Dxxf}}
	\sidesubfloat[]{\includegraphics[width = 0.22 \columnwidth]{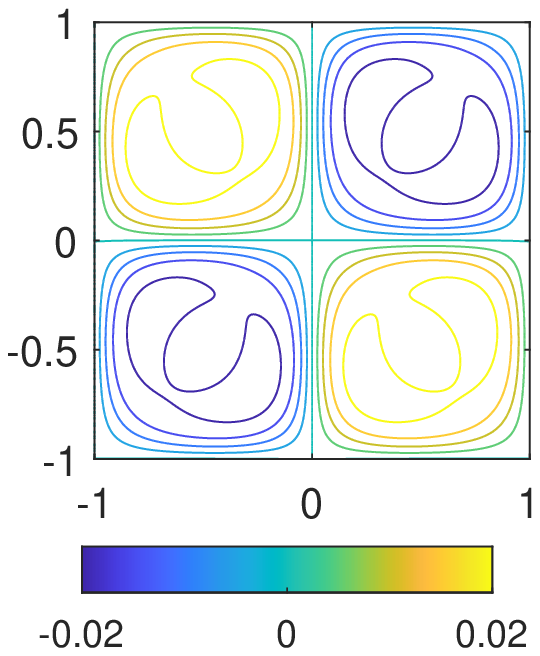}\label{fig:2D0.21beta_0.31B_Dxzf}}
	\sidesubfloat[]{\includegraphics[width = 0.22 \columnwidth]{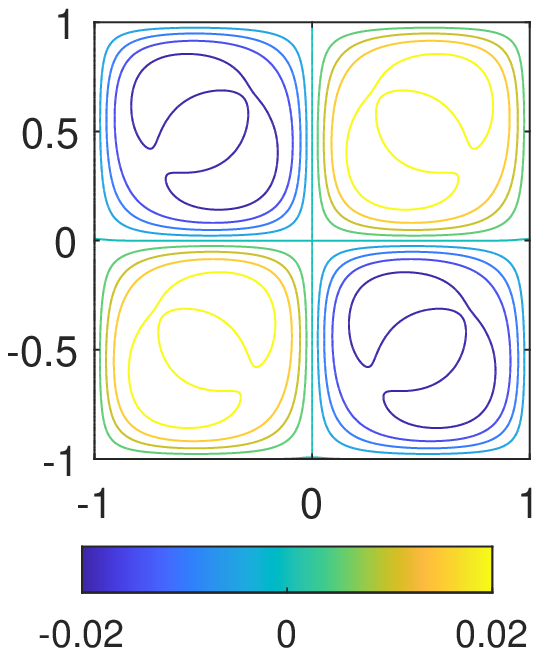}\label{fig:2D0.21beta_0.31B_Dzxf}}
	\sidesubfloat[]{\includegraphics[width = 0.22 \columnwidth]{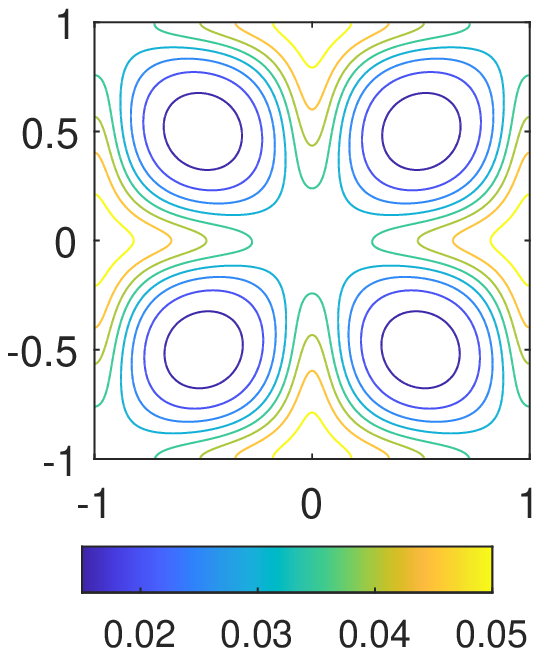}\label{fig:2D0.21beta_0.31B_Dzzf}}\\
	\sidesubfloat[]{\includegraphics[width = 0.22 \columnwidth]{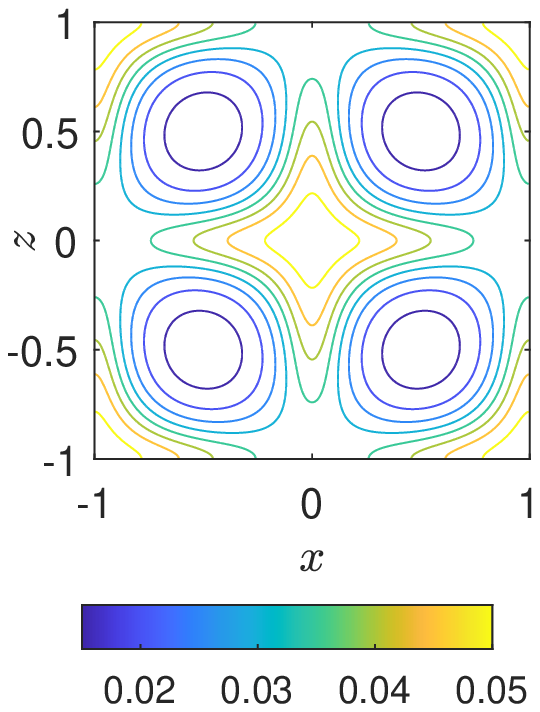}\label{fig:2D0.21beta_0.31B_Dxxg}}
	\sidesubfloat[]{\includegraphics[width = 0.22 \columnwidth]{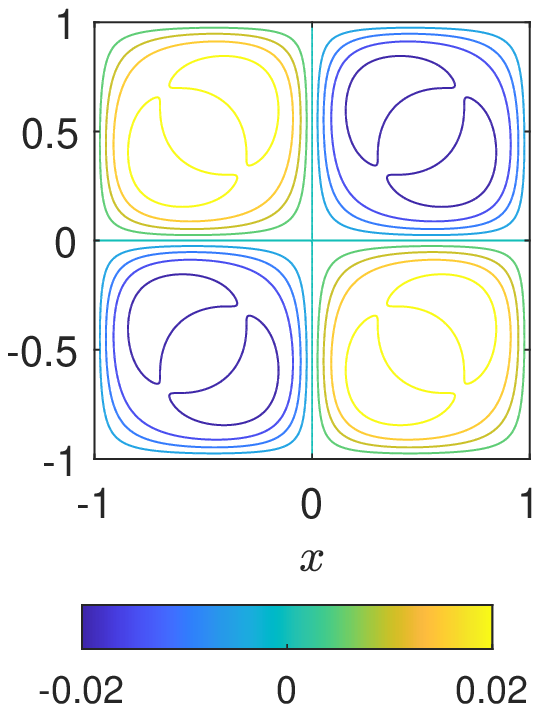}\label{fig:2D0.21beta_0.31B_Dxzg}}
	\sidesubfloat[]{\includegraphics[width = 0.22 \columnwidth]{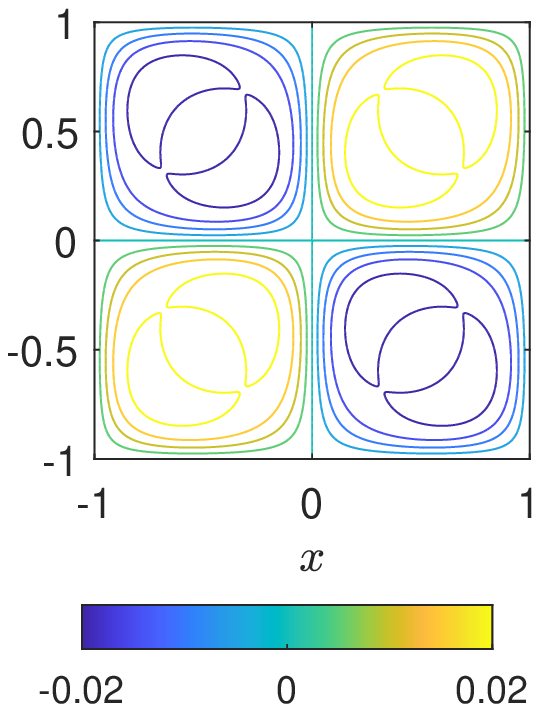}\label{fig:2D0.21beta_0.31B_Dzxg}}
	\sidesubfloat[]{\includegraphics[width = 0.22 \columnwidth]{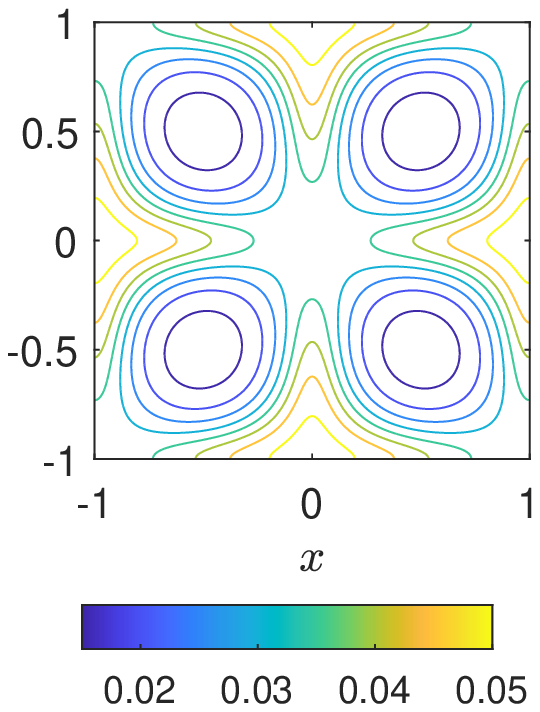}\label{fig:2D0.21beta_0.31B_Dzzg}}	
	\caption{Comparison of $(a-d)$ the dispersion $\mathsfbi{D}_c$ and $(e-h)$ its approximation $Pe_s\mathsfbi{D}_{g,c}$ at the steady state. The plots show the $(a,e)$ $xx$, $(b,f)$ $xz$, $(c,g)$ $zx$, $(d,h)$ $zz$ components of dispersion tensors. The suspension is weakly gyrotactic ($\beta=0.21,\alpha_0=0.31$) and subjected to the convective flow (\ref{eq:conv_cells}) with $\Pe_s=0.25$ and $\Pe_f=0.5$.  \label{fig:2D_beta0.21_D}}
\end{figure}

In the weakly gyrotactic suspension, the local approximation more accurately captures both the number density (\cref{fig:2D_beta0.21_B0.31_ngs}) and the dispersions (\cref{fig:2D_beta0.21_D}), consistent with the previous examples shown. Although the approximations to the drifts $\mathbf{V}_{c}$ and $\mathbf{V}_{u}$ are not as accurate, they are relatively negligible compared to $\pavg_g$ (\cref{fig:2D_beta0.21_V}). 

\section{Discussion: Physical implication of the transformation \label{sec:Discussion}}\label{sec:Discussion_trans}
This work sets out to seek a model transport equation that can predict the number density given by the Smoluchowski equation without solving the equation directly.
To achieve the goal, in \S\ref{sec:exact_transform}, we have shown how the Smoluchowski equation (\ref{eq:smol}) can be transformed into a transport equation by expanding $\pavg_f n$ in (\ref{eq:smol_inte}) in terms of $\pavg_g n$ and other drifts $\mathbf{V}_\star n$ and dispersions/diffusions $\mathsfbi{D}_\star \cdot \delx n$. We note that the averaged orientation $\pavg_g$ of individual particles is often taken directly as the only drift in the previous quasi-homogeneous approximation model of the GTD theory.
However, as $\pavg_f$ is formally expanded thought the transformation, it now clearly reveals how $\pavg_g$ is different from $\pavg_f$.

To better show the implication of this transformation, here we rewrite the procedures in \S\ref{sec:exact_transform} for the examples in \S\ref{sec:example_upright} and \S\ref{sec:example_HS}, which are under the assumptions of a parallel flow. For simplicity, we further assume $D_T=0$. Therefore, we can rewrite (\ref{eq:smol_inte}) as 
\begin{equation}
	\partial_t n(\mathbf{x},t)+  \bnabla_{\mathbf{x}} \bcdot [(\Pe_s \pavg_f(\mathbf{x},t) \, n(\mathbf{x},t)] = 0, \label{eq:smol_inte_noDT}
\end{equation}
in which  $\pavg_f n$ can be expanded through (\ref{eq:smol_full_coarseG}) or   
\begin{equation}
	\pavg_f= \pavg_g - \mathbf{V}_{c} - \mathbf{V}_{\partial t} - \mathsfbi{D}_c \bcdot \frac{\delx n}{n}, \label{eq:pall}
\end{equation}
into the transport equation
\begin{equation}
	\partial_t n + \Pe_s \delx \bcdot \left[(\pavg_g-\mathbf{V}_{c}- \mathbf{V}_{\partial t}) n \right] = \Pe_s \delx \bcdot \mathsfbi{D}_{c} \bcdot \delx n. \label{eq:transport_all}
\end{equation}
Note that equations (\ref{eq:smol_inte_noDT}-\ref{eq:transport_all}) are equivalent to (\ref{eq:smol_inte_x}-\ref{eq:transport_x}) in the full three-dimensional coordinates. Now, equation (\ref{eq:smol_inte_noDT}) and the rewritten equation (\ref{eq:transport_all}) yield two different interpretations of ABP transport. In (\ref{eq:smol_inte_noDT}), particles are purely advected by the motility flux  $\Pe_s \pavg_f n$, which is the ensemble-averaged flux of particles coming in and out of a certain control volume at position $\mathbf{x}$ due to the motility of the particle. In this sense, $\Pe_s \pavg_f n$ may be interpreted as the `Eulerian' motility flux. 
The flux depends on the orientational and spatial distribution of particles inside and at the vicinity of the control volume. However, in (\ref{eq:transport_all}), the average Eulerian motility flux $\Pe_s \pavg_f n$ is decomposed into the flux from the average motility of individual particles $\Pe_s \pavg_g n$, the advective flux due to unsteadiness in particles' orientational dynamics $-\Pe_s \mathbf{V}_{\partial t} n$, the shear trapping flux $-\Pe_s \mathbf{V}_{c} n$ and the dispersion flux $-\Pe_s \mathsfbi{D}_{c} \bcdot \delx n$.

It is evident from (\ref{eq:pall}) that the average Eulerian motility flux $\Pe_s \pavg_f n$ is different from the flux obtained using the average motility of individual particles, $\Pe_s \pavg_g n$. However, it might also be counterintuitive at first glance to decipher their differences.
Here, the average motility of individual particles $\Pe_s \pavg_g$ is defined as the ensemble average of the self-propelling velocity of individual particles when subject to the local velocity gradient and other local factors that may influence their orientation (e.g. taxes). The average motility of individual particles $\Pe_s \pavg_g$ is based on the average orientation of individual particles $\pavg_g$, which is calculated from the homogeneous solution ($g$) to the operator $\mathcal{L}_\mathbf{p}$, representing the orientational dynamics of individual particles. It is a function of the local velocity gradient and the particles' property only, and is explicitly independent of $(\mathbf{x},t)$. In other words, $\pavg_g$ is calculated by decoupling the orientational dynamics ($\mathcal{L}_\mathbf{p}$) from the spatio-temporal dynamics of the Smoluchowski equation. As such, the resulting average motility $\Pe_s \pavg_g$ may be viewed to provide a `Lagrangian' description of each individual particle's motility after being averaged in the local $\mathbf{p}$-space. 

By contrast, the average Eulerian motility flux $\Pe_s \pavg_f n$ considers the distribution of particles at the nearby location in the $(\mathbf{x},t)$-space. It is the result of directly averaging the particles' motility $\Pe_s \mathbf{p} \Psi$ in the Smoluchowski equation (\ref{eq:smol}).
Physically, (\ref{eq:pall}) implies that the averaged Lagrangian motility of individual particles $\Pe_s \pavg_g$ only contributes to part of the overall Eulerian drift $\Pe_s \pavg_f$. It also reveals that the particle dispersion flux $\Pe_s \mathsfbi{D}_c \bcdot \delx n$ originates from the Eulerian motility flux $\Pe_s \pavg_f n$.
The other fluxes from drifts and dispersions arising from the interaction between the orientational dynamics ($\mathcal{L}_\mathbf{p}$) and the rest of the Smoluchowski equation are also included in the average Eulerian motility flux $\Pe_s \pavg_f n$. 
For example, it includes the effect of the different orientation distribution at the nearby location, which gives rise to the extra shear trapping flux $-\Pe_s \mathbf{V}_c n$, even when the average motility $\Pe_s \pavg_g$ is zero (as demonstrated in \S\ref{sec:V_compare}, figure \ref{fig:DVpavg_nongyro}). It also includes the effect of the changing orientation over time 
described by the extra flux $-\Pe_s \mathbf{V}_{\partial t}n$. 
Extending the decomposition to a more general ABP suspension, the passive advection and translation diffusion of particles also give rise to extra drifts and dispersion through the particles' motility, and they emerge with $\mathbf{V}_u$, $\mathbf{V}_{D_T}$ and $\mathsfbi{D}_{D_T}$ (see also table \ref{tab:phys_smol_coarse}). 

\section{Concluding remarks \label{sec:conclusion}}
In this study, we have proposed a new method to reduce the Smoluchowski equation into a simpler transport equation. The Smoluchowski equation governs the statistics of the position and orientation of biased or non-biased ABPs, whose orientational trajectories are described by the Jeffery orbit in the presence of rotational random noise. The framework is directly applicable to dilute suspensions of biased or non-biased ABPs in a large-scale system with strong flow, such as microalgae in the ocean. 
It can also be extended to the flow regime where the long-range hydrodynamic contribution of swimming motion of individual particles can be represented by averaged forces and stress tensors \citep[e.g.][]{Batchelor1970,Hinch1972a,Hinch1972,Pedley1990}.

We have presented a method to transform the Smoluchowski equation into a transport equation exactly for a given flow field. The method involves decomposing the average Eulerian motility flux $\Pe_s \pavg_f n$ at a fixed location into the flux from the average Langrangian motility flux of individual particles $\Pe_s \pavg_g n$ and other contributions. The transformation has shown that $\Pe_s \pavg_g$ is different from $\Pe_s \pavg_f$ and only constitutes part of $\Pe_s \pavg_f$. The transformation also unveils the explicit form of the other drift and dispersion terms contributing to the overall average Eulerian motility. These terms include the shear trapping drift $\mathbf{V}_c$ and the particle dispersion $\mathsfbi{D}_c$ due to rotational diffusion.
In addition, we have also discovered the drift $\mathbf{V}_{\partial t}$ due to the interaction between the unsteadiness in orientation and the orientational dynamics itself, the drift $\mathbf{V}_{D_T}$ and dispersion $\mathsfbi{D}_{D_T}$ arise from the interaction between translational diffusion and the orientational dynamics, and the drift $\mathbf{V}_{u}$ from the interaction between passive advection of orientational distribution and the orientational dynamics.

Although these new physical drifts and dispersions revealed by the transformation are easily interpretable in a transport equation, they cannot be directly used as a model due to the prerequisite of first obtaining $\Psi(\mathbf{x},\mathbf{p},t)$ by solving the Smoluchowski equation directly. In this regard, this work has presented a new model based on the local approximation of the transformation, which only relies on the local flow information instead of the global flow configuration. By assuming that the time scale of the orientational dynamics is much faster than that of the spatial dynamics (or that the ABP run length is much smaller than the characteristic lengthscale of the flow field), we have approximated the orientational space probability density function $f(\mathbf{x},\mathbf{p},t)=\Psi/n$ by the homogeneous solution $g(\mathbf{x},t;\mathbf{p})$ of the orientational space operator $\mathcal{L}_p$, thereby circumventing the need to solve for $\Psi$.
The approximation gives the same shear trapping drift $\mathbf{V}_{g,c}$ and particle dispersion $\mathsfbi{D}_{g,c}$ as that of \cite{Bearon2015} and \cite{Vennamneni2020} when the particles have no taxes and diffusion, but it can also be extended to particles with biased motility (i.e.~taxes) or translational diffusion.

The numerical examples of suspensions in horizontal and vertical shear flows showed that the new model works reasonably well in approximating the full solution of the Smoluchowski equation, especially when the particles are weakly or not biased. The model can capture the shear trapping of non-spherical motile particles because of its applicability in inhomogeneous flow. It can also capture the more dominant flux due to biased motility, which was not accounted for in previous methods for unbiased ABPs \citep[e.g.][]{Bearon2015,Vennamneni2020}. Meanwhile, the examples in the periodic convective cells demonstrated the general applicability of the local approximation in complex flow.
The local approximation has yielded accurate predictions of the number density in weakly gyrotactic suspensions. Although the approximation is not as accurate in strongly gyrotactic suspensions, the result remains qualitatively satisfactory. Overall, this work has shown that the local approximation method has maintained its accuracy in most of the cases considered while being applicable to more general flows.

As pointed out in a recent review \citep{Bees2020}, there is a gap between complex models of individual particles and their equivalent continuum modelling at the population level. 
The general applicability of the presented method for the transport of biased or non-biased ABPs in any flow field is perhaps the most important result especially from a practical perspective. In our numerical examples, we can also see that the presented method can quite accurately approximate the number density from the full solution of the Smoluchowski equation. While the presented examples focused mainly on gyrotactic ABPs, the framework presented can also be extended to other types of taxes, such as phototaxis and chemotaxis, as well as to other types of particle motions, such as the orientation-dependent sedimentation of elongated particles \cite[e.g.][]{Ardekani2017,Clifton2018,Lovecchio2019}.
Hence, the potential application of the framework presented in this work is vast. 

However, the current framework needs further developments and analysis. In particular, more work can be done to improve the inaccuracy in strongly gyrotactic suspensions, where the strong gradient in the number density might affect the accuracy of the approximated drifts and dispersions. For example, the accuracy of the local approximation can be improved by truncating the model at a higher order of $\epsilon$ (see Appendix \ref{app:asymp_details}). Alternatively, perhaps one can also consider applying a different asymptotic limit to the transformation in \S\ref{sec:exact_transform} to derive a different model.
Beyond the examples with periodic boundary conditions shown in this work, one also need a good boundary condition at the wall. For example, \cite{Ezhilan2015} have demonstrated the important role of translational diffusion $D_T$ in the wall accumulation near a no-flux boundary, which this work has yet to demonstrate. On the other hand, the microscopic interactions between the wall and individual ABPs remain a subject of future work. Even with the knowledge of microscopic interactions between the particles and the wall, translating the interactions into suitable boundary conditions at the continuum level remains an important challenge. To this end, the recent work by \cite{Chen2021} offers some insight into how one can account for the non-trivial and shape-dependent nematic alignment with the wall. 

\backsection[Supplementary data]{\label{SupMat}Supplementary movies are available online.}

\backsection[Funding]{R.B. acknowledges support from the Engineering and Physical Sciences Research Council (EP/S033211/1, `Shape, shear, search \& strife; mathematical models of bacteria'). L.F. is funded by the President's PhD Scholarship of Imperial College London.}

\backsection[Declaration of interests]{The authors report no conflict of interest.}

\backsection[Author ORCID]{L. Fung, https://orcid.org/0000-0002-1775-5093;\\ Y. Hwang, https://orcid.org/0000-0001-8814-0822}

\backsection[Author contributions]{R.B. had the original idea of the transformation technique under an asymptotic approximation. L.F. and Y.H. developed the theory and L.F. performed the simulations. All authors contributed to reaching conclusions and in writing the paper.}

\appendix
\section{Derivation of the local approximation \label{app:asymp_details}}
Following the expansion of $\Psi=\Psi^{(0)}+ \epsilon \Psi^{(1)} + \epsilon^2 \Psi^{(2)}+...$ in \S\ref{sec:asymp}, we substitute the expansion into the Smoluchowski equation (\ref{eq:smol}) and yield the following set of equations at successive orders of $\epsilon$: 
\begin{subequations}
	\begin{eqnarray}
\mathcal{O}(1) & : & \partial_t \Psi^{(0)} + \mathcal{L}_p \Psi^{(0)} =0; \label{eq:asymp_0}\\
 \mathcal{O}(\epsilon) & : & \partial_T \Psi^{(0)} + \mathbf{p} \bcdot \bnabla_{\mathbf{x}} \Psi^{(0)} + \tilde{\Pe}_f \mathbf{u} \bcdot \bnabla_{\mathbf{x}} \Psi^{(0)} + \partial_t \Psi^{(1)} + \mathcal{L}_p \Psi^{(1)} =\tilde{D}_T \nabla_x^2 \Psi^{(0)}; \label{eq:asymp_1}\\
\mathcal{O}(\epsilon^2) & : & \partial_T \Psi^{(1)} + \mathbf{p} \bcdot \bnabla_{\mathbf{x}} \Psi^{(1)} + \tilde{\Pe}_f \mathbf{u} \bcdot \bnabla_{\mathbf{x}} \Psi^{(1)} + \partial_t \Psi^{(2)} + \mathcal{L}_p \Psi^{(2)} =\tilde{D}_T \nabla_x^2 \Psi^{(1)}; \mbox{etc..} \label{eq:asymp_2}
	\end{eqnarray} \label{eq:asymp}
\end{subequations}
Integrating over $\mathbf{p}$-space, (\ref{eq:asymp}) becomes
\begin{subequations}
\begin{eqnarray}
\mathcal{O}(1) & : & \partial_t n^{(0)} =0; \label{eq:asymp_n_0}\\
\mathcal{O}(\epsilon) & : & \partial_T n^{(0)} +\partial_t n^{(1)} + \bnabla_{\mathbf{x}} \bcdot \left[ (\tilde{\Pe}_f \mathbf{u}+\pavg^{(0)}) n^{(0)} \right]  =\tilde{D}_T \nabla_x^2 n^{(0)}; \label{eq:asymp_n_1}\\
\mathcal{O}(\epsilon^2) & : & \partial_T n^{(1)} + \partial_t n^{(2)} +  \bnabla_{\mathbf{x}} \bcdot \left[ (\tilde{\Pe}_f \mathbf{u}+\pavg^{(1)}) n^{(1)} \right]  =\tilde{D}_T \nabla_x^2 n^{(1)}; \mbox{etc..}  \label{eq:asymp_n_2}
\end{eqnarray} \label{eq:asymp_n}
\end{subequations}

At the transient time $t \gtrsim \mathcal{O}(1)$ and each order of $\epsilon$, we assume the time dependency of $\Psi^{(i)}$ in $\mathbf{p}$-space has reached quasi-equilibrium, while the time dependency of $\Psi^{(i)}$ in $\mathbf{x}$-space is slow. In other words, we assume that, at each order, $f^{(i)}$ is independent of $t$ as it has reached quasi-equilibrium and $n^{(i)}$ independent of $t$ because it only varies at the slow time scale $T$. 
Therefore, equation (\ref{eq:asymp_0}) now becomes
\begin{equation}
\mathcal{L}_p f^{(0)} =0, \label{eq:asymp_f0}
\end{equation} 
which implies that the leading-order orientational distribution $f^{(0)}$ takes the homogeneous solution of $\mathcal{L}_p(\mathbf{x},t)$ as the solution, i.e. $f^{(0)}=g(\mathbf{x},T;\mathbf{p})$.
Meanwhile, we multiply (\ref{eq:asymp_n_1}) by $f^{(0)}$ and subtract it from (\ref{eq:asymp_1}). This operation is equivalent to the steps towards (\ref{eq:smol_subtracted}) in \S\ref{sec:exact_transform}. The operation yields
\begin{eqnarray}
& & n^{(0)}\partial_T f^{(0)}  \nonumber   \\
&+& (\tilde{\Pe}_f \mathbf{u} \bcdot \bnabla_{\mathbf{x}} f^{(0)} - \tilde{D}_T \nabla_x^2 f^{(0)}) n^{(0)} - 2 {D}_T (\bnabla_{\mathbf{x}} f^{(0)}) \bcdot (\bnabla_{\mathbf{x}} n^{(0)}) \nonumber \\
&+&(\mathbf{p} - \pavg^{(0)}) f^{(0)} \bcdot \bnabla_{\mathbf{x}} n^{(0)} + n^{(0)} (\mathbf{p}\bcdot \bnabla_{\mathbf{x}} f^{(0)} - f^{(0)} \bnabla_{\mathbf{x}}\bcdot \pavg^{(0)}) \nonumber \\
&+& n^{(1)}  \mathcal{L}_p f^{(1)} =0.\label{eq:asymp_1sub}
\end{eqnarray}
Now, (\ref{eq:asymp_1sub}) can be rewritten as
\begin{equation}\label{eq:asymp_substituted}
\left[\mathcal{L}_p (\mathbf{b}_{g,D_T} + \mathbf{b}_{g,c}) \right] \bcdot \bnabla_{\mathbf{x}} n^{(0)}
+ n^{(0)}  \mathcal{L}_p \left[f_{g,u}+f_{g,D_T}+f_{g,c}+f_{g,\partial T} \right]+ n^{(1)} \mathcal{L}_p f^{(1)}=0, 
\end{equation}
where $f_{g,\star}$ and $\mathbf{b}_{g,\star}$ are defined by (\ref{eq:f_b_g}). 
Equations (\ref{eq:f_b_g}) and (\ref{eq:asymp_substituted}) are the equivalent of (\ref{eq:f_b}) and (\ref{eq:smol_substituted}), respectively. We can then follow the same derivation as in \S\ref{sec:exact_transform}, which would lead to 
\begin{align}
& \partial_T n^{(1)}  + \bnabla_{\mathbf{x}} \bcdot \left[ (\pavg_{g}+\tilde{\Pe}_f \mathbf{u}) n^{(1)}  \right] \nonumber \\ 
& =  \tilde{D}_T \nabla^2_x n^{(1)}  + \bnabla_{\mathbf{x}} \bcdot  \left[ (\mathsfbi{D}_{g,c}+\mathsfbi{D}_{g,D_T}) \bcdot  \bnabla_{\mathbf{x}}  n^{(0)} + (\mathbf{V}_{g,u}+\mathbf{V}_{g,D_T}+\mathbf{V}_{g,c}+\mathbf{V}_{g,\partial T}) n^{(0)}  \right], \label{eq:asymp_coarse_semifinal}
\end{align}
where $\mathbf{V}_{g,\star}$ and $\mathsfbi{D}_{g,\star}$ are defined according to (\ref{eq:V_ast}-\ref{eq:D_ast}).

Now, equation (\ref{eq:asymp_coarse_semifinal}) is at $\mathcal{O}(\epsilon^2)$. If we are to recover how $n$ evolves over the long time $T$, we can recompose $\partial_T n=\partial_T n^{(0)}+\epsilon \partial_T n^{(1)}+...$, by summing up (\ref{eq:asymp_n_0}-\ref{eq:asymp_n_2}) with the corresponding  $\epsilon$ scaling while substituting (\ref{eq:asymp_n_2}) with (\ref{eq:asymp_coarse_semifinal}). Hence,
\begin{align}
& \Pe_s \partial_T n  + \bnabla_{\mathbf{x}} \bcdot \left[ (\Pe_s \pavg_{g}+{\Pe}_f \mathbf{u}) n \right] \nonumber \\ 
& \approx  {D}_T \nabla^2_x n + \Pe_s^2 \bnabla_{\mathbf{x}} \bcdot  \left[ (\mathsfbi{D}_{g,c}+\mathsfbi{D}_{g,D_T}) \bcdot \bnabla_{\mathbf{x}}  n^{(0)} + (\mathbf{V}_{g,u}+\mathbf{V}_{g,D_T}+\mathbf{V}_{g,c}+\mathbf{V}_{g,\partial T}) n^{(0)}  \right]. \label{eq:GTD_w_Vi}
\end{align}
Note that we have only included $n^{(0)}$ and $n^{(1)}$ when recomposing $n$ in this example as we are closing the problem at $\mathcal{O}(\epsilon^2)$. Therefore, (\ref{eq:GTD_w_Vi}) is accurate up to $\mathcal{O}(\epsilon^2)$. However, if we close the problem at a higher order, we can repeat a similar process from (\ref{eq:asymp_1sub}) to (\ref{eq:asymp_coarse_semifinal}) at a higher order. 

Here, we would argue that at the transient time $t \rightarrow \infty$, $\partial_t \approx \Pe_s \partial_T$ and $n \approx n^{(0)}$. Because (\ref{eq:GTD_w_Vi}) is accurate up to $\mathcal{O}(\epsilon^2)$ while replacing $\Pe_s^2 n^{(0)}$ with $\Pe_s^2 n$ would only introduce an error at $\mathcal{O}(\epsilon^3)$, the substitution of $n^{(0)}$ by $n$ would not have a tremendous impact on the accuracy of (\ref{eq:GTD_w_Vi}). Under these approximations, we recover the approximated equation (\ref{eq:asymp_coarse_final}).

\section{Analytical solution to a suspension of spherical gyrotactic active particles in a vertical flow}\label{app:analytical_sol}
 If the gyrotactic active particles in a vertical flow are spherical, the steady solution of (\ref{eq:smol}) can be written analytically as $\Psi(\mathbf{x},\mathbf{p},\infty)=n_s(x)f_s(\mathbf{p})$, where
 \begin{subequations}
 \begin{equation}
 f_s(\mathbf{p}) = \frac{\beta}{4 \pi \sinh{\beta}} \exp{(\beta \cos{\theta})}, \label{eq:gyro_analytical}
 \end{equation}{}
 and
 \begin{equation}
 n_s(x)=A \exp{(-\frac{\beta \Pe_f W(x)}{2 \Pe_s})}, \label{eq:smol_analytic_cellden}
 \end{equation}\label{eq:sphere_analytical}
\end{subequations}
where $A$ is the normalisation factor determined by the integral condition $\int_{-1}^{1} n(x) dx=1$. Equation (\ref{eq:sphere_analytical}) may also explain the results of \cite{Jiang2020}, who showed that the number density is strongly dependent on $\beta$ and the ratio between the two P\'{e}clet numbers ($\Pe_f/\Pe_s$).

If we substitute the corresponding parameters of this example into (\ref{eq:smol_full_coarseG}-\ref{eq:smol_coarse_final}), we can recover 
 \begin{equation}
 \Pe_s \mathsfi{D}_{xx,c} \partial_x n_s
 =n_s \psth{x}_{g}, \label{eq:coarseG_app1}
 \end{equation}
 which represents the equilibrium between a dispersion flux and the net drift that is responsible for gyrotactic focusing. Note that $V_{x,c}=0$ in this example because $f_s$ is independent of $\mathbf{x}$. Here, to recover $\mathsfi{D}_{xx,c}$, we can substitute $f_s(\mathbf{p})$ into (\ref{eq:bc}) to get
 \begin{equation}
	b_{x,c}(\mathbf{x};\mathbf{p})=-\frac{\Pe_s}{\beta S(x)} \left( f_s(\mathbf{p})-g(\mathbf{p}) \right),   
\end{equation}
and therefore
 \begin{eqnarray}
	\mathsfi{D}_{xx,c}  & = & -\frac{\Pe_s}{\beta S(x)}\left( \int_{S_p} f_s(\mathbf{p}) p_x d^2 \mathbf{p} - \int_{S_p} g(\mathbf{p}) p_x d^2 \mathbf{p} \right) \nonumber \\
	& = & \frac{\Pe_s}{\beta S(x)} \psth{x}_{g}.
\end{eqnarray}

\bibliographystyle{jfm}
\bibliography{references}

\end{document}